\documentclass[aps,prd,a4paper,onecolumn,amsmath,showpacs,superscriptaddress,nofootinbib,preprintnumbers,notitlepage]{revtex4-1}
 
\usepackage{verbatim}
\usepackage[T1]{fontenc}
\usepackage[utf8]{inputenc}
\usepackage[american]{babel}
\usepackage{epsfig}
\usepackage{graphicx,subcaption,caption}
\usepackage{pgfplots}
\usepackage{booktabs}
\usepackage{multirow}
\usepackage{dcolumn}
\usepackage{amsmath}
\usepackage{mathtools}
\usepackage{amsfonts}
\usepackage{amssymb}
\usepackage{epstopdf}
\usepackage{bm}
\usepackage{siunitx}
\usepackage{braket}
\usepackage{enumitem}
\usepackage{soul}
\usepackage{ulem}
\usepackage{color}
\usepackage{transparent}
\usepackage{pifont}
\usepackage[font={small}]{caption}

\definecolor{navyblue}{rgb}{0.0, 0.0, 0.5}
\definecolor{royalblue}{rgb}{0.25, 0.41, 0.88}
\definecolor{cadmiumgreen}{rgb}{0.0, 0.42, 0.24}
\definecolor{blue-violet}{rgb}{0.54, 0.17, 0.89}
\definecolor{darkviolet}{rgb}{0.58, 0.0, 0.83}
\definecolor{orange(colorwheel)}{rgb}{1.0, 0.5, 0.0}

\usepackage{hyperref}
\hypersetup{
    colorlinks=true, 
    linkcolor=royalblue, 
    citecolor=magenta}

\newcommand\be{\begin{equation}}
\newcommand\ee{\end{equation}}

\newcommand\bea{\begin{eqnarray}}
\newcommand\eea{\end{eqnarray}}

\renewcommand\[{\left[}
\renewcommand\]{\right]}

\usepackage{booktabs}
\usepackage{multirow}
\usepackage{dcolumn}
\usepackage{colortbl}

\definecolor{magenta(process)}{rgb}{1.0, 0.0, 0.56}

\definecolor{darkspringgreen}{rgb}{0.09, 0.45, 0.27}

\definecolor{royalblue(web)}{rgb}{0.25, 0.41, 0.88}

\begin{document}

\title{Warm inflation with bulk viscous pressure for different solutions of an anisotropic universe}

\author{Mehdi Shokri}
\email{mehdishokriphysics@gmail.com}
\affiliation{School of Physics, Damghan University, P. O. Box 3671641167, Damghan, Iran}

\author{Jafar Sadeghi}
\email{pouriya@ipm.ir}
\affiliation{Department of Physics, University of Mazandaran, P. O. Box 47416-95447, Babolsar, Iran}
\affiliation{School of Physics, Damghan University, P. O. Box 3671641167, Damghan, Iran}

\author{Ram\'{o}n Herrera}
\email{ramon.herrera@pucv.cl}
\affiliation{Instituto de F\'{\i}sica, Pontificia Universidad Cat\'{o}lica de Valpara\'{\i}so, Avenida Brasil 2950, Casilla 4059, Valpara\'{\i}so, Chile.
}

\author{Saeed Noori Gashti}
\email{saeed.noorigashti@stu.umz.ac.ir}
\affiliation{Department of Physics, University of Mazandaran, P. O. Box 47416-95447, Babolsar, Iran}
\preprint{}
\begin{abstract}
We study a warm inflationary model for different expansions assuming an anisotropic universe described by Bianchi I metric. The universe is filled with a scalar field or inflaton, radiation, and bulk viscous pressure. We carry out the inflationary analysis for different solutions of such universe in two different cases of the bulk viscosity coefficient $\xi$ and the dissipation coefficient $\Gamma$ as constant and variable parameters, respectively. We compare the obtained 
results with the recent observations, in order to find the observational constraints on the parameters space of the models. Moreover, we attempt to present a better judgment among the considered models by calculation of the non-linear parameter $f_{NL}$ describing the non-Gaussianity property of the models. Additionally, we investigate the warm inflationary models with viscous pressure from the Weak Gravity Conjecture approach, considering the swampland criteria. 
\end{abstract}

\maketitle
\section{Introduction}
As is well known, the primordial acceleration phase of the universe or cosmic inflation corresponds to an unavoidable part of modern cosmology and it was first proposed by Guth, in order to remove the shortcomings of the Hot Big Bang model \cite{guth}. On the other hand, the density perturbations generated during the inflationary era have been known as the most important candidate for the formation of the structures on large scale. Also, the tensor perturbations are the main responsible for producing the primordial gravitational waves that can be traced by B-mode polarized cosmic microwave background (CMB) photons \cite{Linde:1981my,Albrecht:1982wi,Lyth:1998xn}. The simplest description of inflation or ''cold'' inflation is based on a  single scalar field (\textit{inflaton}), that slowly rolls down from a maximum point to a minimum point of the potential under the \textit{slow-roll} approximation. At the end of the inflationary epoch, the inflaton oscillates around the minimum point and then starts to decay to the particles of the standard model in order to reheat a super-cooled cosmos under the \textit{reheating} mechanism \cite{Kofman2,Shtanov}. In a more complete approach, some inflationary models consider a \textit{preheating} stage, before the reheating process so that first, the classical inflaton decays into massive particles (in particular, into $\varphi$-particles) due to a rapid and broad parametric resonance. Then, these particles decay to the particles of the standard model, and eventually, it follows by a thermalization stage for the produced particles. Besides the mentioned approach to inflation (\textit{cold inflation}), we encounter with a more interesting viewpoint to describe the inflationary era, so-that called \textit{warm inflation}, in which dissipative effects of inflaton into particles and energy are non-negligible \cite{ber1,ber2,ber3,ber4,ber5}. In this context, the interactions between inflaton and other types of matter produce a thermal bath of particles continuously during the inflationary era. Thus, the universe smoothly transits to the radiation-dominated phase without any reheating process. Here, we mention that an inflationary model with a dissipative inflaton to heat and particles was also developed in refs. \cite{be1,be2,be3,be4,be5}. Also, some developments about warm inflation can be found in refs. \cite{c1,c2,c3,c4,c5,c6,c7,c8,c9,c10,c11,c12,c13,c14,c15,c16,c17,c18,c19,c20,c21,c22}.

In relation to the viscous pressure introduced in  models of warm inflation, this dissipative effect appears 
naturally due  to the decay of particles within the fluid. An analysis of the  dynamics  of warm inflation with viscous pressure
showed that the viscous pressure facilitates the inflationary stage \cite{Mi}. However, the dissipative effects from the bulk pressure in warm inflation appears at both the background and the perturbation level. In this respect, the amplitude of the spectrum was obtained in ref. \cite{c24} and  for some viscous pressures, this spectrum induces a small variation in the amplitude ($\sim 4\%$) in relation to the case without bulk pressure. Additionally, in ref. \cite{Arj} was found the primordial curvature spectrum generated in the framework of warm inflation, including shear
viscous effects. Also, by  assuming the Hamilton-Jacobi formalism together with the presence the viscous pressure in the context of warm inflation was studied in ref. \cite{HJ}. For a review of different warm inflationary models with bulk pressure, see refs. \cite{c12,c31,marr1,Setare:2013dd,marr2}.

In relation to the exact inflationary solutions in the frame of General Relativity (GR), the most well-known solution belongs to the de-Sitter universe in which the scalar factor varies exponentially in terms of cosmic time \textit{i.e.} $a(t)\propto e^{t}$. Also, a power-law expansion of the scale factor in cosmic time \textit{i.e.} $a(t)\propto t^{p}$ with $p>1$ corresponds to an exact inflationary solution coming from an exponential inflationary potential. We also have the intermediate solution derived from an effective theory at low dimensions of string theory, in which the evolution of the scale factor changes with the cosmic time as $a(t)\propto \exp({At^{f}})$ with $A>0$ and $0<f<1$, giving origin to the power-law potential \cite{c23,c24,c25}. The solution is named the intermediate model since it is slower than the de-Sitter solution but faster than the power-law solution. Beside the discussed inflationary models, we deal with a generalized form for the scale factor, characterized by a logarithmic term, so-that called the logamediate inflationary model, in which the scale factor evolves as $a(t)\propto \exp({B\ln (t)^{\lambda}})$ with $B>0$ and $\lambda>1$ \cite{c26}. Obviously, this model reduces to the power-law solution for $\lambda=1$ and $B=p$ \cite{c27}. Despite the interesting features of such exact solutions, they are ruled out by observations in the context of cold inflation. For instance, the intermediate model shows an observationally-disfavored value of the spectral index $n_{s}=1$ and also a tensor-to-scalar ratio $r$ much bigger than its observational value in the context of GR. By going beyond the cold inflation and working with warm inflation, these excluded models may be situated in good agreement with observations. Hence, one can find a wide range of warm inflationary models investigated in the intermediate and logamediate regimes \cite{c28,c29,c30,c31,c32,c33}.

Monitoring CMB photons has been known as the main source for the observational test of inflationary models. To be more precise, scalar and tensor perturbations generated during inflation are main responsible for temperature and polarization anisotropies of CMB photons, respectively. On the other hand, we build our cosmological models based on a homogeneous and isotropic universe described by the Friedmann Robertson Walker (FRW) metric. This feature is a consequence of the cosmological principle that says the universe is \textit{almost} homogeneous and isotropic in large-scale structures. The important key here is that the existence of temperature and polarization anisotropies in CMB photons motivates us to abandon the formal attitude to the universe and to approach a real universe using some anisotropic metrics like Bianchi models instead of the FRW model. In this sense, the Bianchi models are a viable  alternative to the FRW models to explicate  the anisotropies and the large angle anomalies (low quadrupole moment) of CMB. In particular, the great amount of power suppression at large scales of the low quadrupole moment could be explained by small deviations of the an universe homogeneous but with an  anisotropic (Bianch) background metric. Additionally, the primordial gravitational wave and its possible direct detection from the quantity so–called
spectral energy density, in the context of an anisotropic (Bianchi type–I) metric was recently studied in ref. \cite{AB}. Here, the spectral graphs for spectral energy density considering the anisotropic Bianchi metric has the ability to adapt to the observational data and in particular to the big-bang nucleosynthesis bounds, see also ref. \cite{AB1}. From these observational motivations, different inflationary models in the framework of anisotropic background metrics have been considered in the literature \cite{AB2,AB3}. In particular, we distinguish refs. \cite{c35,c36}, in which the authors Sharif and Saleem studied a model of warm inflation assuming an anisotropic universe described by Bianchi I (BI) metric. Also, see   \cite{AB4} in which inflation is recently analyzed in the framework of warm anisotropic inflation.

The goal of this research is to analyze a warm inflationary model with the presence of bulk pressure, in the framework of an  
anisotropic universe described by Bianchi metric. In this scenario, we investigate how
the anisotropic warm inflationary model that includes different viscous pressures, modifies the dynamics of the background variables as well the perturbations cosmological. In this form, in the present work, we investigate a model of warm inflation for different expansions described by the scale factor and assuming an anisotropic universe described by Bianchi metric. We consider the universe is filled with inflaton, matter-radiation fluid and bulk viscous pressure. We carry out the inflationary analysis for different solutions, separately. Then, we compare the obtained results with the observational data coming from CMB anisotropies, in order to find the observational constraints on the parameters space of the model. Also, we study the non-Gaussianity feature of the models by using the non-linear parameter $f_{NL}$ introduced under the $\delta N$ formalism. Finally, we examine the models from viewpoint of the Weak Gravity Conjecture using the swampland criteria. The above discussion motivates us to arrange the paper as follows. In \S II, we introduce the warm inflationary mechanism in the context of an anisotropic universe filled with bulk viscous pressure. In \S III, we present a detailed study of anisotropic warm inflation for the intermediate model in the presence of bulk viscous pressure by comparing the obtained results with the observational datasets from the Planck and the BICEP2/Keck array satellites. Moreover, we calculate the non-linear parameter $f_{NL}$ of the model in order to search the non-Gaussianity feature of the model. The section is finalized by testing the model from the WGC approach using the swampland criteria. We repeat the inflationary analysis discussed in \S III for the logamediate and exponential models in \S IV and \S V, respectively. In §VI, we conclude the analysis of the models and draw the possible outlooks. We choose units so that $\kappa^2=8\pi G=1$.
\section{Warm inflationary model in an anisotropic background}
Let us start with flat BI metric describing an anisotropic universe with the line element \cite{me} 
\begin{equation}
ds^{2}=-dt^{2}+a^{2}(t)dx^{2}+b^{2}(t)dy^{2}+c^{2}(t)dz^{2},
\label{1}    
\end{equation}
where $a(t)$, $b(t)$, $c(t)$ are the scale factor along $x$, $y$ and $z$ directions, respectively. Therefore, we deal with an average Hubble parameter defined as
\begin{equation}
H=\frac{H_{1}+H_{2}+H_{3}}{3}, 
\label{2}
\end{equation}
where $H_{1}=\dot{a}/{a}$, $H_{2}=\dot{b}/{b}$ and $H_{3}=\dot{c}/{c}$ are directional Hubble parameters, respectively. In the following, we will consider that a dot depicts derivative with respect to cosmic time $t$. By using the Locally Rotationally Symmetric (LRS) BI model where $b(t)=c(t)$ and also a linear relationship $a=b^{m}$ ($m\neq1$), the metric (\ref{1}) takes the following form
\begin{equation}
ds^{2}=-dt^{2}+b^{2m}(t)dx^{2}+b^{2}(t)(dy^{2}+dz^{2}),
\label{3}    
\end{equation}
where $m$ is a constant parameter referring anisotropy property of the universe and it is called factor of anisotropicity \cite{AB4}. We note that for the special case in which $m=1$, it recovers the flat FRW metric. Also, the average Hubble parameter (\ref{2}) takes the form $H=(m+2)H_{2}/3$.

In relation to the matter sector of our warm inflationary model, we deal with inflaton as the main matter component of the universe described by the following energy density and pressure
\begin{equation}
\rho_{\varphi}=\frac{\dot{\varphi}^{2}}{2}+V(\varphi),\hspace{1cm} P_{\varphi}=\frac{\dot{\varphi}^{2}}{2}-V(\varphi),
\label{4}    
\end{equation}
where $V$ is the corresponding potential of inflaton. Besides the scalar field, we have a fluid having radiation and particles created by decaying inflaton during the warm inflationary era. If the number density of the particle is negligible, we can treat the particle as radiation and so, we involve with a pure thermal fluid. Otherwise, taking into account the role of particles leads to two important effects in the matter-radiation fluid: i) The energy density of matter-radiation fluid shifts from $P=\rho/3$ to an extended form or general expression $P=(\gamma-1)\rho$ where $1\leq\gamma\leq2$ is called the adiabatic index. ii) A non-equilibrium viscous pressure is produced due to the inter-particle interactions or decaying particles existed in the matter-radiation fluid \cite{W1,W2,W3}. In overall, we deal with an  imperfect fluid characterized by energy density $\rho=TS(\varphi,T)$ where $T$ and $S$ are temperature and entropy density of the fluid, respectively. Also, pressure of the fluid is expressed as $P+\Pi$ where $\Pi=-(m+2)\xi H_{2}$ is the bulk viscous pressure, where the coefficient of bulk viscosity is denoted by $\xi$. Also, for the specific case in which m=1, the bulk viscosity corresponds to the usual fluid dynamic expression $\Pi=-3\xi\,H$. Regarding the second thermodynamics law, $\xi$ must be positive definite and depends on $\rho$, \textit{i.e.}, $\xi=\xi(\rho)$.

Let's study the dynamics of a warm inflationary model with bulk viscous pressure in an anisotropic universe. The Friedmann equation of such a model is given by
\begin{equation}
H_{2}^{2}=\frac{\rho_{\varphi}+\rho}{1+2m},
\label{5}    
\end{equation}
where $\rho_{\varphi}$ and $\rho$ are the energy density of inflaton and imperfect fluid, respectively. Due to the conservation law of energy-momentum tensor $\nabla^{\nu}T_{\mu\nu}=0$, the time component of the energy-momentum tensor can be found as 
\begin{equation}
\dot{\rho}+(m+2)H_{2}(\gamma\rho+\Pi)=\Gamma\dot{\varphi}^{2},
\label{6}
\end{equation}
\begin{equation}
\dot{\rho}_{\varphi}+(m+2)H_{2}(\rho_{\varphi}+P_{\varphi})=-\Gamma\dot{\varphi}^{2},
\label{7}
\end{equation}
where $\Gamma$ is the rate of decaying inflaton to imperfect fluid. Its value takes a positive value and this dissipation  coefficient $\Gamma$ has a dependency that  can be considered as a function of $\varphi$ or $T$ or both or simply a constant \cite{ber1,be3,marr3,marr4}. Using the energy density and pressure of inflaton (\ref{4}), the Eq. (\ref{7}) can be rewritten as
\begin{equation}
\ddot{\varphi}+\big((m+2)H_{2}+\Gamma\big)\dot{\varphi}=-V'(\varphi),
\label{8}    
\end{equation}
where prime denotes derivative with respect to scalar field $\varphi$. For entropy density of imperfect fluid, we use the thermodynamics relation 
\begin{equation}
S(\varphi, T)=-\frac{\partial f}{\partial T}=\frac{\partial V}{\partial T},
\label{9}    
\end{equation}
where the Helmholtz free energy $f=\rho_{\varphi}+\rho-TS$ is dominated by $V$. Using the above expression, the Eq. (\ref{6}) is rewritten as
\begin{equation}
T\dot{S}+(m+2)H_{2}(\gamma TS+\Pi)=\Gamma\dot{\varphi}^{2},
\label{10}    
\end{equation}
where we have considered that the term $\dot{T}$ is negligible. Here the energy transfer to the matter-radiation fluid is related with an increase of the  entropy density associated to fluid.

To fulfil our purposes, we consider a series of conditions: i) $\rho_{\varphi}\approx V(\varphi)$, $\rho_{\varphi}>\varphi$ ii) slow-roll approximation $\dot{\varphi}^{2}\ll V(\varphi), \ddot{\varphi}^{2}\ll ((m+2)H_{2}+\gamma)\dot{\varphi}$ iii)  $\dot{\rho}\ll(m+2)H_{2}(\Gamma\rho+\Pi), \dot{\rho}\ll\Gamma\dot{\varphi}^{2}$, see refs. \cite{c35,c36}. In this form, the dynamical equations are reduced to the 
\begin{equation}
H_{2}^{2}\simeq\frac{V(\varphi)}{1+2m},\hspace{1cm} (m+2)H_{2}(1+r)\dot{\varphi}\simeq-V'(\varphi),\hspace{1cm}
\gamma\rho\simeq r\dot{\varphi}^{2}-\Pi,
\label{11}
\end{equation}
where $r=\frac{\Gamma}{(m+2)H_{2}}$ refers to the dissipation ratio. In this context, we can study the evolution of the warm inflation in two dissipative regimes; the weak regime characterized by $r\ll1$ and strong regime in which $r\gg1$. In the next sections, we will implement the analysis of our warm anisotropic model for strong dissipative regime. The reason is that in a low dissipative model, radiation and particles released from inflaton decay scatter by inflationary expansion and consequently they have less chance for interacting and producing bulk viscous pressure. 

The slow-roll parameters of a warm inflationary model with bulk viscous pressure for anisotropic universe are given by \cite{c36}
\begin{equation}
\epsilon=\frac{3(1+2m)}{2(1+r)(m+2)^{2}}\Big(\frac{V'}{V}\Big)^{2},\hspace{1cm}\eta=\frac{3(1+2m)}{(1+r)(m+2)^{2}}\bigg(\Big(\frac{V''}{V}\Big)-\frac{1}{2}\Big(\frac{V'}{V}\Big)^{2}\bigg),
\label{12}
\end{equation}
respectively.

 In relation to the theoretical foundations of the early universe, we can contrast our warm anisotropic model from the viewpoint of Weak Gravity Conjecture (WGC) using the conditions of the swampland de Sitter conjecture \cite{Vafa,Brennan,Ooguri,Palti,Palti2,Obied} given by
\begin{equation}
\sqrt{2\epsilon}\geq c,\hspace{1cm}|\eta|\leq -c',
\label{13}
\end{equation}
where $c$ and $c'$ are constant parameters. 

Besides, we can define the number of $e$-folds for our warm anisotropic model and it takes the following form
\begin{equation}
N=-\frac{(m+2)^{2}}{3(1+2m)}\int_{\varphi_{i}}^{\varphi_{f}}(1+r)\bigg(\frac{V}{V'}\bigg)d\varphi,
\label{14}
\end{equation}
\begin{figure*}[!hbtp]
	\centering
	\includegraphics[width=.32\textwidth,keepaspectratio]{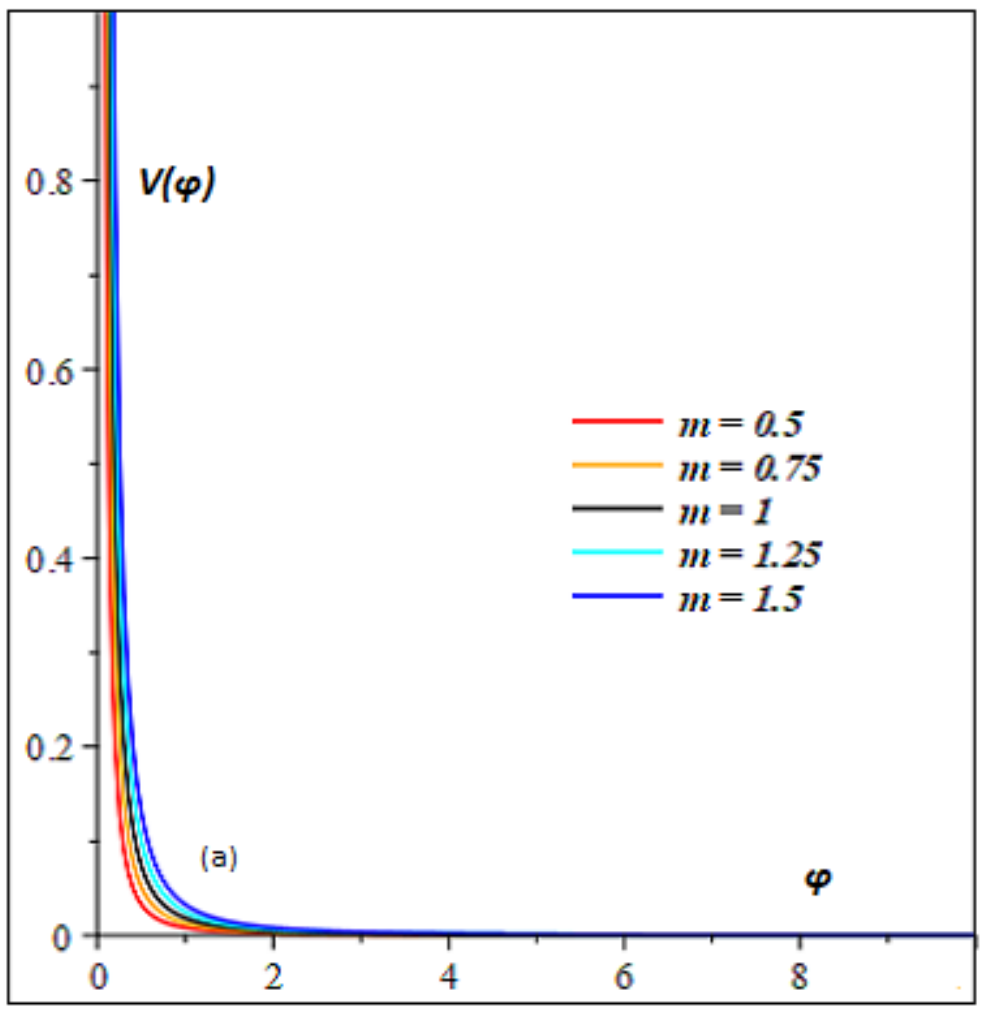}
	\includegraphics[width=.32\textwidth,keepaspectratio]{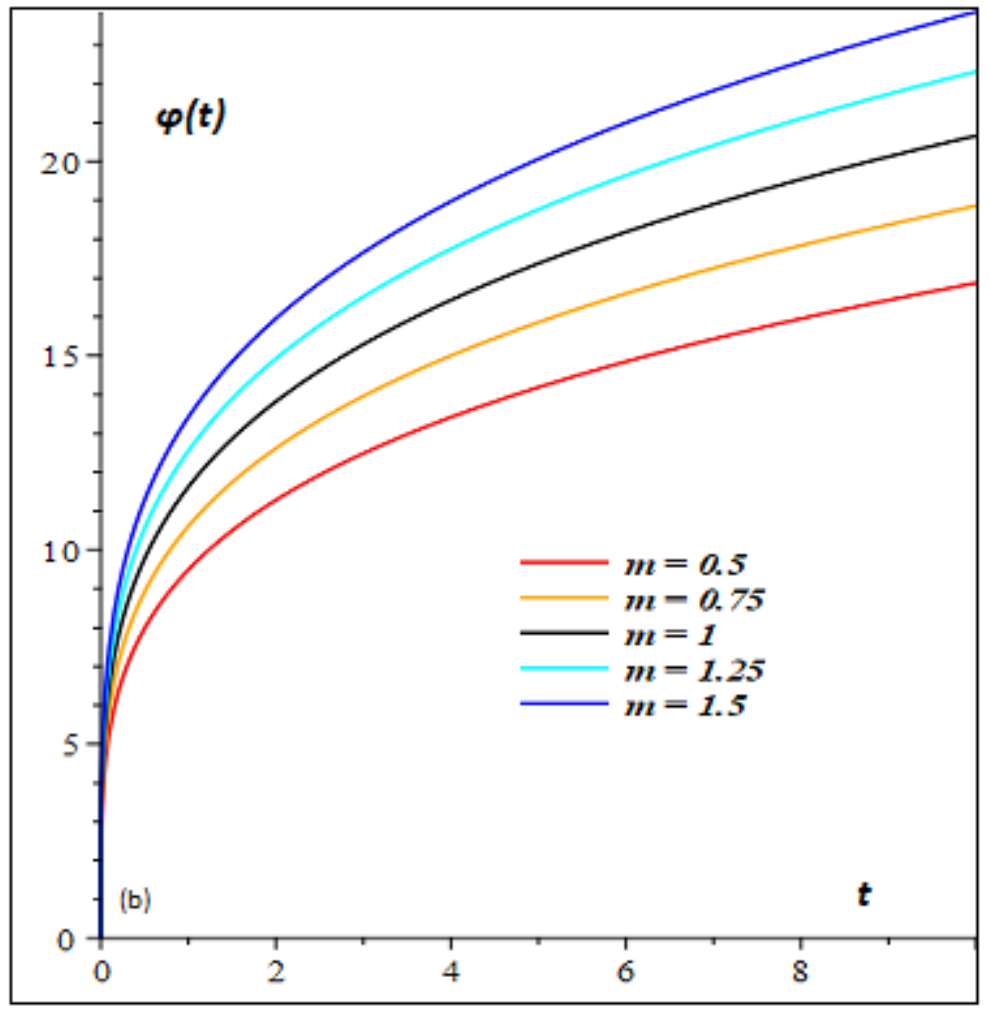}
	\includegraphics[width=.32\textwidth,keepaspectratio]{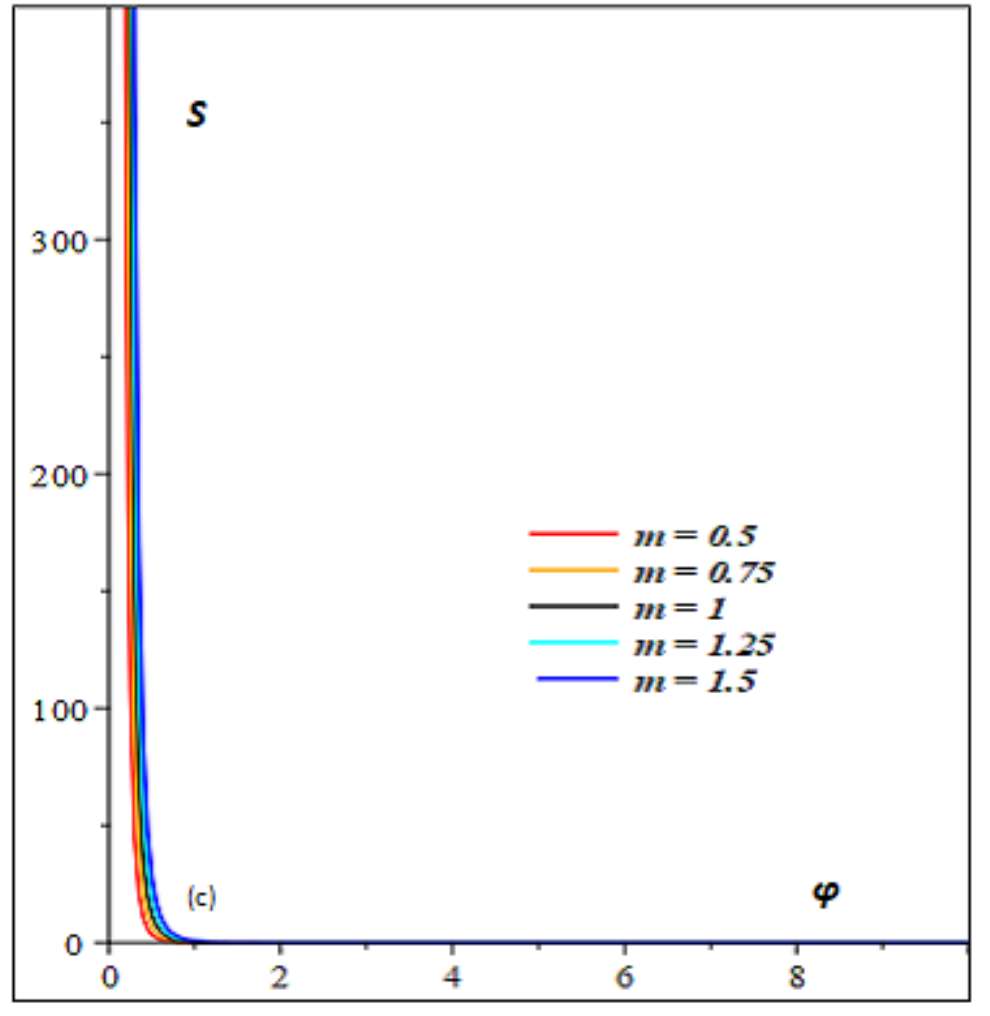}
	\caption{Panel (a): The reconstruction of the effective potential (\ref{24}) plotted versus inflaton $\varphi$ when $\alpha=0.03$, $\beta=0.75$ and $\Gamma_{0}\sim10^{-3}$. Panel (b): The  evolution of inflaton (\ref{24}) plotted versus cosmic time $t$ for different values of $m$ when $\alpha=0.1$, $\beta=0.75$ and $\Gamma_{0}\sim10^{-3}$. Panel (c): The entropy (\ref{25}) plotted versus inflaton $\varphi$ for different values of $m$ when $\alpha=0.003$, $\beta=0.75$, $\gamma=1.5$, $\xi_{0}\sim 10^{-6}$, $\Gamma_{0}\sim10^{-3}$ and $T\sim10^{-5}$ \cite{caption}.}
	\label{fig1}
\end{figure*}
\begin{figure*}[!hbtp]
	\centering
	\includegraphics[width=.65\textwidth,keepaspectratio]{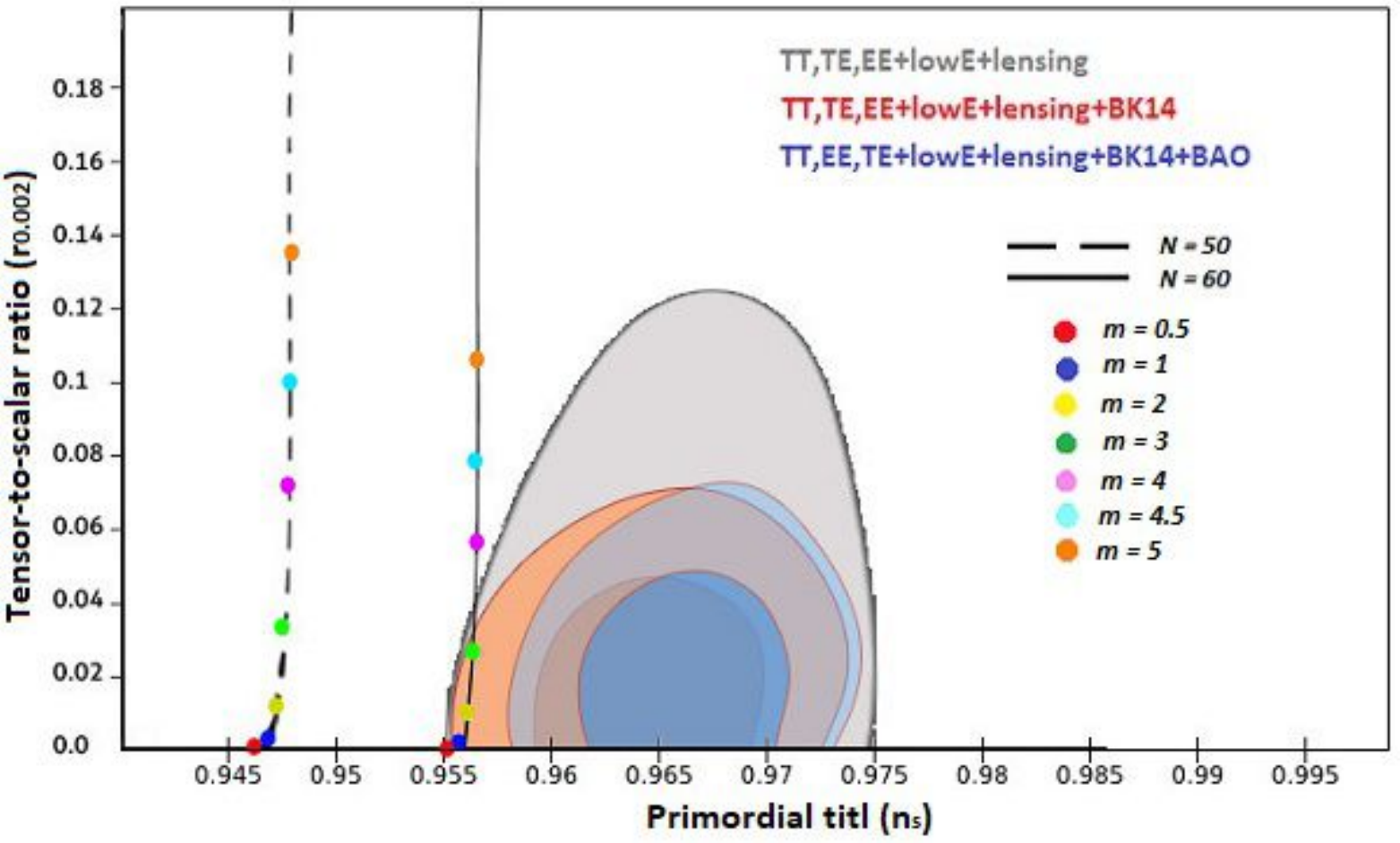}
	\caption{The marginalized joint 68\% and 95\% CL regions for $n_{s}$ and $r$ at $k = 0.002$ Mpc$^{-1}$ from Planck 2018 in combination with BK14+BAO data \cite{cmb} and the $n_{s}-r$ constraints on the intermediate model (\ref{22}) in the case of $\Gamma$ and $\xi$ as constant parameters. The dashed and solid lines represent $N=50$ and $N=60$, respectively. The panel is plotted for different values of $m$ when $\alpha=1$, $\beta=0.9$, $\gamma=1.5$, $\xi_{0}\sim10^{-6}$, $\Gamma_{0}\sim10^{-3}$ and $T_{r}\sim10^{-5}$ \cite{caption}.}
	\label{fig2}
\end{figure*}
where $\varphi_{i}$ and $\varphi_{f}$ are the values of inflaton at start and end of inflation. Here we have considered that the number of $e-$ folds at the initial of the inflationary epoch is $N_i=N(\varphi_i)=0$.

On the other hand, to find the spectral parameters in the strong regime ($r\gg1$), we consider the scalar power spectrum defined as \cite{c4}
\begin{equation}
P_{R}=\frac{T_{r}}{2\pi^{2}\epsilon\sqrt{rV^{3}}}\exp{[-2\mathcal{N(\varphi)}]},
\label{15}    
\end{equation}
where $T_{r}$ is the temperature of thermal bath. Also, the function $\mathcal{N}$ in strong dissipative regime is given by
\begin{equation}
\mathcal{N}(\varphi)=\int\bigg[-\frac{\Gamma'}{(m+2)rH_2}-\frac{3}{8G(\varphi)}\bigg(1-\bigg\{(\gamma-1)+\frac{\Pi}{\xi}\frac{d\xi}{d\rho}\bigg\}\frac{\Gamma' V'}{(m+2)^{2}r\gamma H_{2}^{2}}\bigg)\frac{V'}{V}\bigg]d\varphi,
\label{16}    
\end{equation}
where the term $G(\varphi)$ associated to the matter-radiation fluid is given by
\begin{figure*}[!hbtp]
	\centering
	\includegraphics[width=.32\textwidth,keepaspectratio]{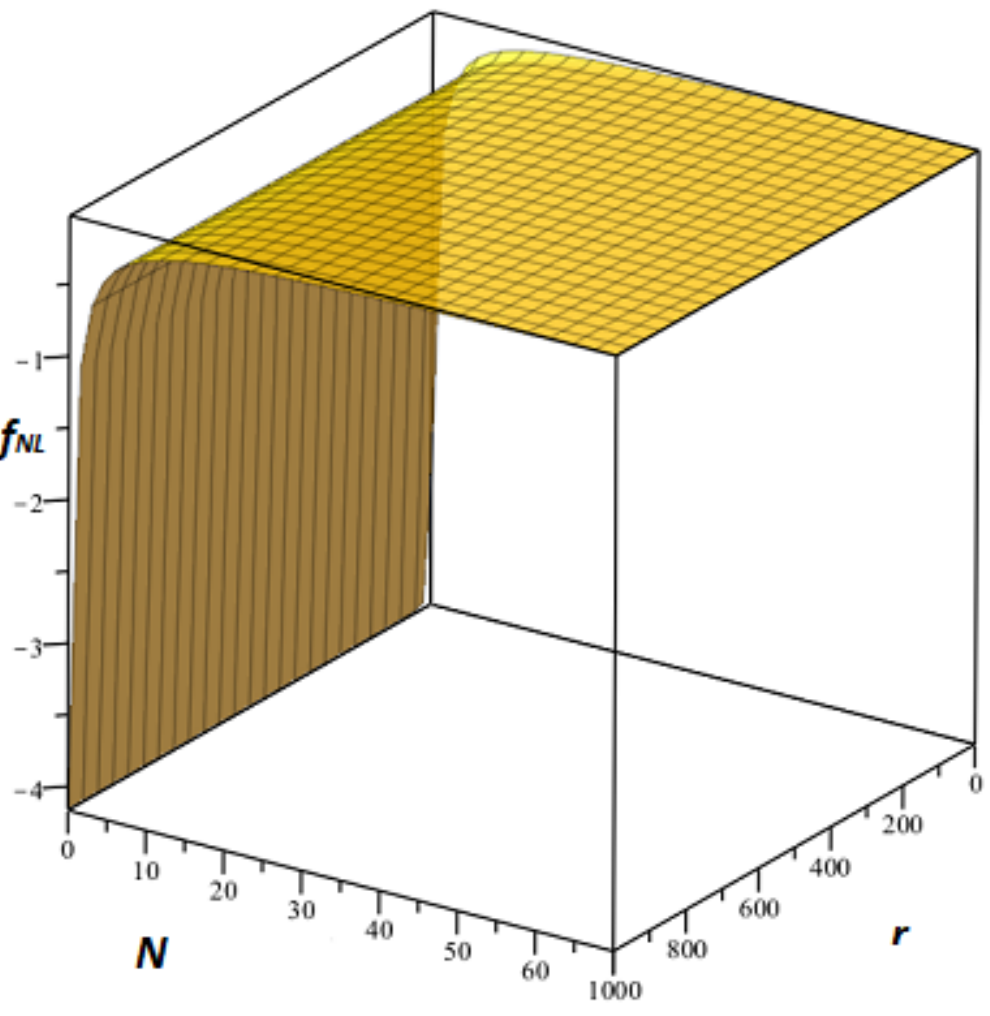}
	\caption{The non-linear parameter $f_{NL}$ versus number of e-folds $N$ and dissipation strength $r$ in intermediate model in the absence of the second term of the Eq. (\ref{31}). The figure is plotted for $\alpha=1$, $\beta=0.9$, $\gamma=1.5$, $\Gamma_{0}\sim 10^{-3}$ and $T_{r}\sim10^{-5}$ \cite{caption}.}
	\label{fig3}
\end{figure*}
\begin{figure*}[!hbtp]
	\centering
	\includegraphics[width=.32\textwidth,keepaspectratio]{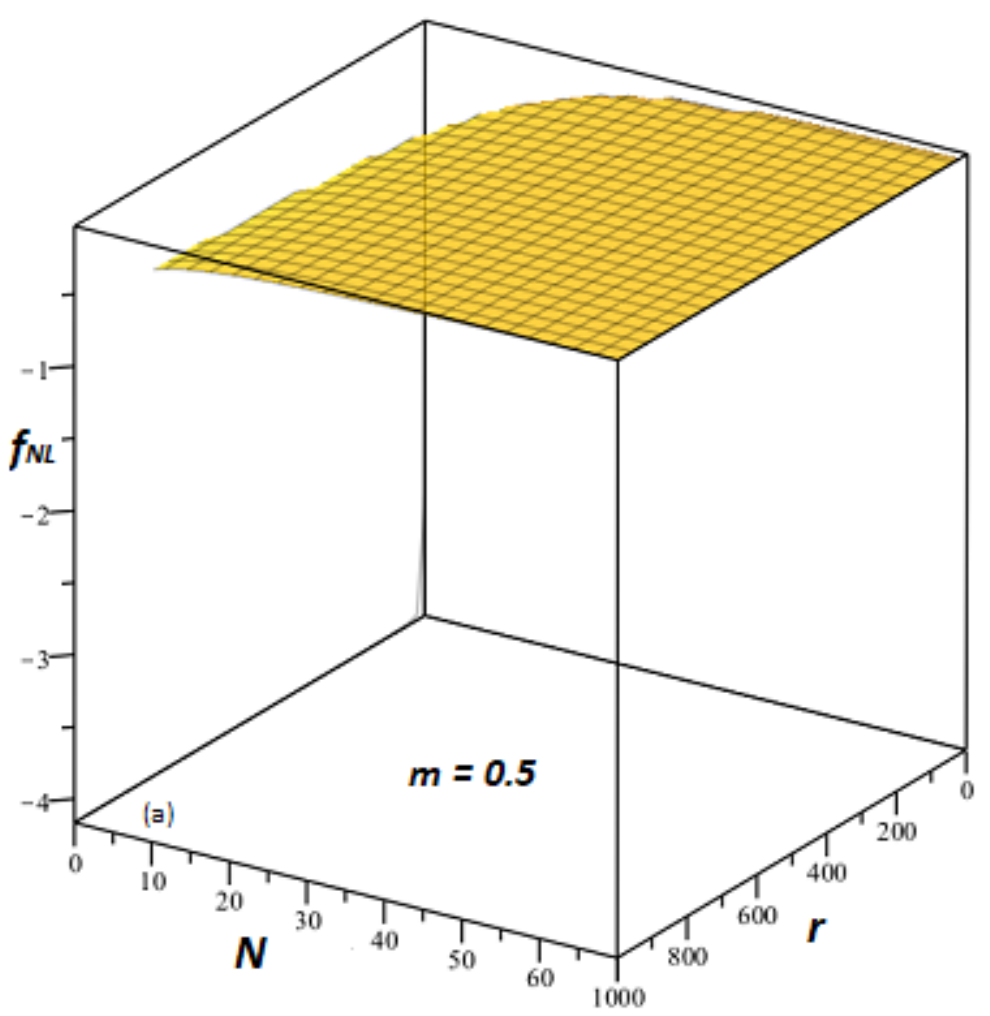}
	\includegraphics[width=.32\textwidth,keepaspectratio]{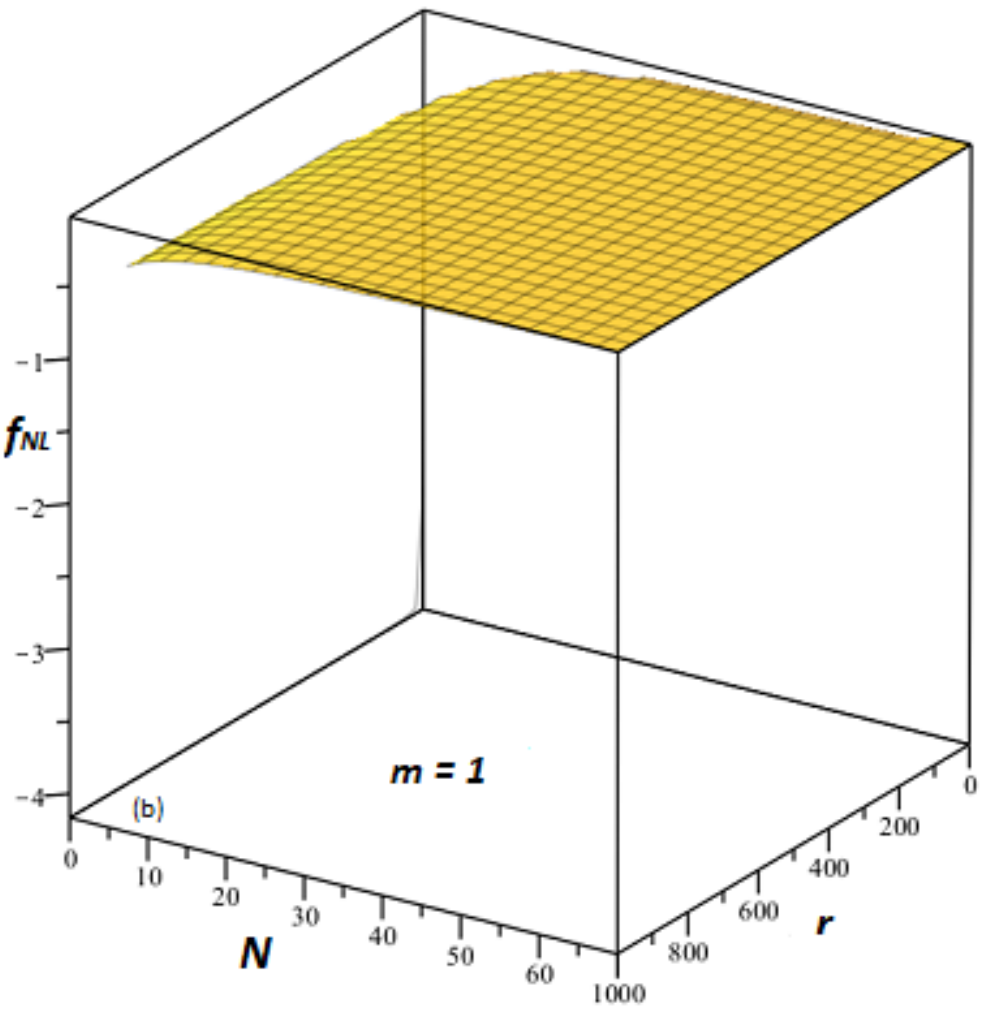}
	\includegraphics[width=.32\textwidth,keepaspectratio]{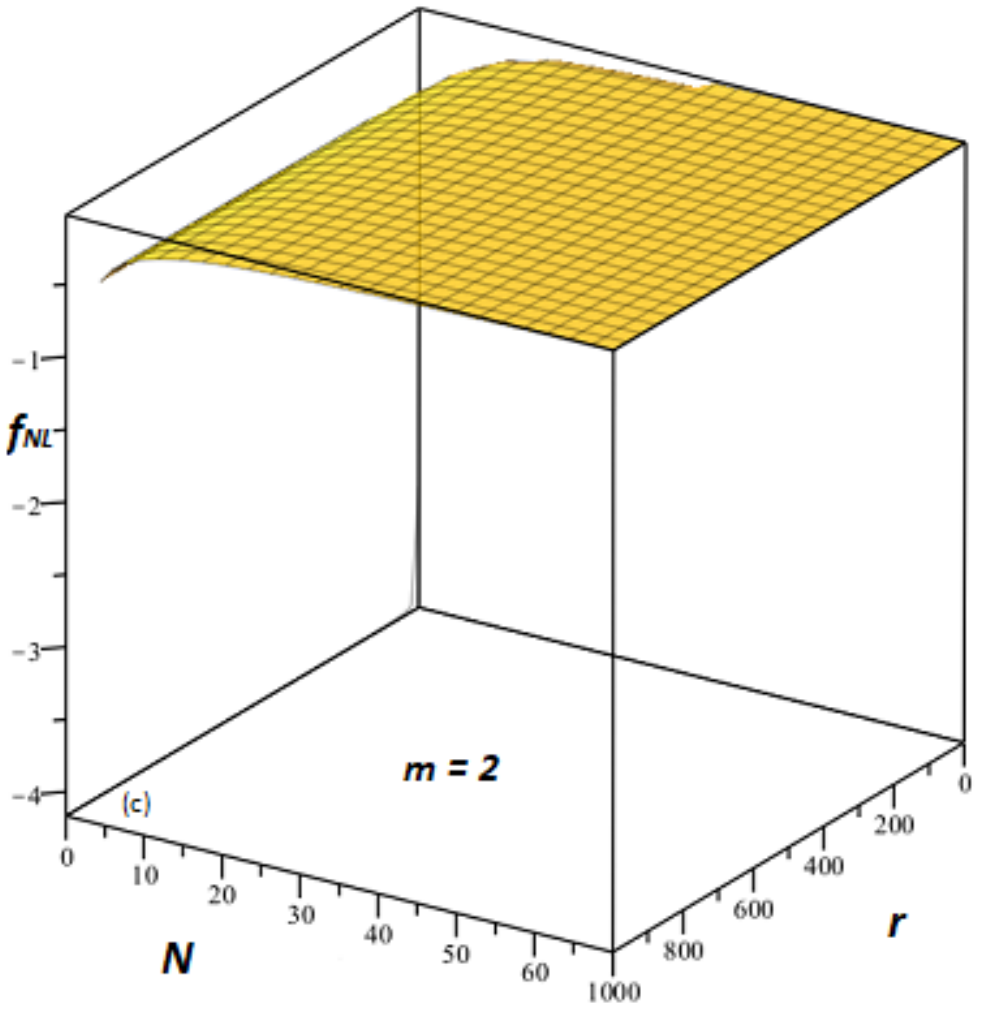}
	\includegraphics[width=.32\textwidth,keepaspectratio]{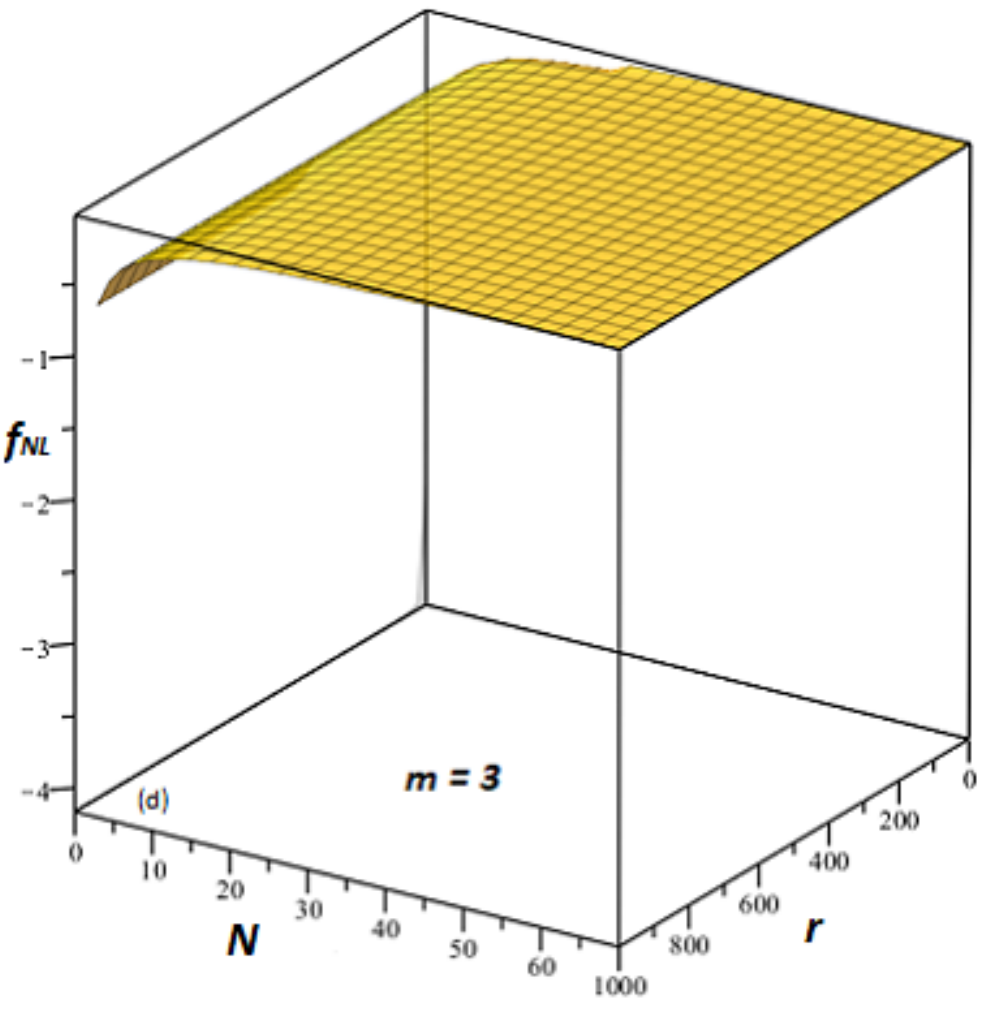}
	\includegraphics[width=.32\textwidth,keepaspectratio]{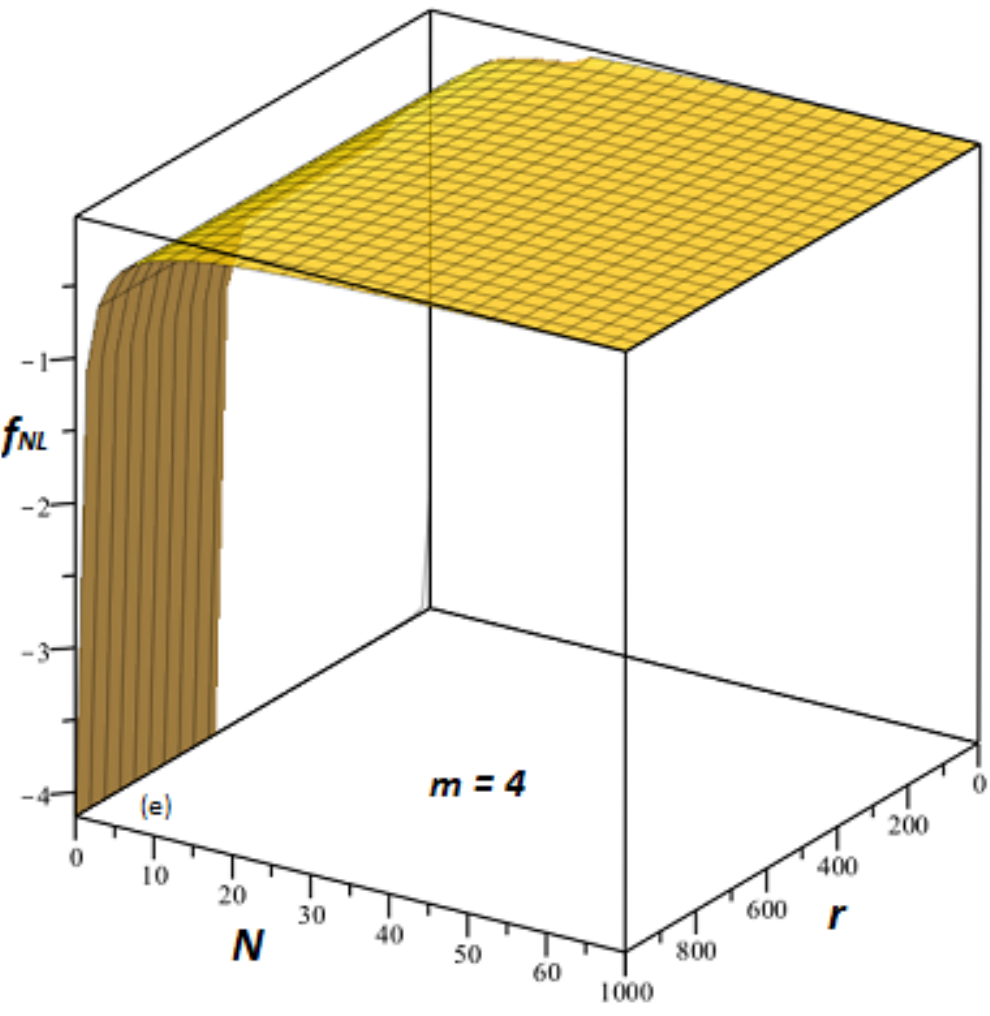}
	\includegraphics[width=.32\textwidth,keepaspectratio]{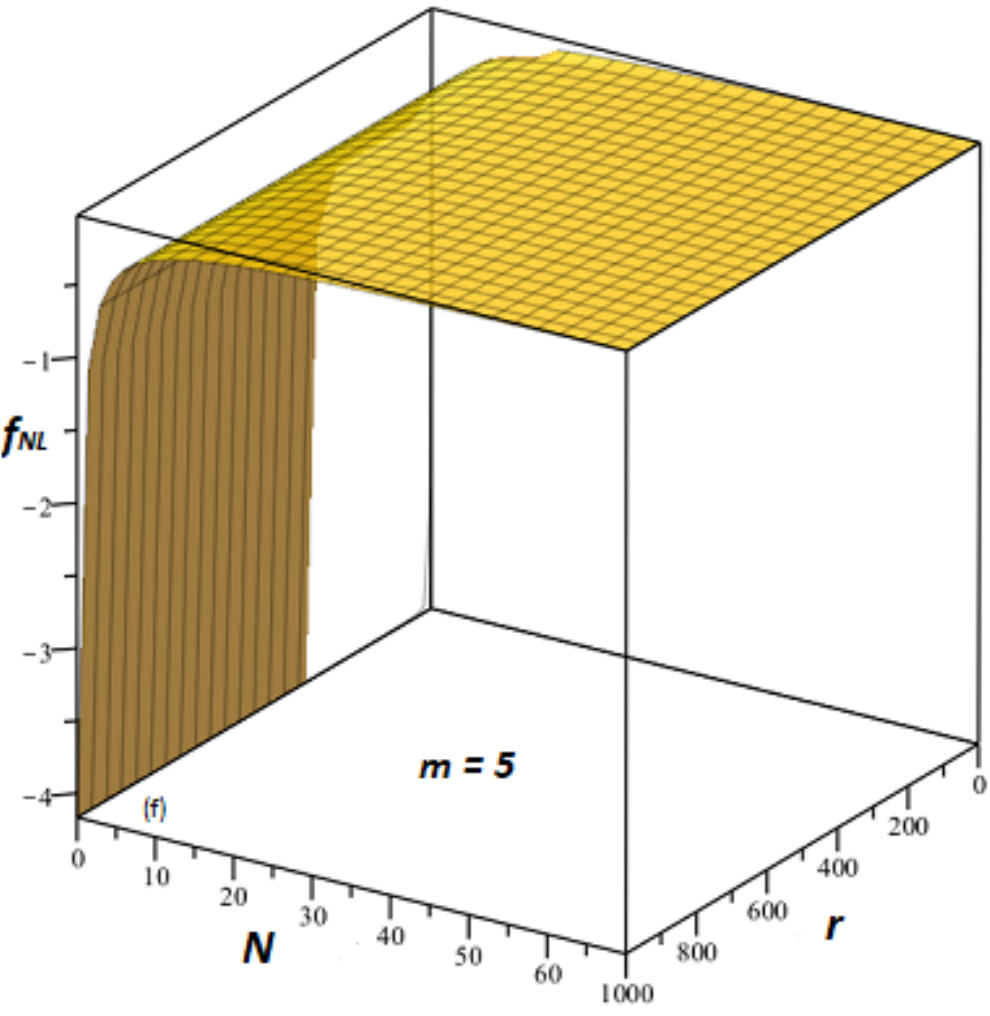}
	\caption{The non-linear parameter $f_{NL}$ versus number of e-folds $N$ and dissipation strength $r$ for different values of $m$ in intermediate model in the presence of the second term of the Eq. (\ref{31}). The figure is plotted for $\alpha=1$, $\beta=0.9$, $\gamma=1.5$, $\Gamma_{0}\sim 10^{-3}$ and $T_{r}\sim10^{-5}$ \cite{caption}.}
	\label{fig4}
\end{figure*}
\begin{equation}
G(\varphi)=1-\frac{1}{8H_{2}^{2}}\bigg(2\gamma\rho+3\Pi+\frac{\gamma\rho+\Pi}{\gamma}\big(\frac{\Pi}{\xi}\frac{d\xi}{d\rho}-1\big)\bigg).
\label{17}
\end{equation}
Additionally, the scalar spectral index $n_{s}$ associated to the power spectrum is expressed by \cite{c4}
\begin{figure*}[!hbtp]
	\centering
	\includegraphics[width=.32\textwidth,keepaspectratio]{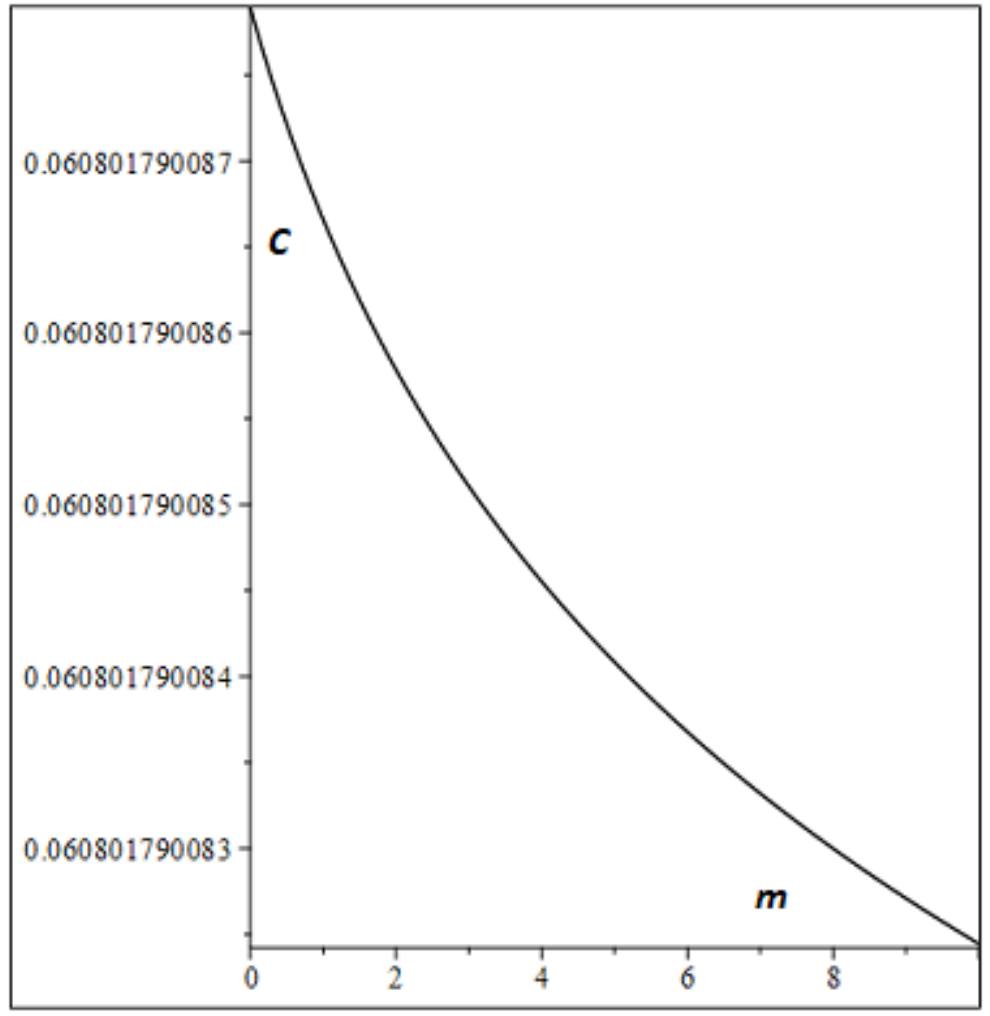}
	\hspace{0.5cm}
	\includegraphics[width=.32\textwidth,keepaspectratio]{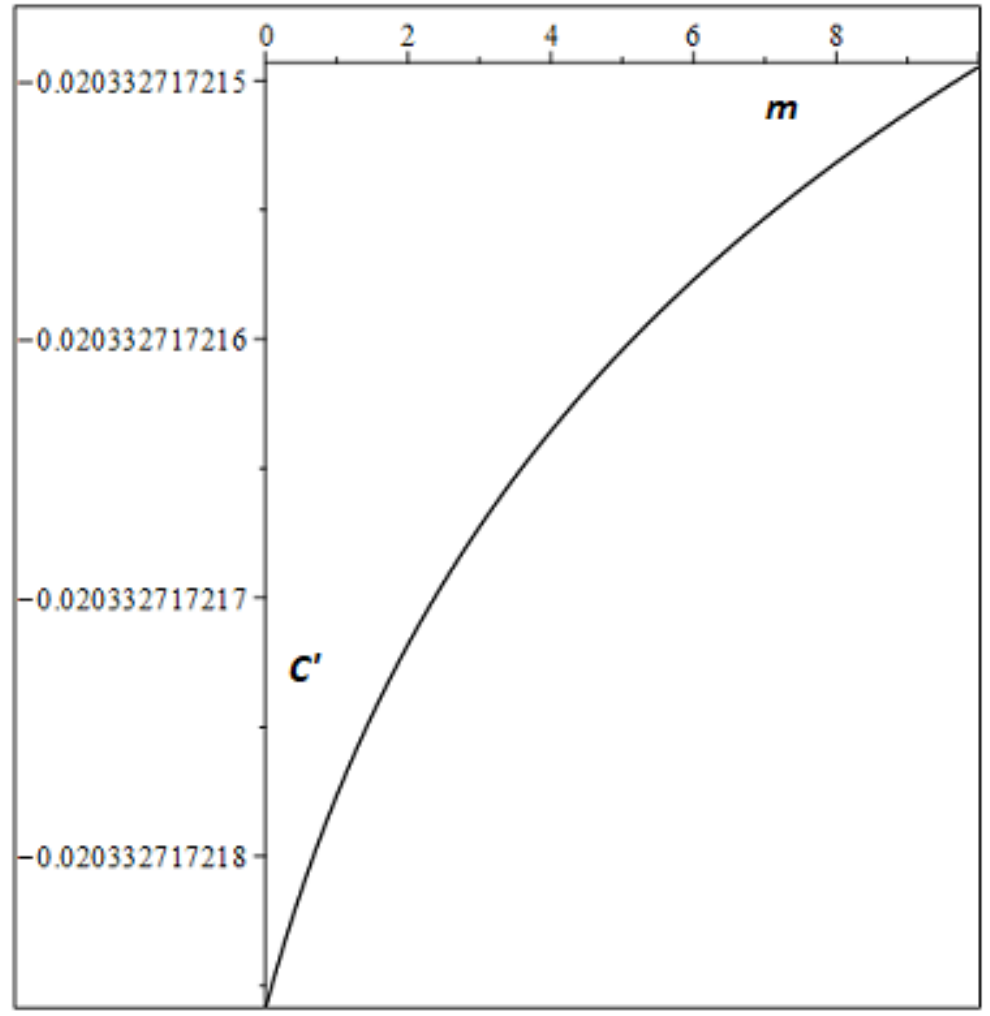}
	\caption{The behaviour of the swampland parameters $c$ and $c'$ versus $m$ in intermediate model (\ref{22}) in the case of $\Gamma$ and $\xi$ as constant parameters when $\alpha=1$, $\beta=0.9$ and $N=60$.}
	\label{fig5}
\end{figure*}
\begin{equation}
n_{s}-1=\frac{d\ln P_R}{d\ln k}\simeq-\frac{3}{(m+2)}\sqrt{\frac{2\epsilon}{r}}\Big(2\mathcal{N}'-\frac{r'}{2r}\Big)-\epsilon-2\eta.
\label{18}
\end{equation}
As argued by ref. \cite{caption}, the production of tensor perturbations during the inflationary scenario  gives rise to stimulated emission in the thermal background of gravitational waves. In this sense, we can introduce  another observational parameter and it corresponds to the tensor to scalar ratio $R$ given by
\begin{equation}
R(k_{0})=\frac{2(m+2)^{2}}{9(1+2m)}\Big(\frac{\epsilon\sqrt{r V^{5}}}{T_{r}}\Big)\exp\big(2\mathcal{N}\big)\coth\bigg[\frac{k}{2T}\bigg]\bigg|_{k=k_{0}},
\label{19}
\end{equation}
where $k_{0}=0.002$ Mpc$^{-1}$ is the pivot scale and $T$ denotes the thermal background temperature of gravitational waves and in general $T\neq T_r$, see ref. \cite{c4}.

For the sake of completeness, we study the non-Gaussianity feature of our models in order to discriminate which models work or not. For
this reason, we calculate the non-linear parameter $f_{NL}$ introduced in the context of $\delta N$ formalism as \cite{c37,c38,c39}
\begin{equation}
-\frac{3}{5}f_{NL}=\frac{\sum_{ij}N_{,i}N_{,j}N_{,ij}}{2\[\sum_{i}N_{,i}^{2}\]^{2}}+\ln(kL)\frac{P_{R}}{2}\frac{\sum_{ijk}N_{,ij}N_{,jk}N_{,ki}}{\[\sum_{i}N_{,i}^{2}\]^{3}},
\label{20}    
\end{equation}
where the logarithmic term can be considered to be of order 1 since it involves the size $k^{-1}$ of a typical scale under consideration, relative to the size $L$ of the region within which the stochastic
properties are specified. On the cosmological scales a few Hubble times before horizon entry, observations tell us that the curvature perturbations are almost Gaussian and their spectrum $P_{R}$ is in the order of $\mathcal{O}(10^{-10})$. Hence, the second term of the relation (\ref{20}) is usually neglected in inflationary literature. However, we keep this term since we are concerned to find the role and contribution of bulk viscous pressure in our warm anisotropic inflationary models. Therefore, the non-linear parameter (\ref{20}) for warm  inflation with only one inflaton is reduced to
\begin{equation}
-\frac{3}{5}f_{NL}\simeq\frac{1}{2}\frac{N_{,ii}}{N_{,i}^{2}}+\frac{P_{R}}{2}\frac{N_{,ii}^{3}}{N_{,i}^{6}},
\label{21}    
\end{equation}
where subscribes $i$ and $ii$ denote the first and second derivatives with respect to inflaton $\varphi$, respectively. 
\section{Intermediate model}
In this section, we will  study warm intermediate inflation with bulk viscous pressure for an anisotropic universe. In relation to the intermediate inflation, the scale factor follows the law
\begin{equation}
b(t)=b_{0}\exp(\alpha t^{\beta}),
\label{22}    
\end{equation}
where $\alpha>0$ and $0<\beta<1$ \cite{c24,c25}. In the following we will investigate the model for two different cases in which $\Gamma$, $\xi$ are constant and variable parameters, respectively.
\subsection{$\Gamma$, $\xi$ as constant}
As a simple case, dissipation and bulk viscous coefficients are often introduced as constant parameters by
\begin{figure*}[!hbtp]
	\centering
	\includegraphics[width=.32\textwidth,keepaspectratio]{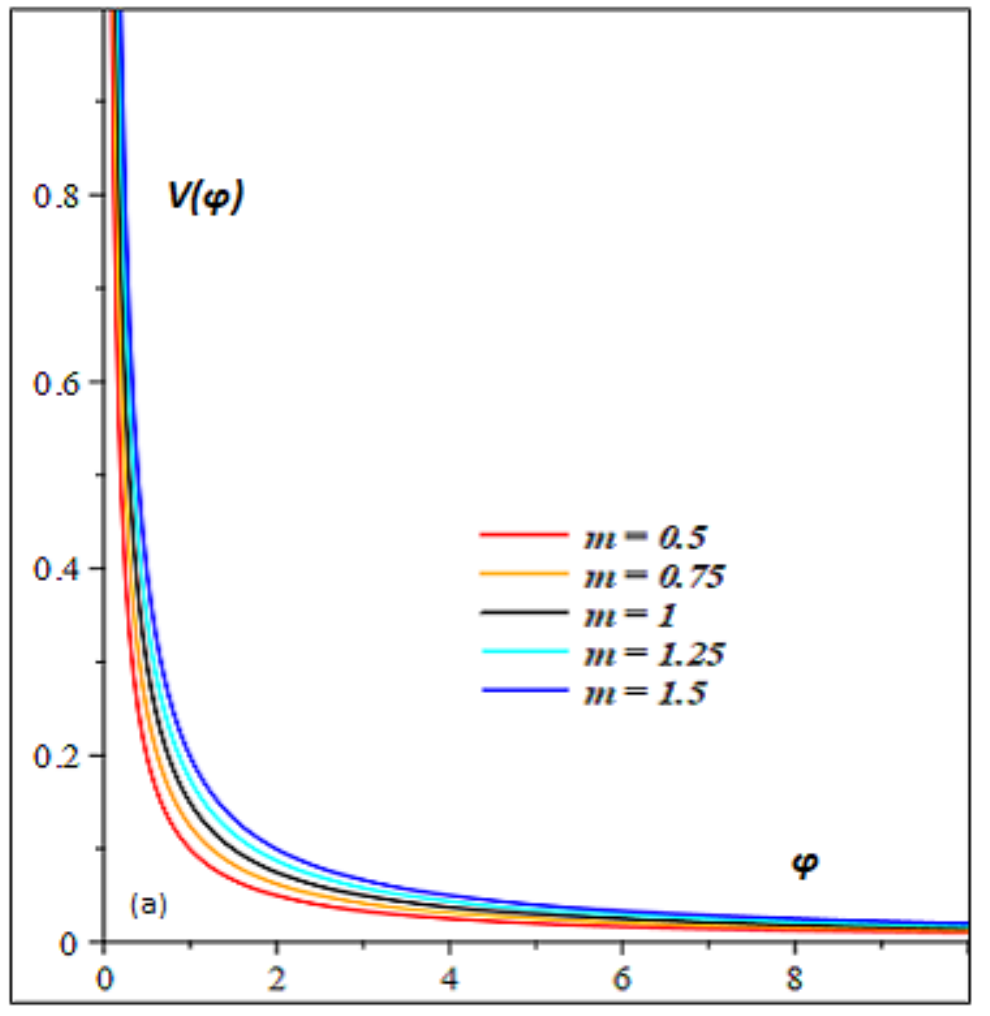}
	\includegraphics[width=.32\textwidth,keepaspectratio]{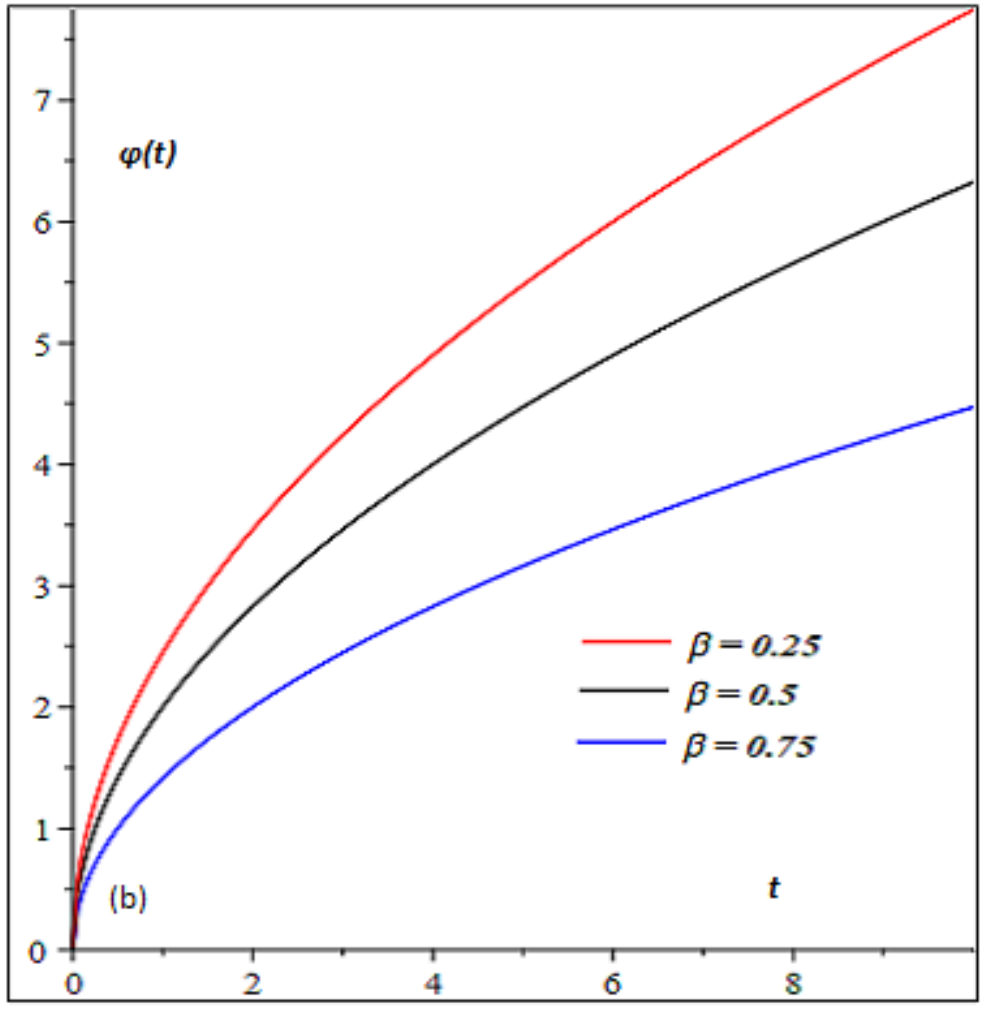}
	\includegraphics[width=.32\textwidth,keepaspectratio]{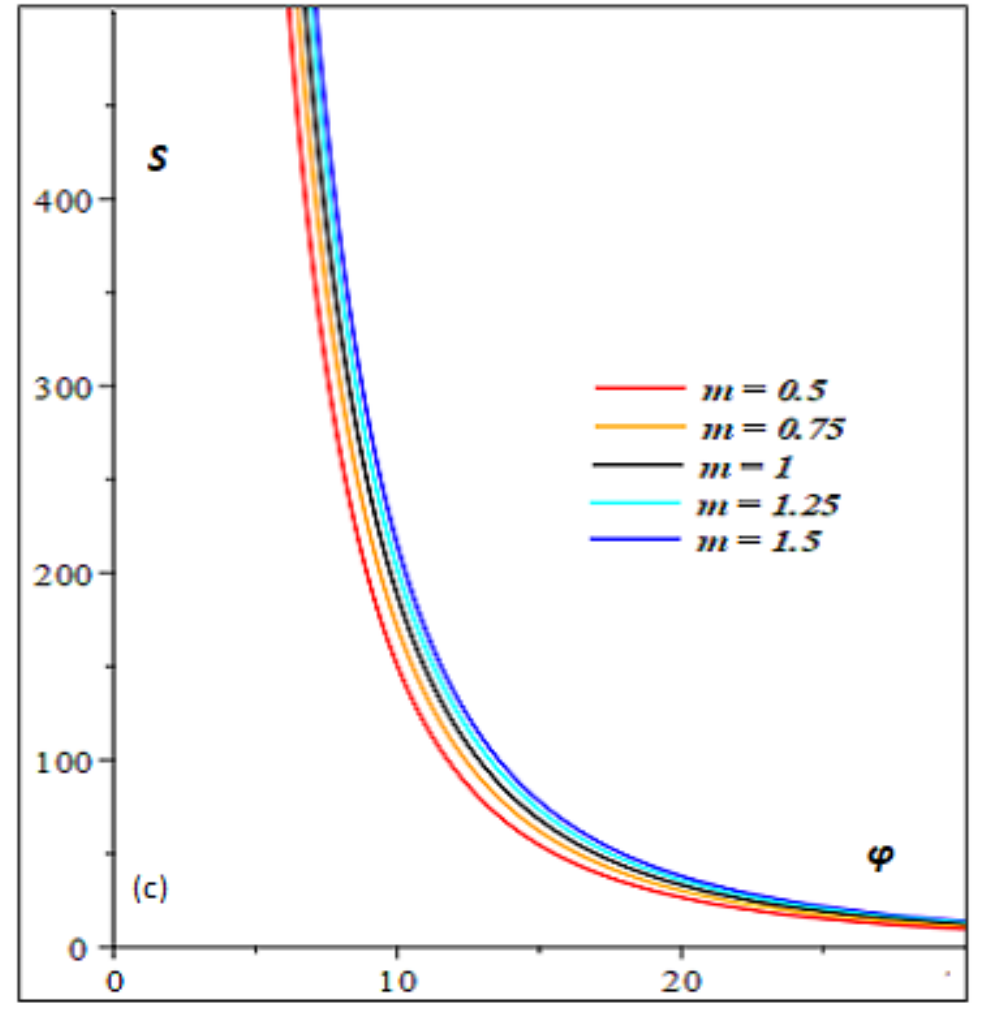}
	\caption{Panel (a): The potential (\ref{33}) plotted versus inflaton $\varphi$ for different values of $m$ when $\alpha=0.25$ and $\beta=0.75$. Panel (b): The evolution of inflaton (\ref{33}) versus cosmic time $t$ for different values of $\beta$. Panel (c): The entropy (\ref{35}) plotted versus inflaton $\varphi$ for different values of $m$ when $\alpha=1$, $\beta=0.75$, $\gamma=1.5$, $\xi_{0}\sim 10^{-8}$ and $T\sim10^{-5}$ \cite{caption}.}
	\label{fig6}
\end{figure*}
\begin{figure*}[!hbtp]
	\centering
	\includegraphics[width=.65\textwidth,keepaspectratio]{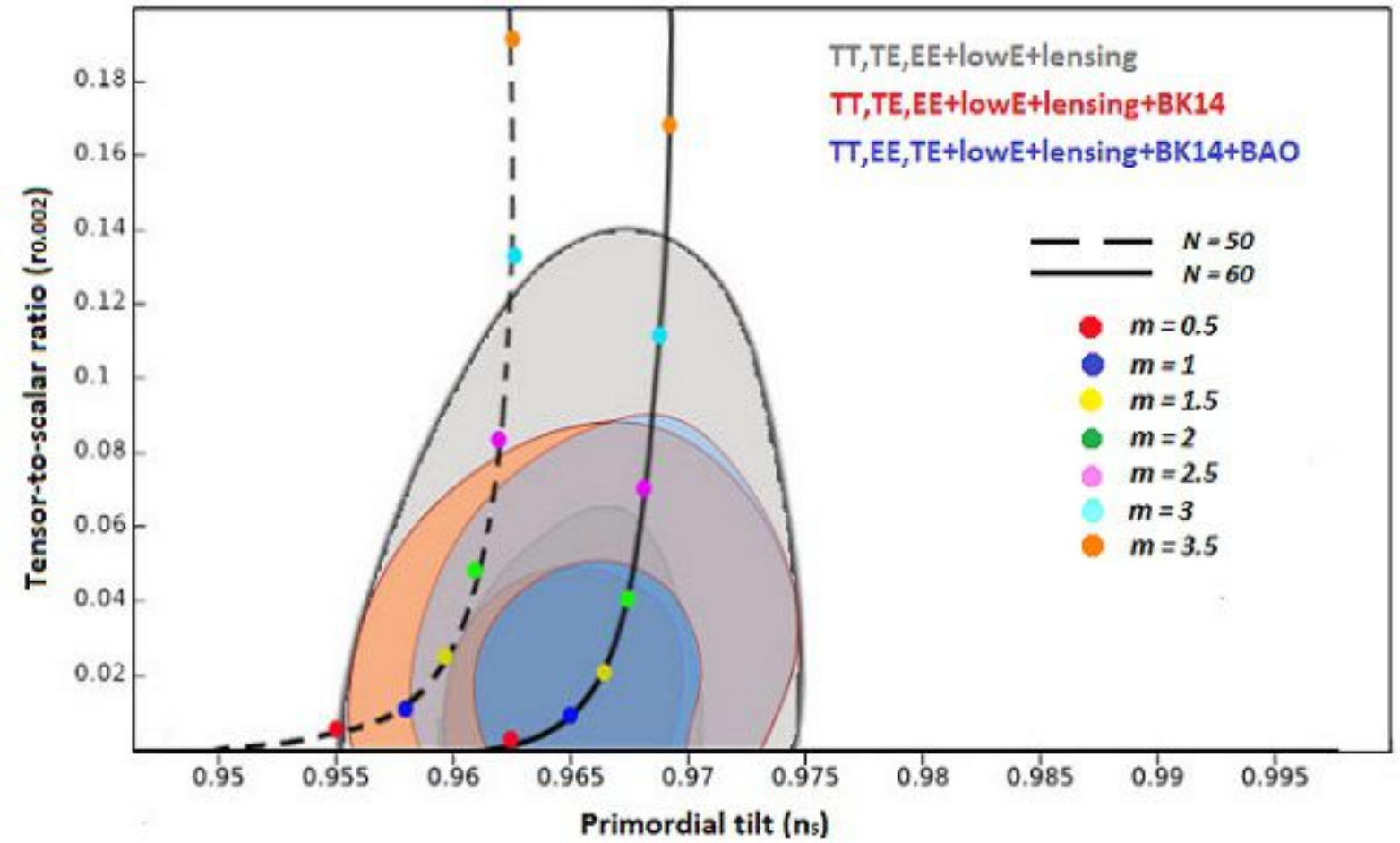}
	\caption{The marginalized joint 68\% and 95\% CL regions for $n_{s}$ and $r$ at $k = 0.002$ Mpc$^{-1}$ from Planck 2018 in combination with BK14+BAO data \cite{cmb} and the $n_{s}-r$ constraints on the intermediate model (\ref{22}) in the case of $\Gamma$ and $\xi$ as variable parameters. The dashed and solid lines represent $N=50$ and $N=60$, respectively. The panel is plotted for different values of $m$ when $\alpha=1$, $\beta=0.9$, $\gamma=1.5$, $\xi_{0}\sim10^{-8}$ and $T_{r}\sim10^{-5}$ \cite{c4,caption}.}
	\label{fig7}
\end{figure*}
\begin{equation}
\Gamma=\Gamma_{0},\hspace{1cm}\xi=\xi_{0}.
\label{23}    
\end{equation}
Then by substitution of the Eq. (\ref{22}) in (\ref{11}), we obtain that the reconstruction of the effective potential and the temporal evolution of inflaton become
\begin{equation}
V(\varphi)=(1+2m)(\alpha\beta)^{2}\Big(\frac{\varphi}{w}\Big)^{\frac{4(\beta-1)}{2\beta-1}},\hspace{1cm}\varphi(t)=wt^{\beta-\frac{1}{2}},
\label{24}    
\end{equation}
where $w=\sqrt{\frac{2(1+2m)(1-\beta)}{\Gamma_{0}}}\,\left[\frac{2\alpha\beta}{2\beta-1}\right]$ and with the parameter $\beta\neq 1/2$. Here, we mention that the results given by Eq. (\ref{24}) for $\varphi(t)$ and $V(\varphi)$ coincide with the found in ref. \cite{c36}.
Also, combining the Eqs. (\ref{24}) and (\ref{11}), the energy density of imperfect fluid is given by
\begin{equation}
\rho=TS=\frac{\alpha\beta}{\gamma(m+2)}\bigg(2(1+2m)(1-\beta)\Big(\frac{\varphi}{w}\Big)^{\frac{2(\beta-2)}{2\beta-1}}+(m+2)^{2}\xi_{0}\Big(\frac{\varphi}{w}\Big)^{\frac{2(\beta-1)}{2\beta-1}}\bigg).  
\label{25}    
\end{equation}

The reconstruction of the effective potential and the evolution of inflaton are plotted in panels (a) and (b) of Figure \ref{fig1}, respectively. The panel (a) is drawn for different values of $m$ when $\alpha=0.03$, $\beta=0.75$ and $\Gamma_{0}\sim10^{-3}$, see refs. \cite{c4,caption}. Also, the panel (b) is drawn for different values of $m$ when $\alpha=0.1$, $\beta=0.75$ and $\Gamma_{0}\sim10^{-3}$. 
\begin{figure*}[!hbtp]
	\centering
	\includegraphics[width=.32\textwidth,keepaspectratio]{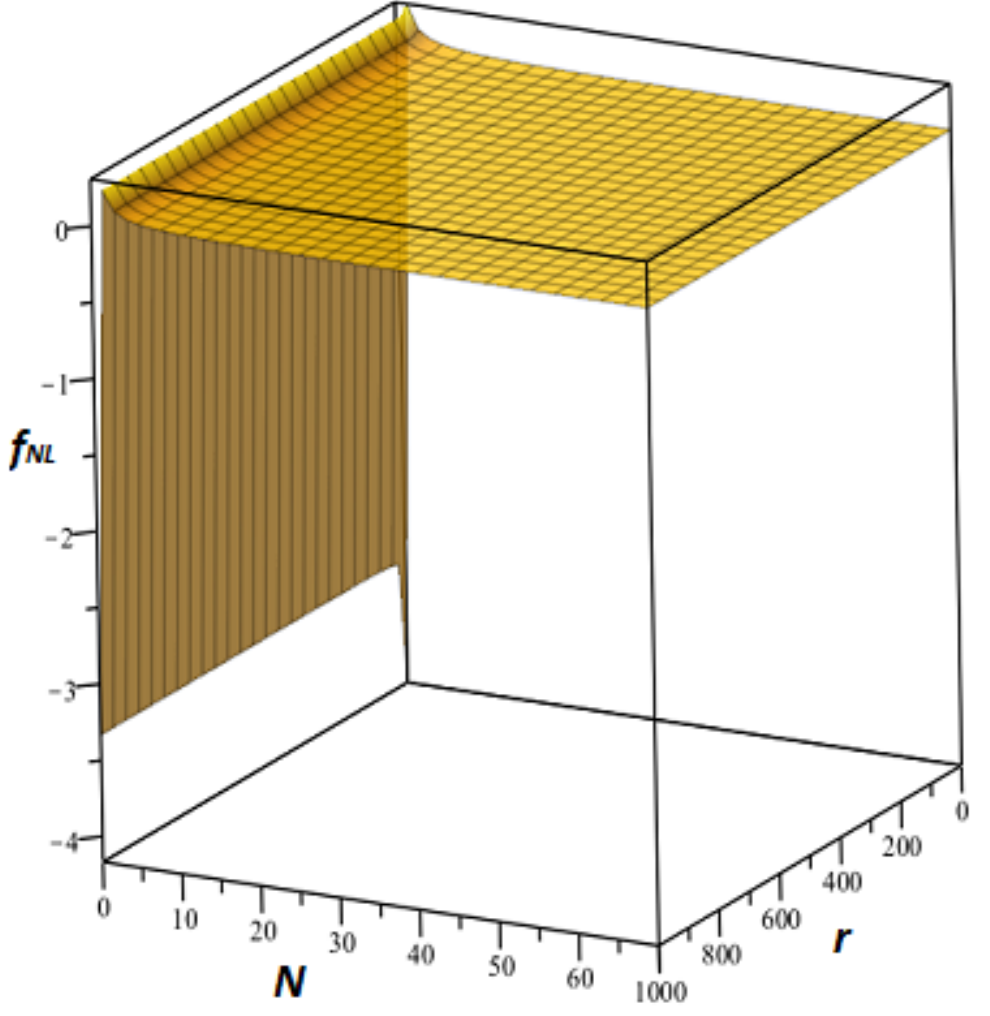}
	\caption{The non-linear parameter $f_{NL}$ versus number of e-folds $N$ and dissipation strength $r$ in intermediate model in the absence of the second term of the Eq. (\ref{41}). The figure is plotted for $\alpha=1$, $\beta=0.9$, $\gamma=1.5$, $\xi_{0}\sim 10^{-8}$ and $T_{r}\sim10^{-5}$ \cite{c4,caption}.}
	\label{fig8}
\end{figure*}
\begin{figure*}[!hbtp]
	\centering
	\includegraphics[width=.32\textwidth,keepaspectratio]{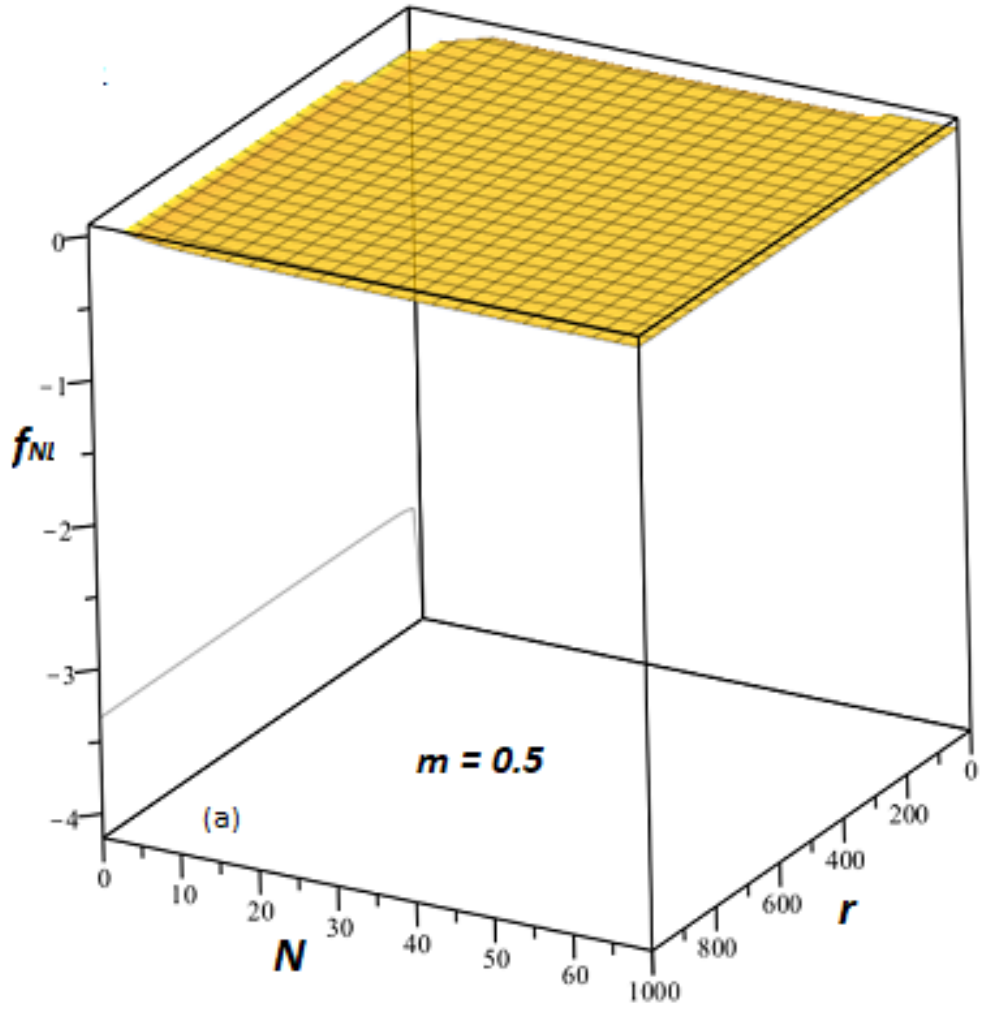}
	\includegraphics[width=.32\textwidth,keepaspectratio]{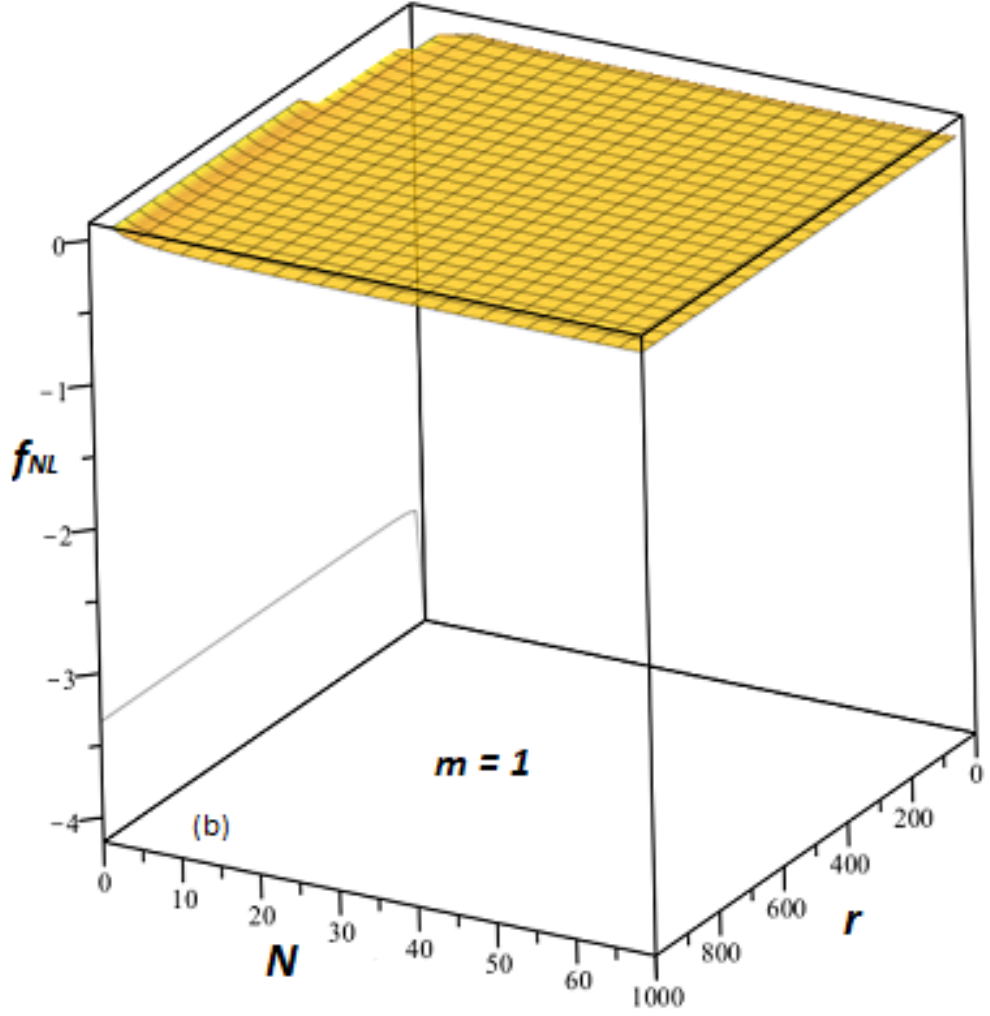}
	\includegraphics[width=.32\textwidth,keepaspectratio]{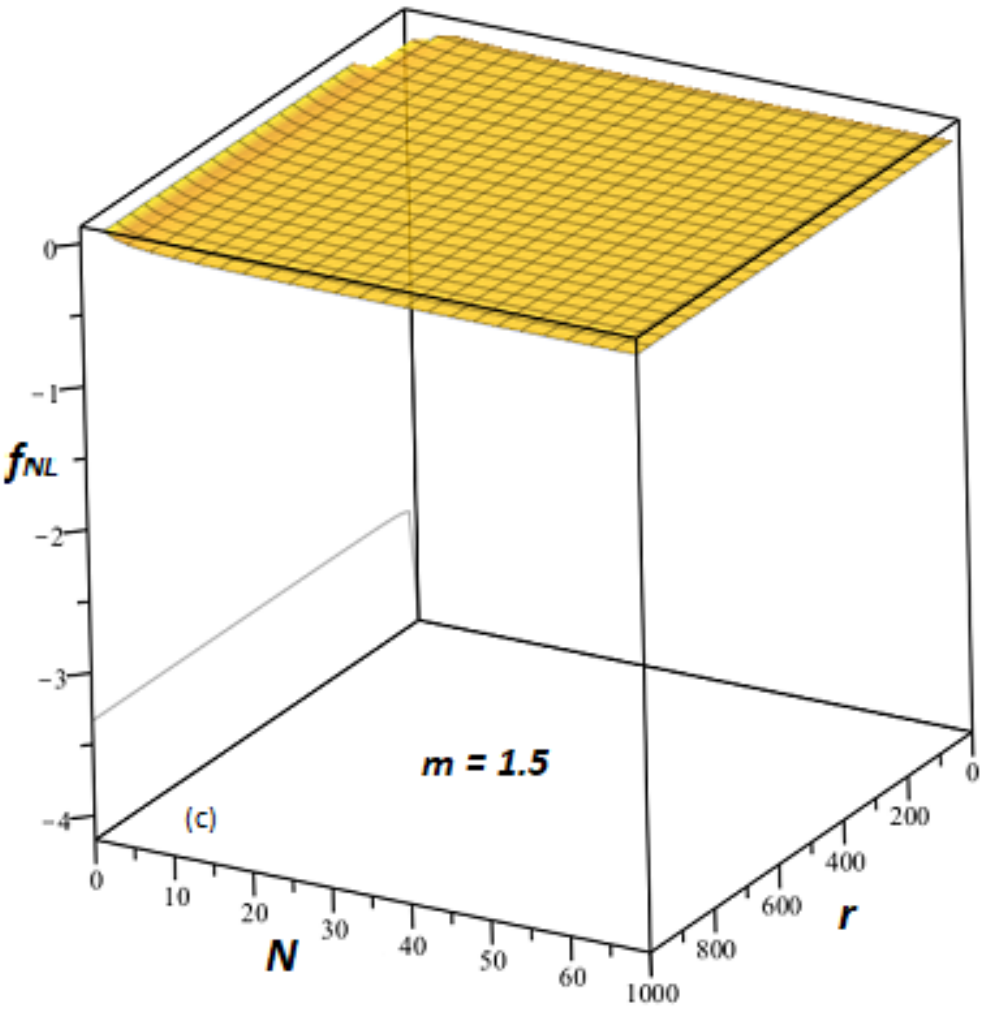}
	\includegraphics[width=.32\textwidth,keepaspectratio]{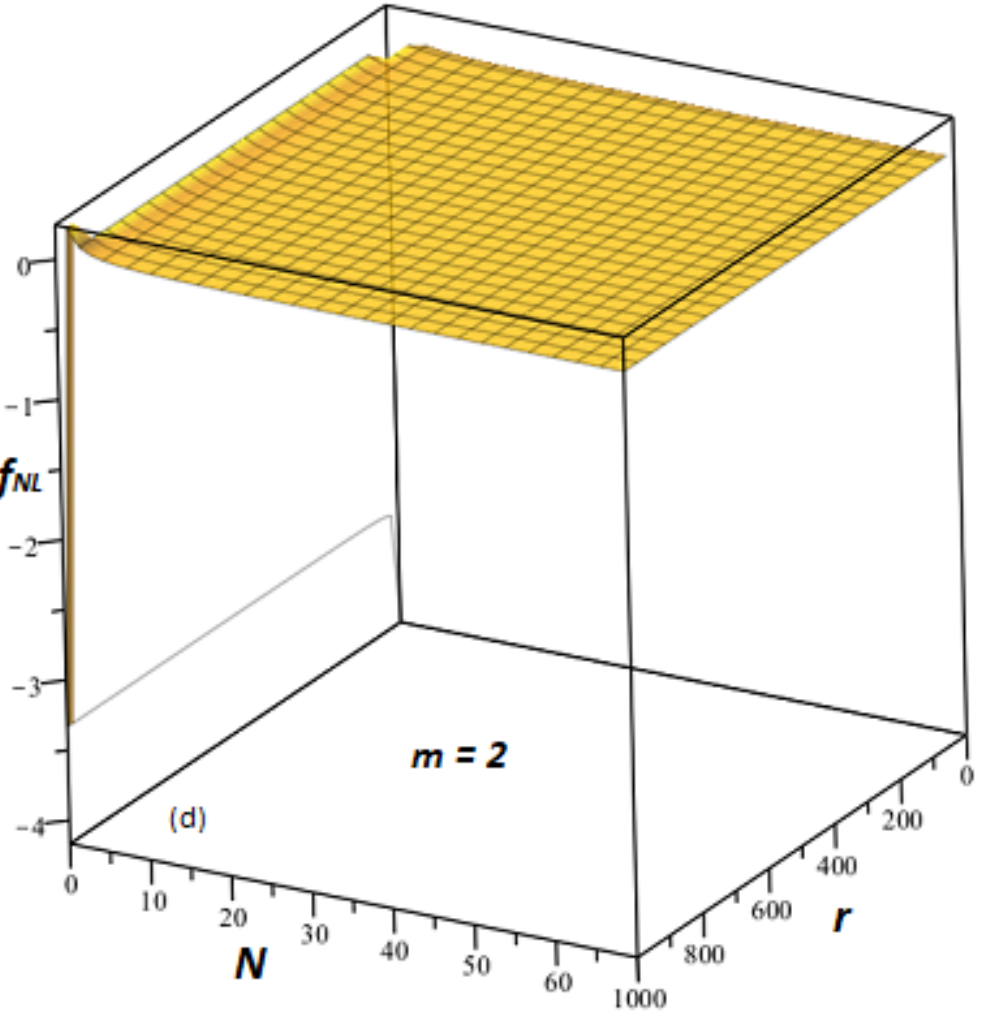}
	\includegraphics[width=.32\textwidth,keepaspectratio]{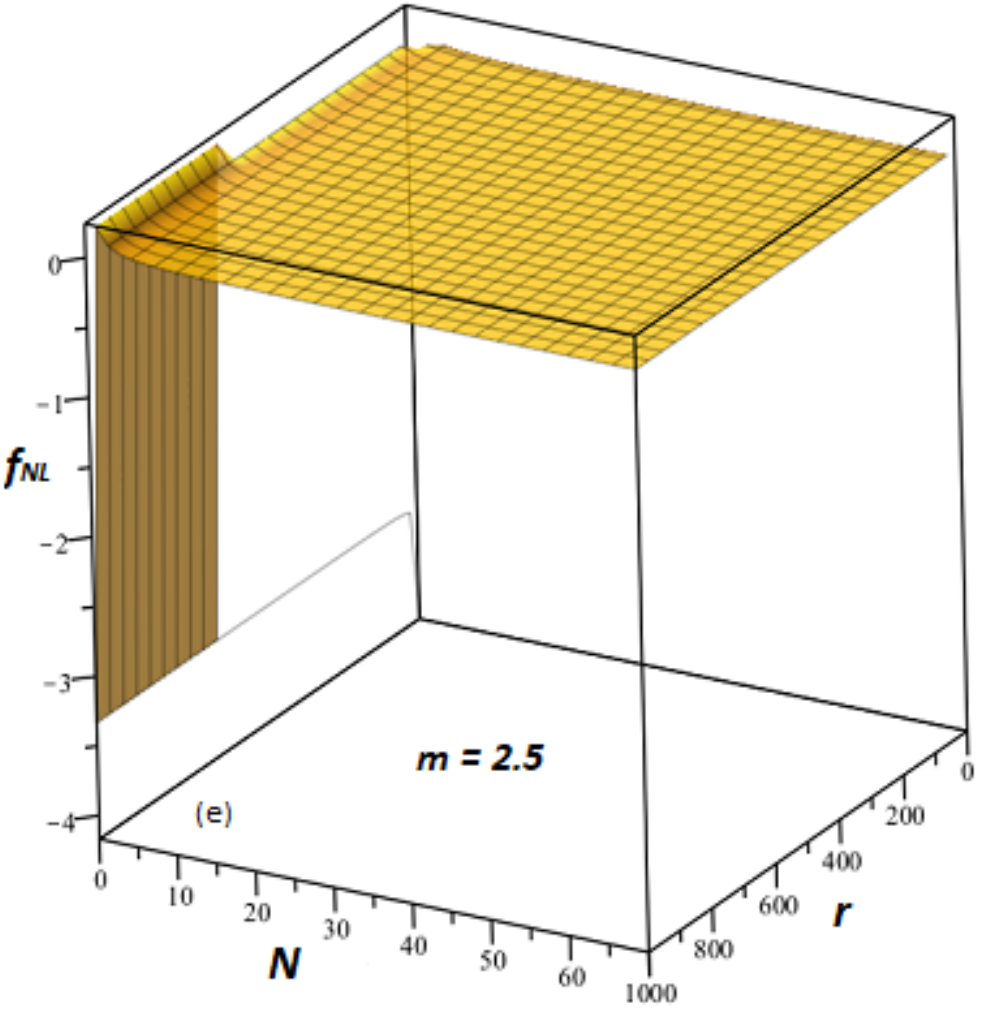}
	\includegraphics[width=.32\textwidth,keepaspectratio]{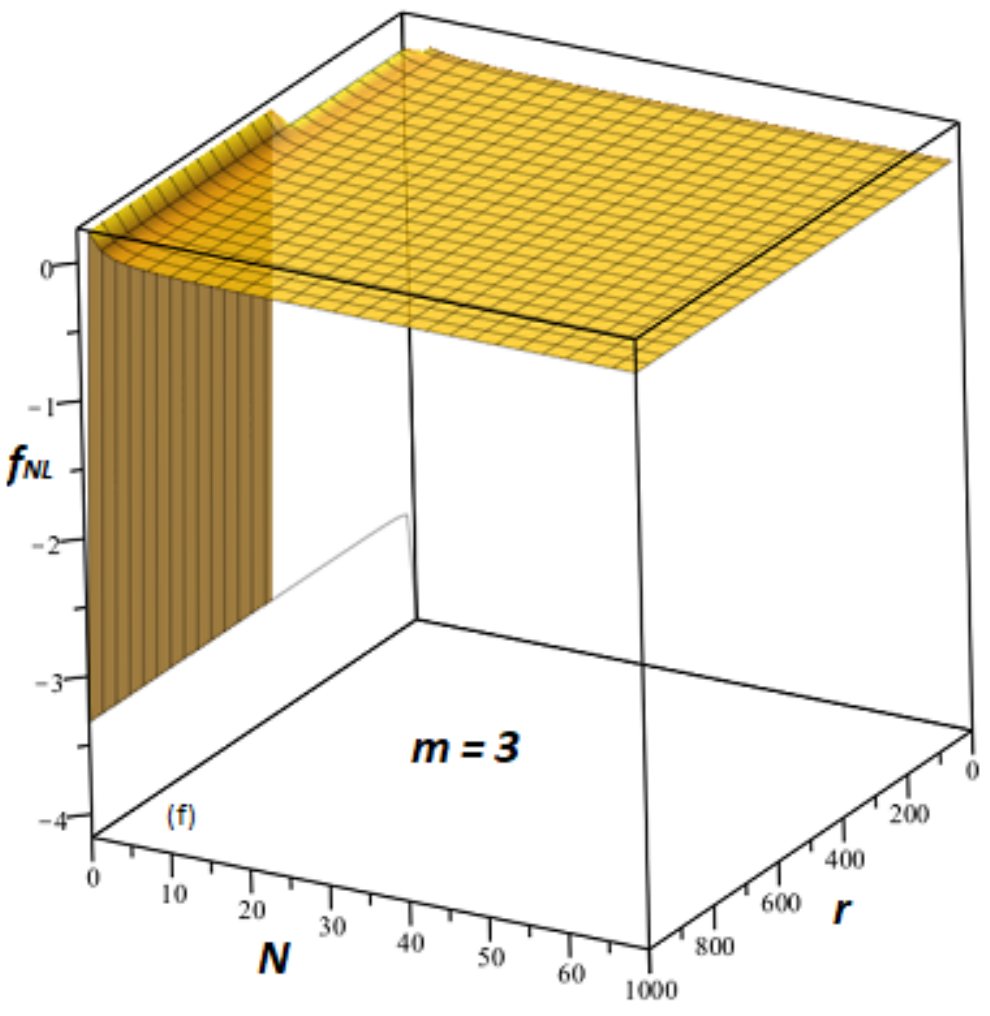}
	\caption{The non-linear parameter $f_{NL}$ versus number of e-folds $N$ and dissipation strength $r$ for different values of $m$ in intermediate model in the presence of the second term of the Eq. (\ref{41}). The figure is plotted for $\alpha=1$, $\beta=0.9$, $\gamma=1.5$, $\xi_{0}\sim 10^{-8}$ and $T_{r}\sim10^{-5}$ \cite{c4,caption}.}
	\label{fig9}
\end{figure*}
The panel (c) shows the behaviour of the entropy of imperfect fluid versus the scalar field for different values of $m$ when $\alpha=0.003$, $\beta=0.75$, $\gamma=1.5$, $\xi_{0}\sim10^{-6}$, $\Gamma_{0}\sim10^{-3}$ and $T\sim10^{-5}$. Here, we note that the entropy increases for values of $m>1$ and decreases for $m<1$ with respect to the case of $m=1$ as the isotropic universe. 

The slow-roll 
\begin{figure*}[!hbtp]
	\centering
	\includegraphics[width=.32\textwidth,keepaspectratio]{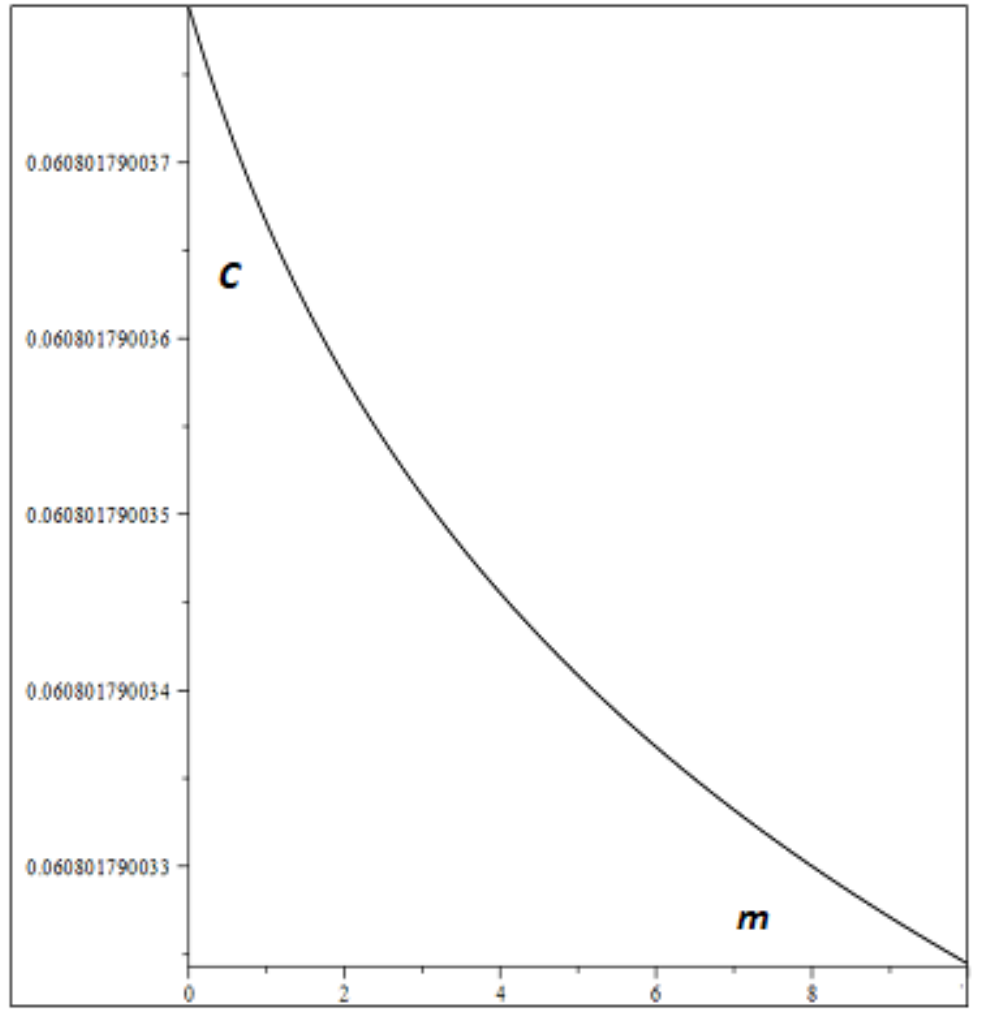}
	\hspace{0.5cm}
	\includegraphics[width=.32\textwidth,keepaspectratio]{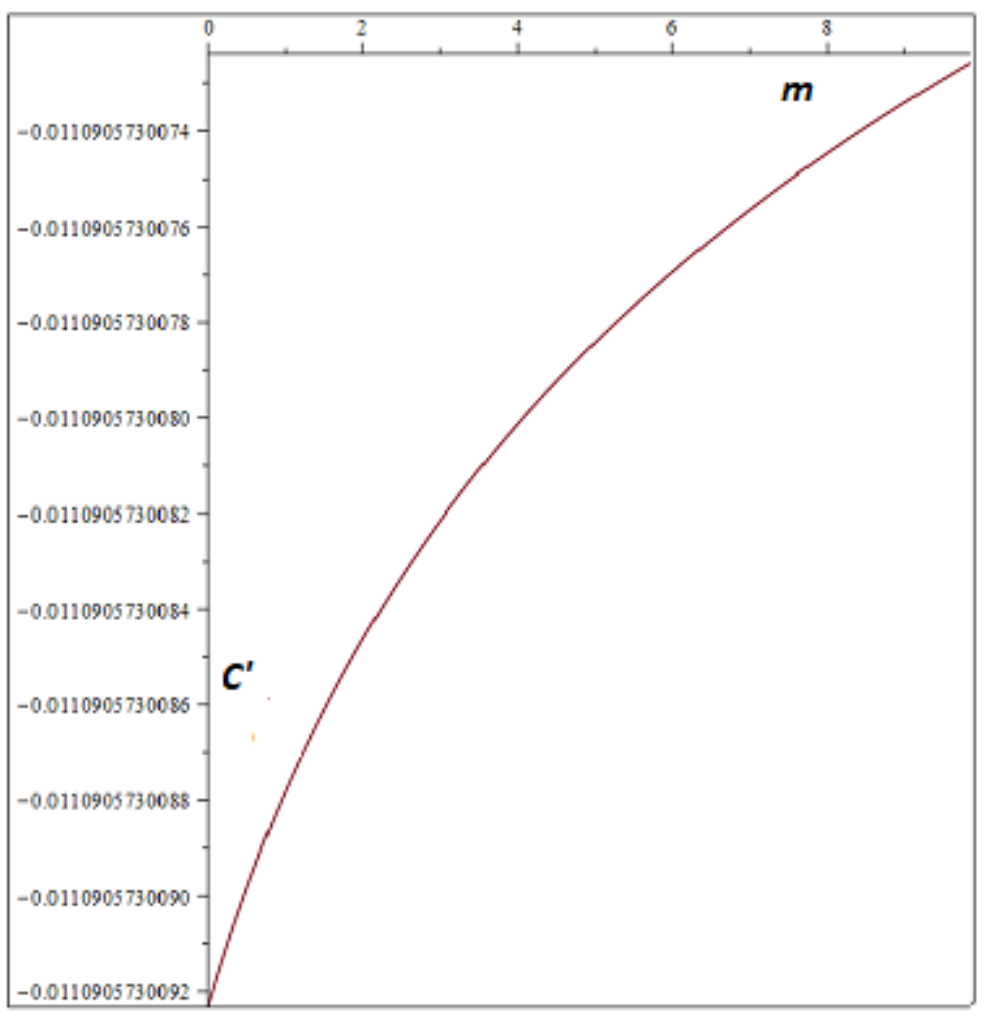}
	\caption{The behaviour of the swampland parameters $c$ and $c'$ versus $m$ in intermediate model (\ref{22}) in the case of $\Gamma$ and $\xi$ as variable parameters when $\alpha=1$, $\beta=0.9$ and $N=60$.}
	\label{fig10}
\end{figure*}
parameters (\ref{12}) of the model are driven by
\begin{equation}
\epsilon=\frac{3(1-\beta)}{\alpha\beta(m+2)}\Big(\frac{\varphi}{w}\Big)^{\frac{2\beta}{1-2\beta}},\hspace{1cm}\eta=\frac{3(2-\beta)}{\alpha\beta(m+2)}\Big(\frac{\varphi}{w}\Big)^{\frac{2\beta}{1-2\beta}}.
\label{26}    
\end{equation}
Also, the number of $e$-folds (\ref{14}) takes the following form 
\begin{equation}
N(\varphi)=\frac{\alpha(m+3)}{3}\bigg(\Big(\frac{\varphi_{f}}{w}\Big)^{\frac{2\beta}{2\beta-1}}-\Big(\frac{\varphi_{i}}{w}\Big)^{\frac{2\beta}{2\beta-1}}\bigg).
\label{27}    
\end{equation}

Following Refs. \cite{c24,c25}, we consider that the inflationary scenario begins at the earliest possible epoch in which the parameter $\epsilon=\epsilon(\varphi=\varphi_i)=1$. In this form, we find that  the initial value of the scalar field $\varphi_i$ results
\begin{equation}
\varphi_{i}=w\bigg(\frac{3(1-\beta)}{\alpha\beta(m+2)}\bigg)^{\frac{2\beta-1}{2\beta}},
\label{28}    
\end{equation}
and then, form the Eqs. (\ref{27}) and (\ref{28}), the value of inflaton at the end of inflation can be obtained by
\begin{equation}
\varphi_{f}=w\bigg(\frac{3}{(m+2)}\Big(\frac{1-\beta}{\alpha\beta}+\frac{N}{\alpha}\Big)\bigg)^{\frac{2\beta-1}{2\beta}}.    
\label{29}     
\end{equation}
Also, we mention that the evolution of the scalar field during the inflationary epoch is given by 
\begin{equation}
\varphi<w\bigg(\frac{3(1-\beta)}{\alpha\beta(m+2)}\bigg)^{\frac{2\beta-1}{2\beta}},    
\end{equation}
here we have considered that during inflation the parameter $\epsilon<1$ (or equivalently $\ddot{a}>0$).

Now we can calculate the spectral index (\ref{18}) and tensor-to-scalar ratio (\ref{19}) of the model using the Eqs. (\ref{26}) and (\ref{29}) as shown (\ref{a1}) and (\ref{a2})
in the Appendix. Figure \ref{2} shows the $n_{s} - r$ constraints coming from the marginalized joint 68\% and 95\% CL regions of the Planck 2018 in combination with BK14+BAO data on the intermediate model (\ref{22}) in the case of $\Gamma$ and $\xi$ as constant parameters \cite{cmb}. The dashed and solid lines represent $N=50$ and $N=60$, respectively. The figure is plotted for different values of $m$ when $\alpha=1$, $\beta=0.9$, $\gamma=1.5$, $\xi_{0}\sim10^{-6}$, $\Gamma_{0}=10^{-3}$ and $T_{r}\sim10^{-5}$. For $N=50$, the values of $n_{s}$ and $r$ related to all considered numbers of $m$ are inconsistent with all three observational datasets. The situation for $N=60$ is different to the case $N=$50, since we find the observational constraint $0.5\leq m<4$ coming from the Planck alone and also in combination with BK14 at only 68\% CL. Moreover, it is obvious that the Planck in combination with BK14+BAO data as a full consideration does not show any constraints on $m$. The non-linear parameter (\ref{21}) of the model is given by 
\begin{equation}
-\frac{3}{5}f_{NL}=\frac{1}{2}\Big(-\epsilon+\frac{\epsilon r}{1+r}+\eta\Big)+\frac{P_{R}}{2}\Big(-\epsilon+\frac{\epsilon r}{1+r}+\eta\Big)^{3},
\label{30}    
\end{equation}
and then using the Eqs. (\ref{12}) and (\ref{14}), we have
\begin{equation}
-\frac{3}{5}f_{NL}=\frac{2r(\beta-1)+(1-2\beta)(1+r)}{4(1+r)(2N(\beta-1)+\beta-1)}+\frac{P_{R}}{2}\bigg(\frac{2r(\beta-1)+(1-2\beta)(1+r)}{2(1+r)(2N(\beta-1)+\beta-1)}\bigg)^{3},
\label{31}    
\end{equation}
\begin{figure*}[!hbtp]
	\centering
	\includegraphics[width=.32\textwidth,keepaspectratio]{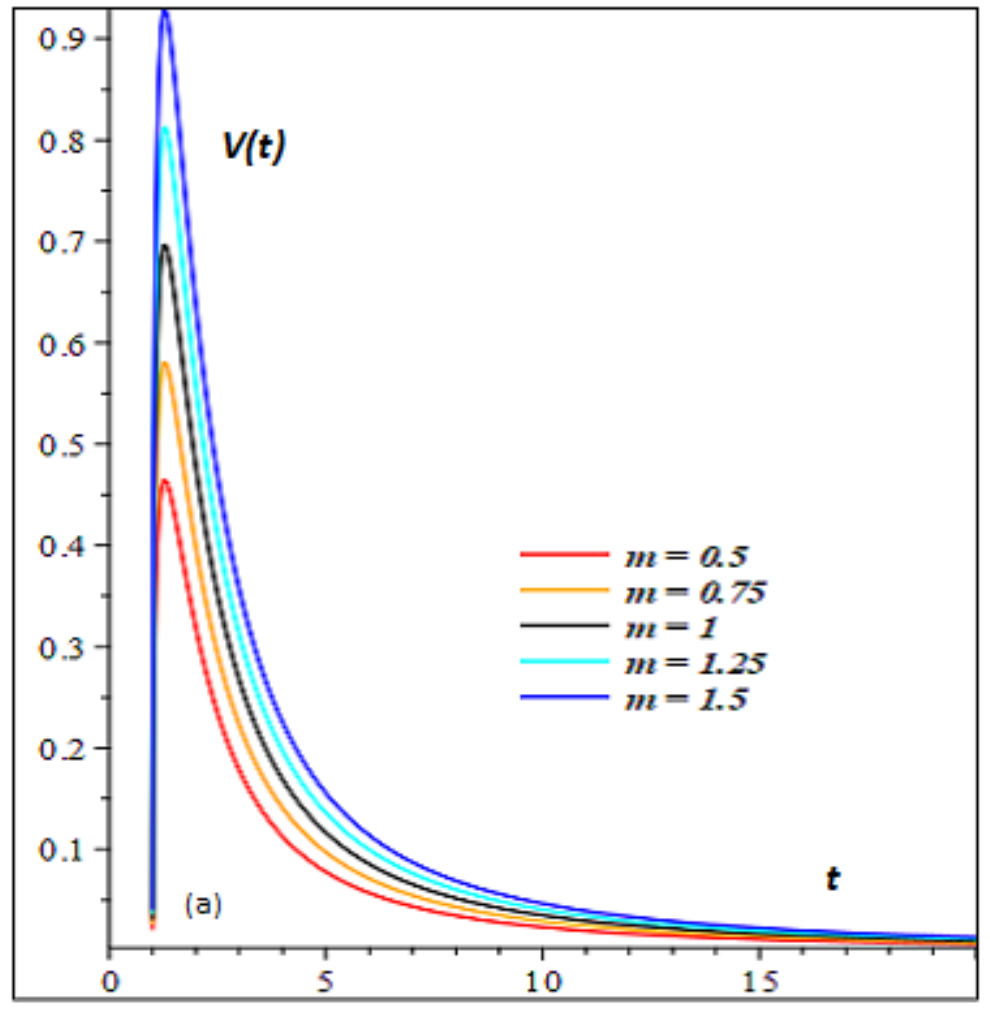}
	\includegraphics[width=.32\textwidth,keepaspectratio]{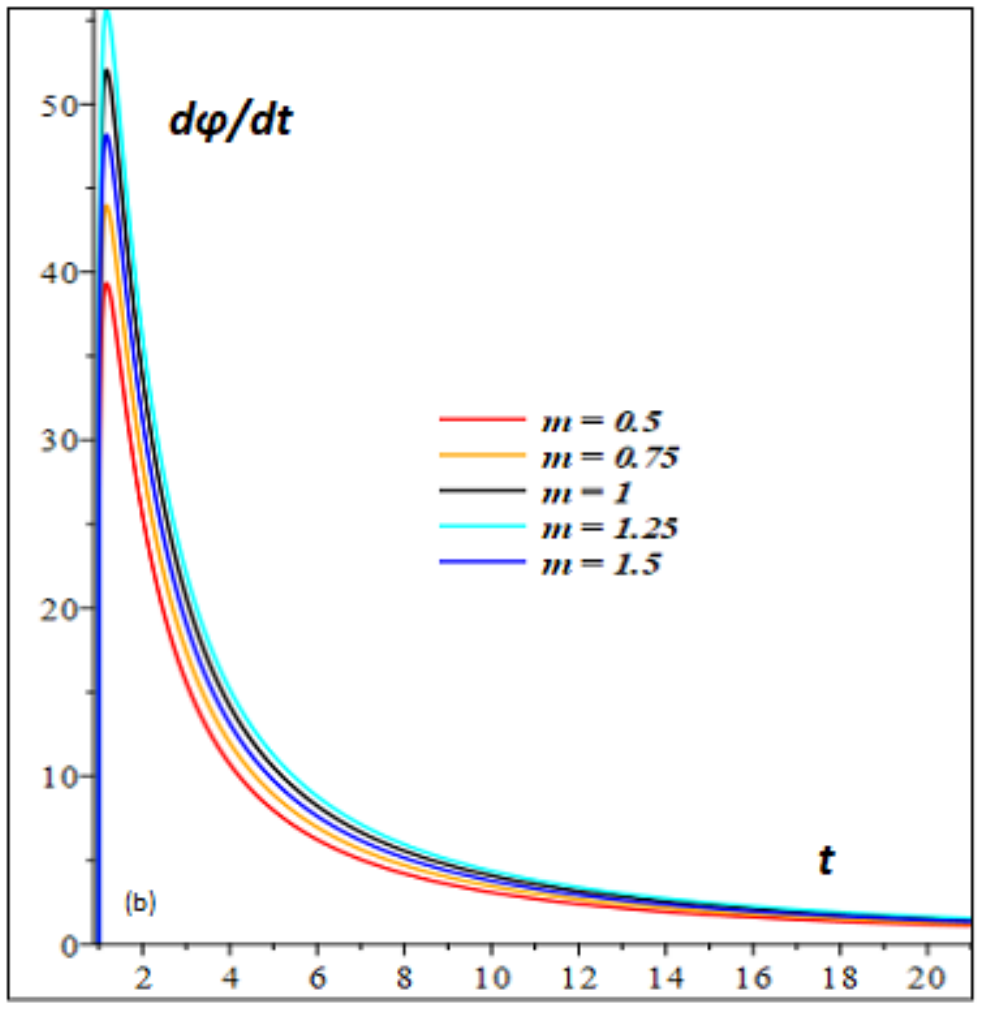}
	\includegraphics[width=.32\textwidth,keepaspectratio]{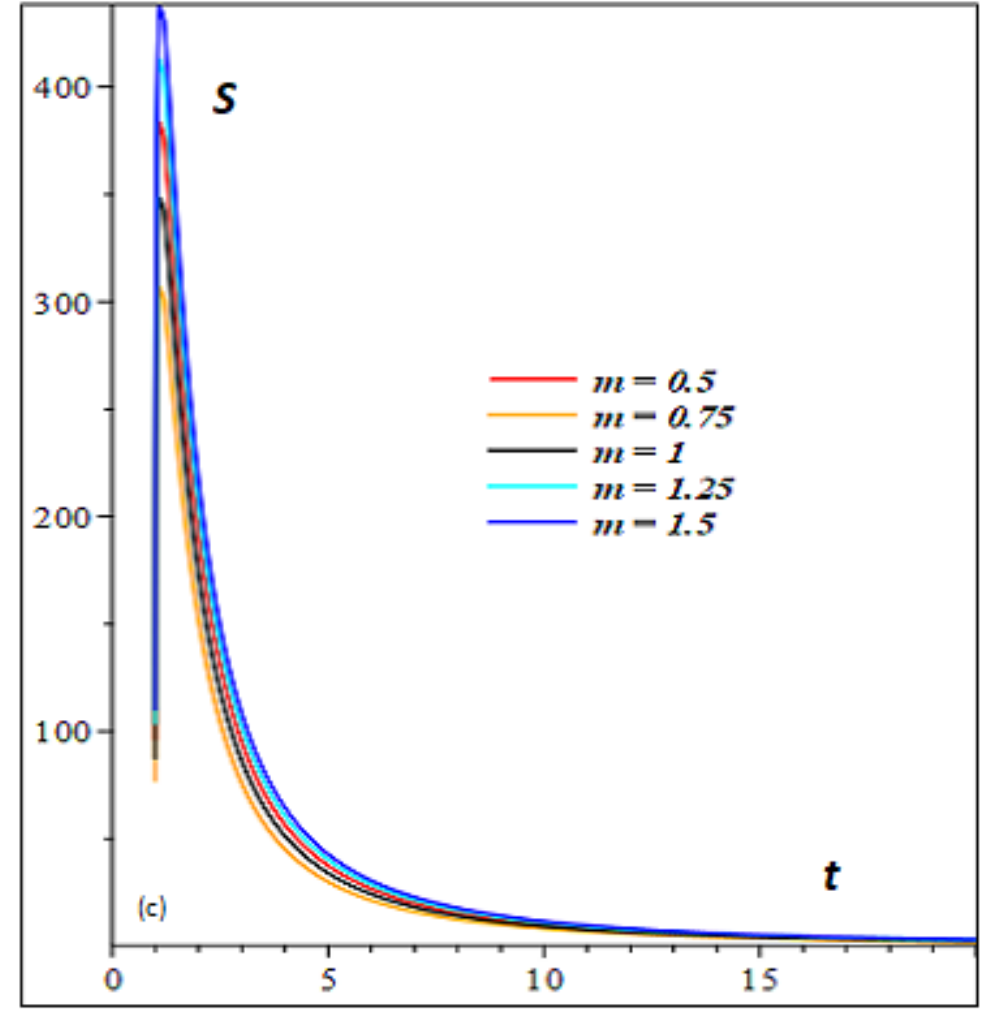}
	\caption{Panel (a): The potential (\ref{43}) plotted versus cosmic time $t$ for different values of $m$ when $\alpha=0.7$ and $\beta=1.25$. Panel (b): The velocity of the scalar field (\ref{43}) plotted versus cosmic time $t$ for different values of $m$ when $\alpha=1$ and $\beta=1.25$. Panel (c): The entropy (\ref{44}) plotted versus cosmic time $t$ for different values of $m$ when $\alpha=0.005$, $\beta=1.25$, $\gamma=1.5$, $\xi_{0}\sim 10^{-6}$ and $T\sim10^{-5}$ \cite{caption}.} 
	\label{fig11}
\end{figure*} 
\begin{figure*}[!hbtp]
	\centering
	\includegraphics[width=.65\textwidth,keepaspectratio]{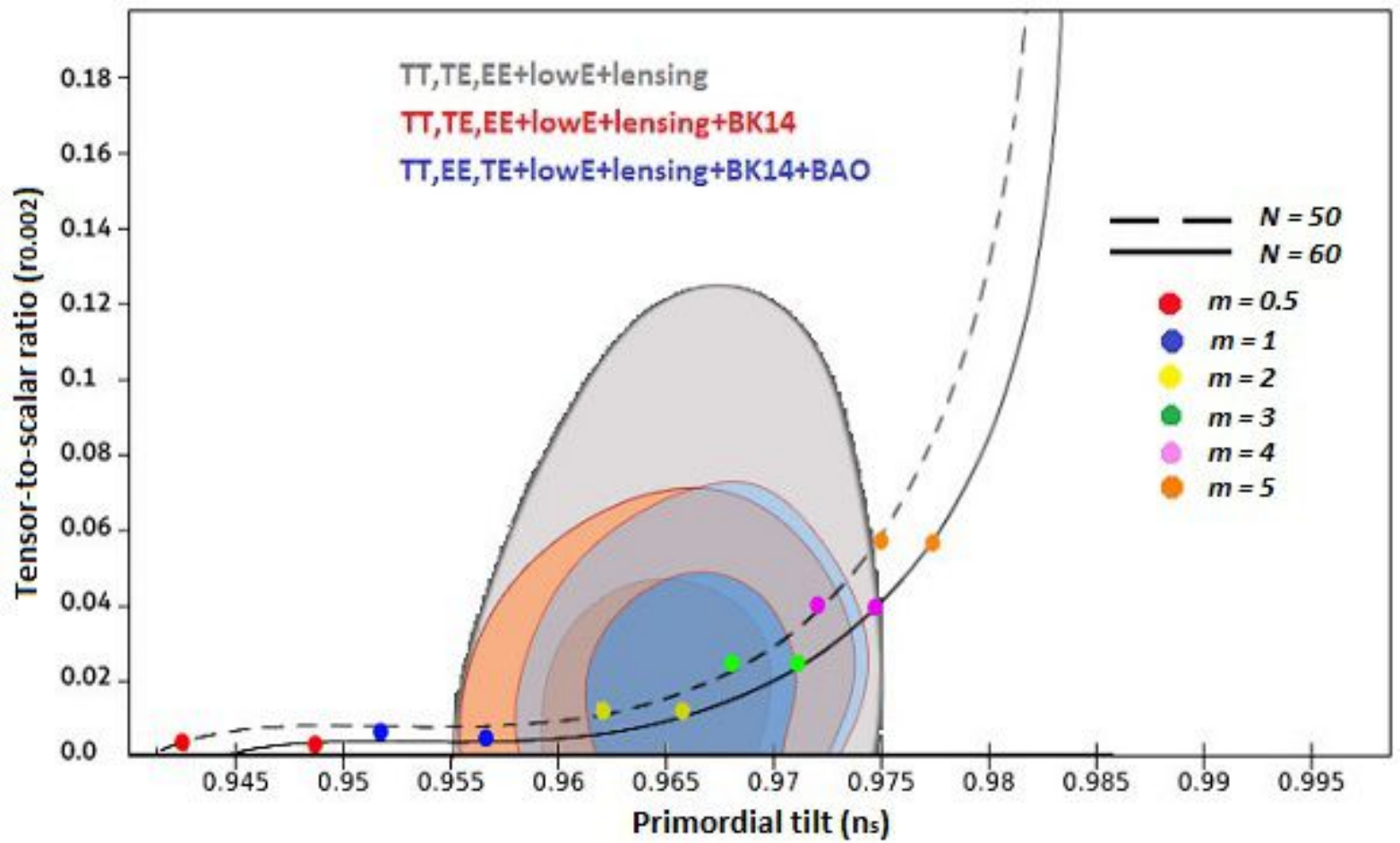}
	\caption{The marginalized joint 68\% and 95\% CL regions for $n_{s}$ and $r$ at $k = 0.002$ Mpc$^{-1}$ from Planck 2018 in combination with BK14+BAO data \cite{cmb} and the $n_{s}-r$ constraints on the logamediate model (\ref{42}) in the case of $\Gamma$ and $\xi$ as constant parameters. The dashed and solid lines represent $N=50$ and $N=60$, respectively. The panel is plotted for different values of $m$ when $\alpha=0.005$, $\beta=6$, $\gamma=1.5$, $\xi_{0}\sim10^{-6}$, $\Gamma_{0}\sim10^{-3}$ and $T_{r}\sim10^{-5}$ \cite{caption}.}
	\label{fig12}
\end{figure*}
where as before $P_{R}$ is the power spectrum of curvature perturbations. 
Figure \ref{3} shows the changes of the non-linear parameter $f_{NL}$ versus the number of $e$-folds $N$ and the dissipation strength $r$ for the intermediate model in the absence of the second term of the Eq. (\ref{31}). In this plot, we have considered the values $\alpha=1$, $\beta=0.9$, $\gamma=1.5$, $\Gamma_{0}\sim10^{-3}$ and $T_{r}\sim10^{-5}$, respectively. As we can see, the parameter $f_{NL}$ changes very fast in both cases of weak ($r\ll1$) and strong ($r\gg1$) dissipation so that it increases and reaches zero by approaching to the end of inflation ($N=70$). Also, we note that the non-lineal parameter $f_{NL}$ in the absence of the second term of the Eq. (\ref{31}) does not depend of the anisotropic parameter $m$. In a more complete case, figure \ref{4} presents the variation of the non-linear parameter $f_{NL}$ versus the number of e-folds $N$ and the dissipation strength $r$ for different values of $m$ when the second term of the Eq.(\ref{31}) is considered in our analysis. The figure reveals that the effects of non-Gaussianity is appeared in strong and weak dissipation for $m=3.5$ and bigger values of $m$, respectively. Also, we observe that for values of the anisotropic parameter $m>3$, the contribution of the  second term of the Eq. (\ref{31}) becomes negligible in the specific case of the strong dissipative regime in which $r\gg$1. Analogous to the previous case, the magnitude of the non-Gaussianity enhances and then reaches zero with the expansion of the universe in both strong and weak dissipative regimes. 

For the sake of completeness, we study the model from the viewpoint of WGC using the conditions of the swampland de Sitter conjecture (\ref{13}). In Figure \ref{5}, we present the behaviour of two swampland parameters $c$ and $c'$ versus $m$ for the intermediate model in the case of $\Gamma$ and $\xi$ as the constant parameters. In these plots, we have used the values $\alpha=1$, $\beta=0.9$ and $N=60$. From the panels, we find that the swampland conditions are given by $0.060817900845< c\leq0.060817900875$ and $-0.0203327185\leq c'<-0.0203327165$ when the observationally favoured values of $m$ are considered. In this form, we find that the range for the parameters associated to the swampland conjecture $c$ and $c'$ is very narrow in order to satisfy the observational data. 
\begin{figure*}[!hbtp]
	\centering
    \includegraphics[width=.32\textwidth,keepaspectratio]{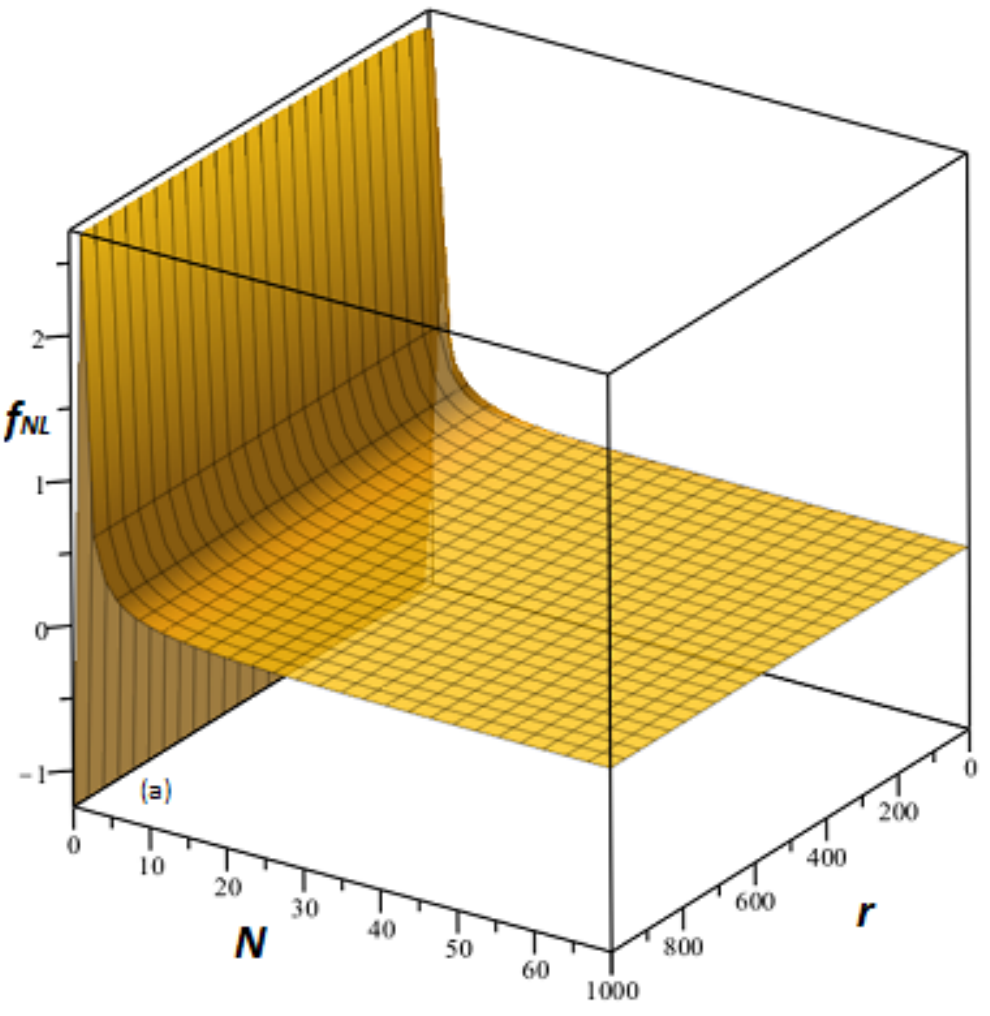}
    \includegraphics[width=.32\textwidth,keepaspectratio]{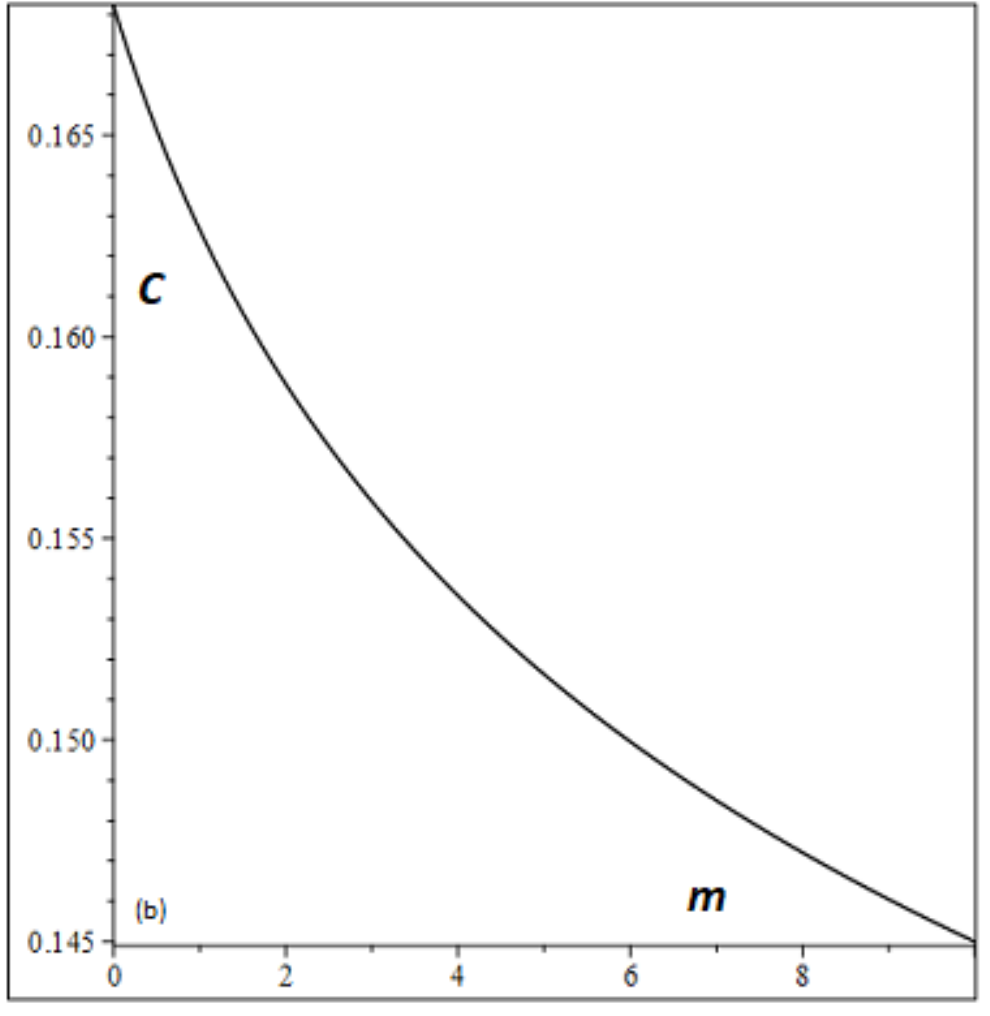}
	\hspace{0.2cm}
	\includegraphics[width=.32\textwidth,keepaspectratio]{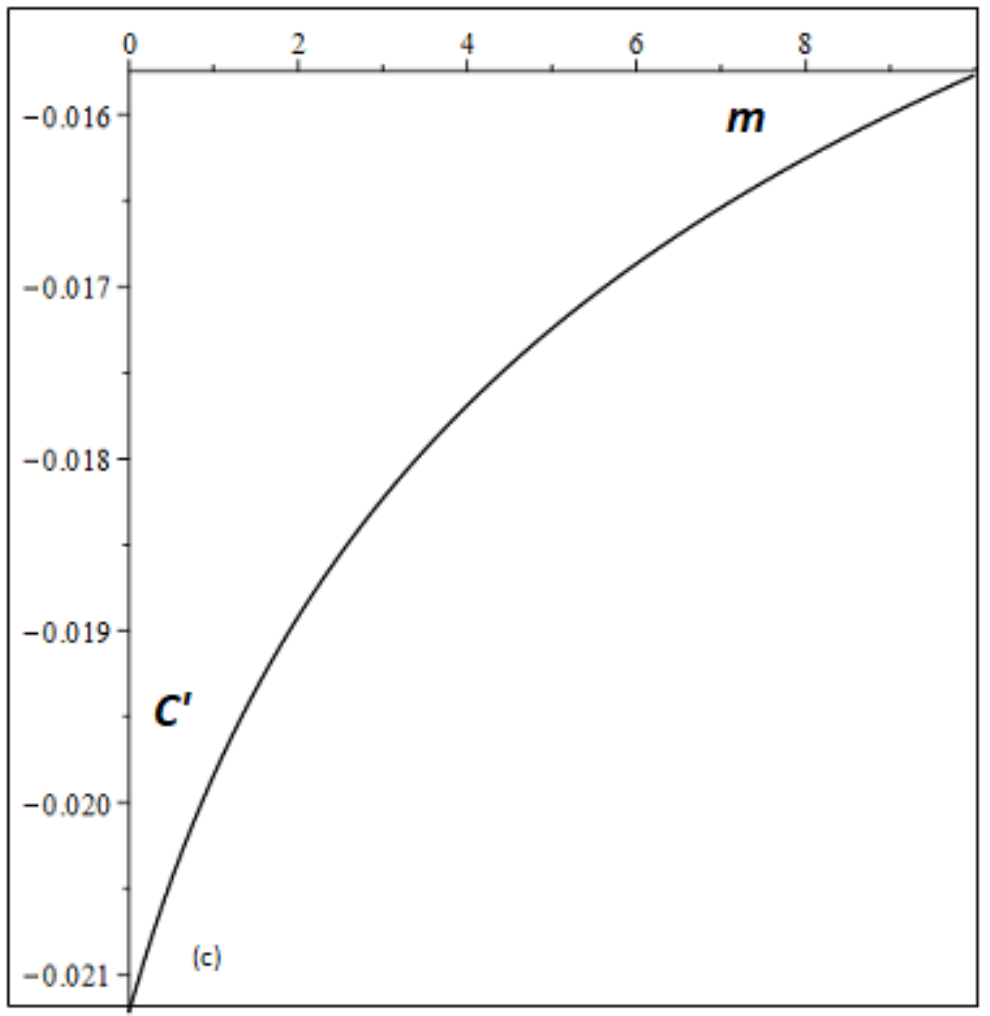}
	\caption{The non-linear parameter $f_{NL}$ versus number of e-folds $N$ and dissipation strength $r$ in logamediate model when $\alpha=1$, $\beta=6$, $\gamma=1.5$, $\xi_{0}\sim 10^{-6}$, $\Gamma_{0}\sim10^{-3}$ and $T_{r}\sim10^{-5}$ (a) \cite{caption}. The behaviour of the swampland parameters $c$ and $c'$ versus $m$ in logamediate model (\ref{42}) in the case of $\Gamma$ and $\xi$ as constant parameters when $\alpha=0.005$, $\beta=6$ and $N=60$ (b and c).}
	\label{fig13}
\end{figure*}

\subsection{$\Gamma$, $\xi$ as variable}

As a more rigorous approach, we consider the dissipation and bulk viscous coefficients as a function of inflaton and imperfect fluid energy 
 density and following refs. \cite{c4,c36} we have
\begin{equation}
\Gamma=\Gamma(\varphi)=\Gamma_{0}V(\varphi),\hspace{1cm} \xi=\xi(\rho)=\xi_{0}\rho.
\label{32}    
\end{equation}
where $\Gamma_{0}$ and $\xi$ are constants. For simplicity, in the following we will consider $\Gamma_{0}=1$. Using the Eqs. (\ref{11}) and (\ref{22}), the temporal evolution of inflaton and the corresponding reconstructed potential are obtained as \cite{c36}
\begin{equation}
\varphi(t)=2\sqrt{2(1-\beta)t},\hspace{1cm}V(\varphi)=(1+2m)(\alpha\beta)^{2}\bigg(\frac{\varphi}{2\sqrt{2(1-\beta)}}\bigg)^{4(\beta-1)}.
\label{33}    
\end{equation}
By taking a look at the obtained potential, we find that it behaves like large field inflationary model $V(\varphi)=V_{0}\varphi^{k}$ so that 
\begin{equation}
V_{0}=(1+2m)(\alpha\beta)^2\Big(2\sqrt{2(1-\beta)}\Big)^{4(1-\beta)},\hspace{1cm}k=4(\beta-1)<0.
\label{34}
\end{equation}
Here, we note that the reconstructed potential   (power law)  decreases for large field and it tends to zero for $\varphi\to \infty$. In Figure (\ref{fig6}), the panel (a) presents the behaviour of the reconstructed potential (\ref{33}) for different values of the anisotropic parameter $m$ when $\alpha=0.25$ and $\beta=0.75$. Panel (b) shows the evolution of inflaton (\ref{33}) versus cosmic time $t$ for the values of $\beta=0.25, 0.5, 0.75$. Also, we show the entropy of imperfect fluid (\ref{35}) versus inflaton in panel (c) of Figure (\ref{fig6}) in which entropy enhances for $m>1$ and declines for $m<1$ with respect to the isotropic universe ($m=1$).

In order to obtain    entropy of imperfect fluid  (Fig. \ref{fig6}, panel (c)), we combine Eqs. (\ref{11}) and (\ref{33}) results
\begin{equation}
S=\frac{2(1+2m)(1-\beta)\alpha\beta\Big(\frac{\varphi}{2\sqrt{2(1-\beta)}}\Big)^{2(\beta-2)}}{(m+2)\bigg(\gamma-(m+2)\alpha\beta\xi_{0}\Big(\frac{\varphi}{2\sqrt{2(1-\beta)}}\Big)^{2(\beta-1)}\bigg)\,T}.
\label{35}    
\end{equation}
 Using the Eq. (\ref{33}), the slow-roll parameters (\ref{12}) are calculated as
\begin{equation}
\epsilon=\frac{3(1-\beta)}{\alpha\beta(m+2)}\Big(\frac{\varphi}{2\sqrt{2(1-\beta)}}\Big)^{-2\beta},\hspace{1cm}\eta=\frac{3(3-2\beta)}{2\alpha\beta(m+2)}\Big(\frac{\varphi}{2\sqrt{2(1-\beta)}}\Big)^{-2\beta}.
\label{36}    
\end{equation}
Also, the number of e-folds (\ref{14}) of the model is given by
\begin{figure*}[!hbtp]
	\centering
	\includegraphics[width=.32\textwidth,keepaspectratio]{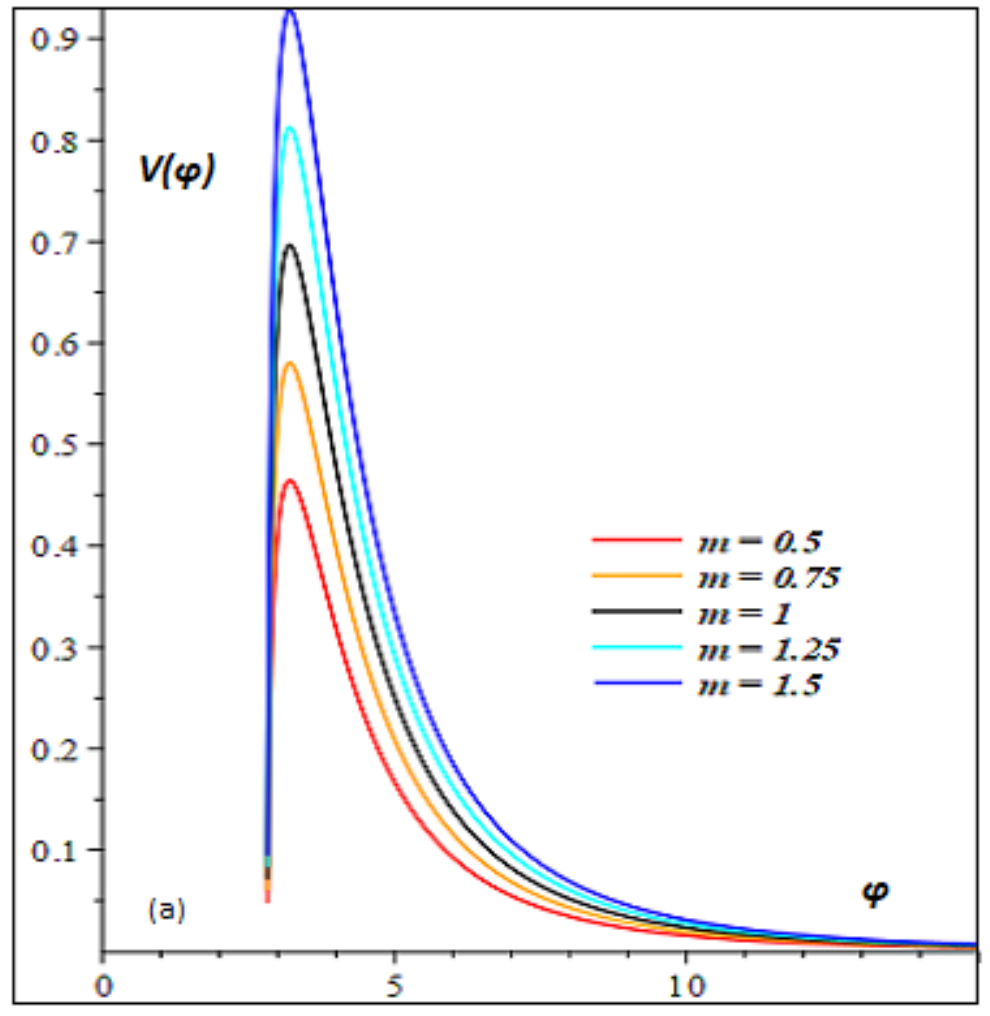}
	\includegraphics[width=.325\textwidth,keepaspectratio]{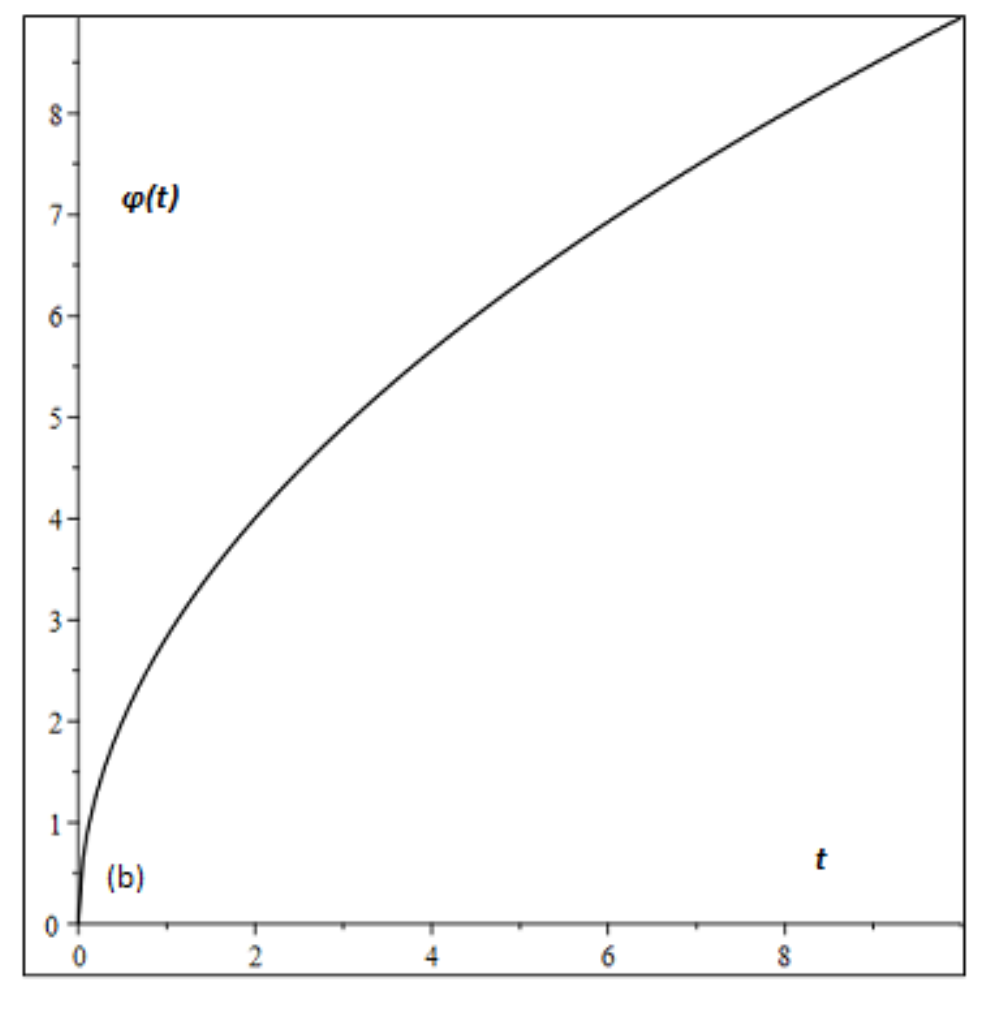}
	\includegraphics[width=.32\textwidth,keepaspectratio]{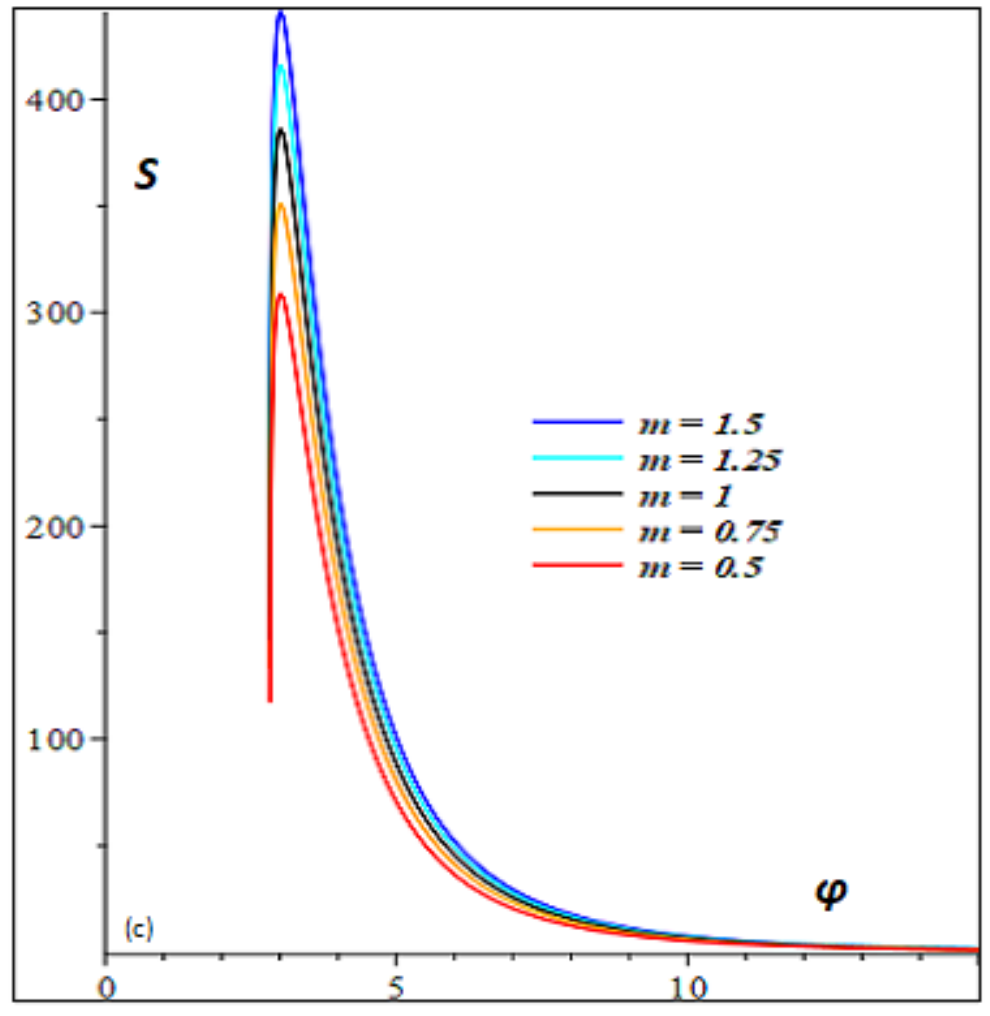}
	\caption{Panel (a): The potential (\ref{50}) plotted versus inflaton $\varphi$ for different values of $m$ when $\alpha=0.7$ and $\beta=1.25$. Panel (b): The evolution of inflaton (\ref{50}) versus cosmic time $t$. Panel (c): The entropy (\ref{51}) plotted versus inflaton $\varphi$ for different values of $m$ when $\alpha=0.005$, $\beta=1.25$, $\gamma=1.5$, $\xi_{0}\sim 10^{-8}$ and $T\sim10^{-5}$ \cite{c4,caption}.} 
	\label{fig14}
\end{figure*} 
\begin{figure*}[!hbtp]
	\centering
	\includegraphics[width=.65\textwidth,keepaspectratio]{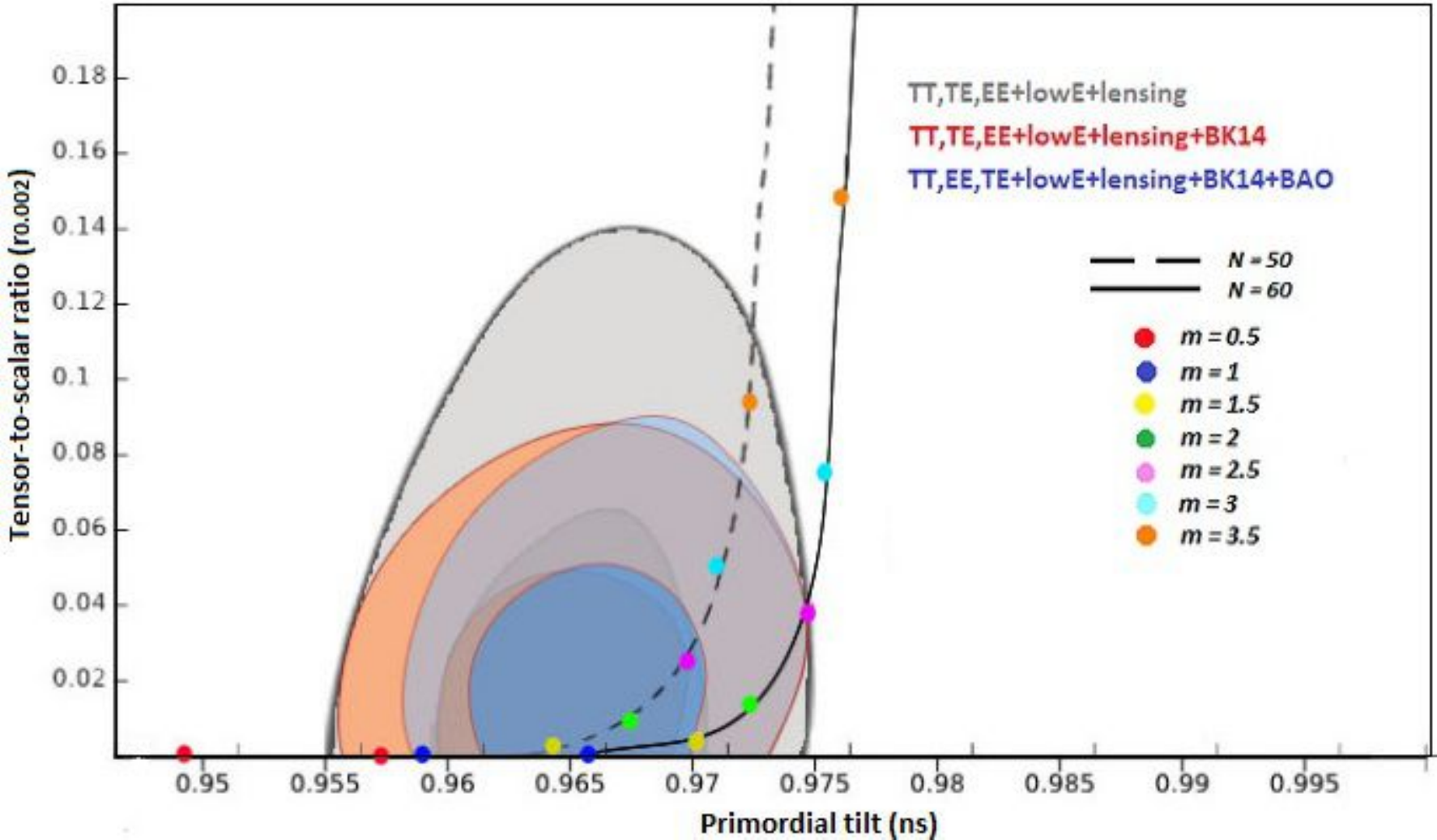}
	\caption{The marginalized joint 68\% and 95\% CL regions for $n_{s}$ and $r$ at $k = 0.002$ Mpc$^{-1}$ from Planck 2018 in combination with BK14+BAO data \cite{cmb} and the $n_{s}-r$ constraints on the logamediate model (\ref{42}) in the case of $\Gamma$ and $\xi$ as variable parameters. The dashed and solid lines represent $N=50$ and $N=60$, respectively. The panel is plotted for different values of anisotropic parameter $m$ when $\alpha=1$, $\beta=6$, $\gamma=1.5$, $\xi_{0}\sim10^{-8}$ and $T_{r}\sim10^{-5}$ \cite{c4,caption}.}
	\label{fig15}
\end{figure*}
\begin{equation}
N=\frac{\alpha(m+2)}{3} \bigg(\Big(\frac{\varphi_{f}}{2\sqrt{2(1-\beta)}}\Big)^{2\beta}-\Big(\frac{\varphi_{i}}{2\sqrt{2(1-\beta)}}\Big)^{2\beta}\bigg).
\label{37}    
\end{equation}
As before from Refs. \cite{c24,c25}, we consider that the inflationary stage begins at the earliest possible epoch where $\epsilon=\epsilon(\varphi=\varphi_i)=1$. Thus, we obtain that the initial value of the scalar field $\varphi_i$ becomes
\begin{equation}
\varphi_{i}=2\sqrt{2(1-\beta)}\Big(\frac{3(1-\beta)}{\alpha\beta(m+2)}\Big)^{\frac{1}{2\beta}}, 
\label{38}
\end{equation}
and by combining (\ref{37}) and (\ref{38}), inflaton value at the end of inflation takes the form 
\begin{equation}
\varphi_{f}=2\sqrt{2(1-\beta)}\bigg(\frac{3}{(m+2)}\Big(\frac{1-\beta}{\alpha\beta}+\frac{N}{\alpha}\Big)\bigg)^{\frac{1}{2\beta}}.
\label{39}    
\end{equation}
\begin{figure*}[!hbtp]
	\centering
	\includegraphics[width=.32\textwidth,keepaspectratio]{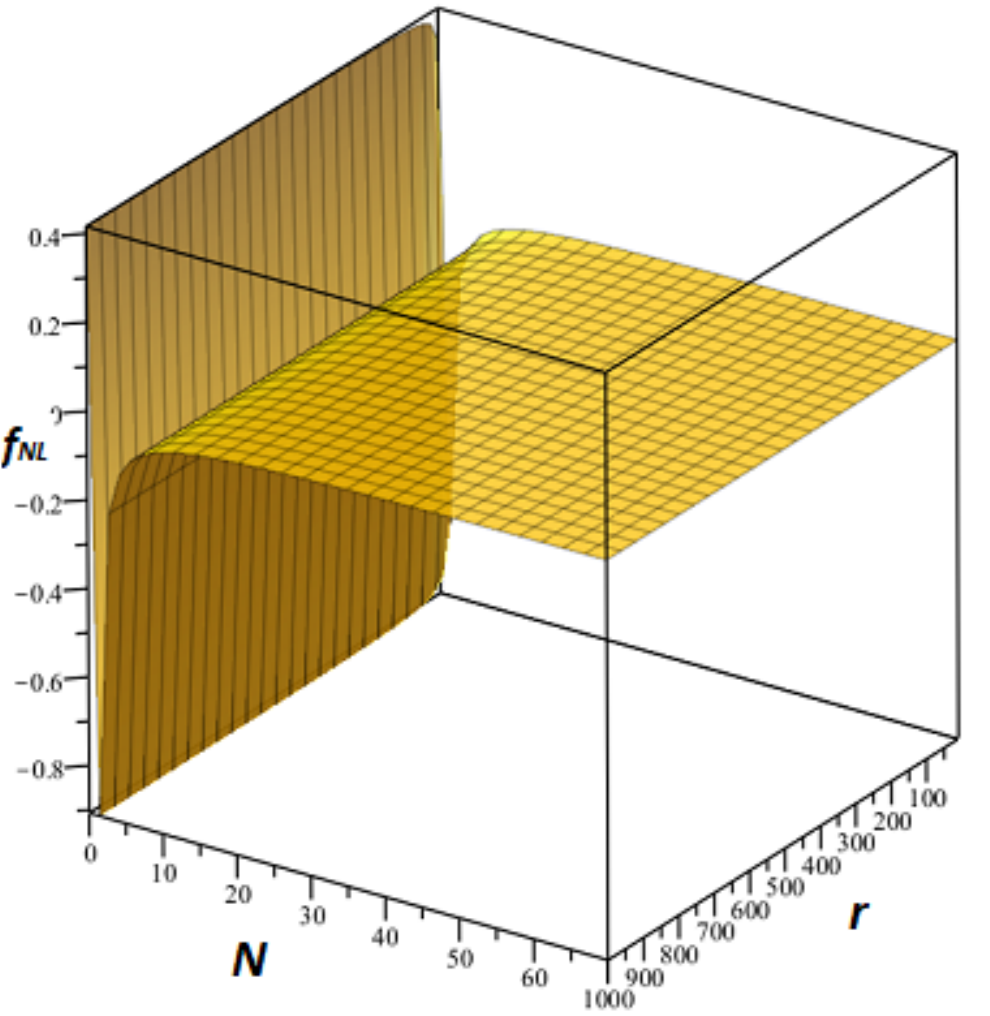}
	\caption{The non-linear parameter $f_{NL}$ versus number of e-folds $N$ and dissipation strength $r$ in logamediate model in the absence of the second term of the Eq. (\ref{56}). The figure is plotted for $\alpha=1$, $\beta=6$, $\gamma=1.5$, $\xi_{0}\sim 10^{-8}$ and $T_{r}\sim10^{-5}$ \cite{c4,caption}.}
	\label{fig16}
\end{figure*}
\begin{figure*}[!hbtp]
	\centering
	\includegraphics[width=.32\textwidth,keepaspectratio]{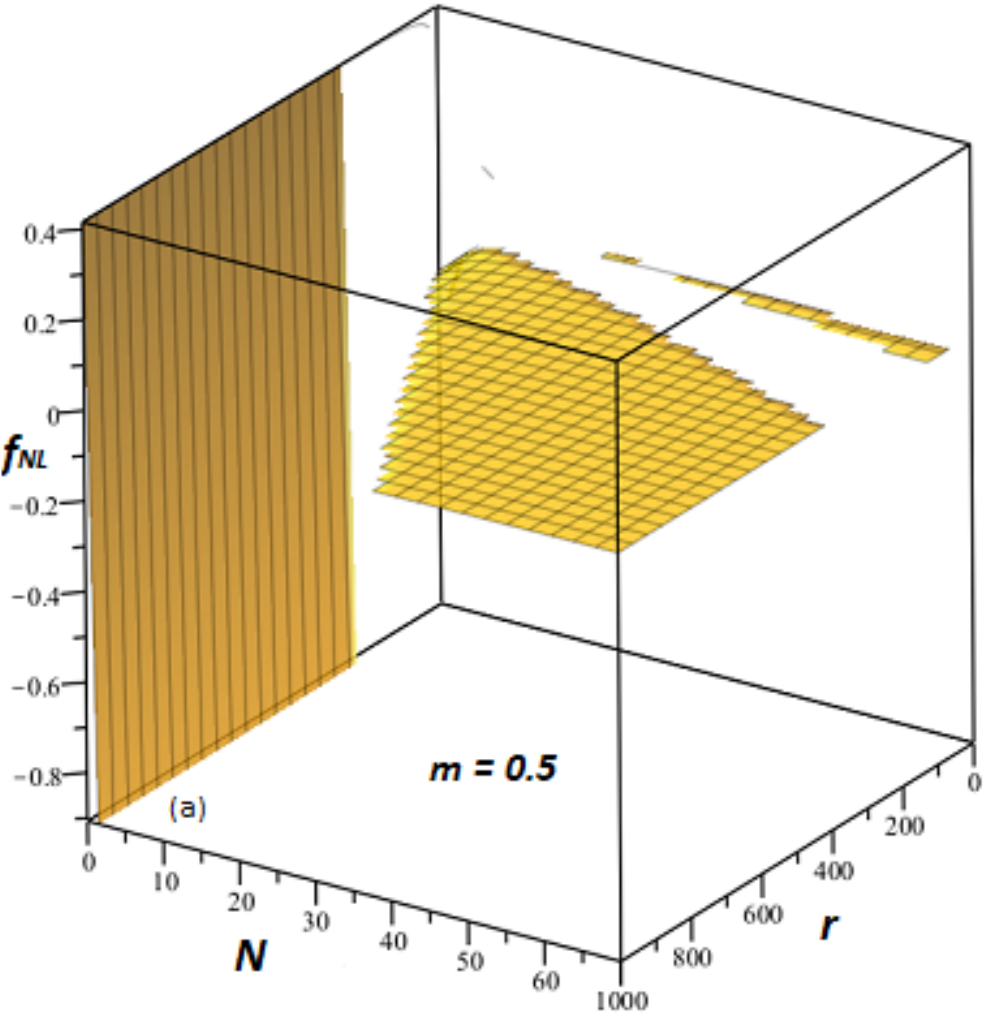}
	\includegraphics[width=.32\textwidth,keepaspectratio]{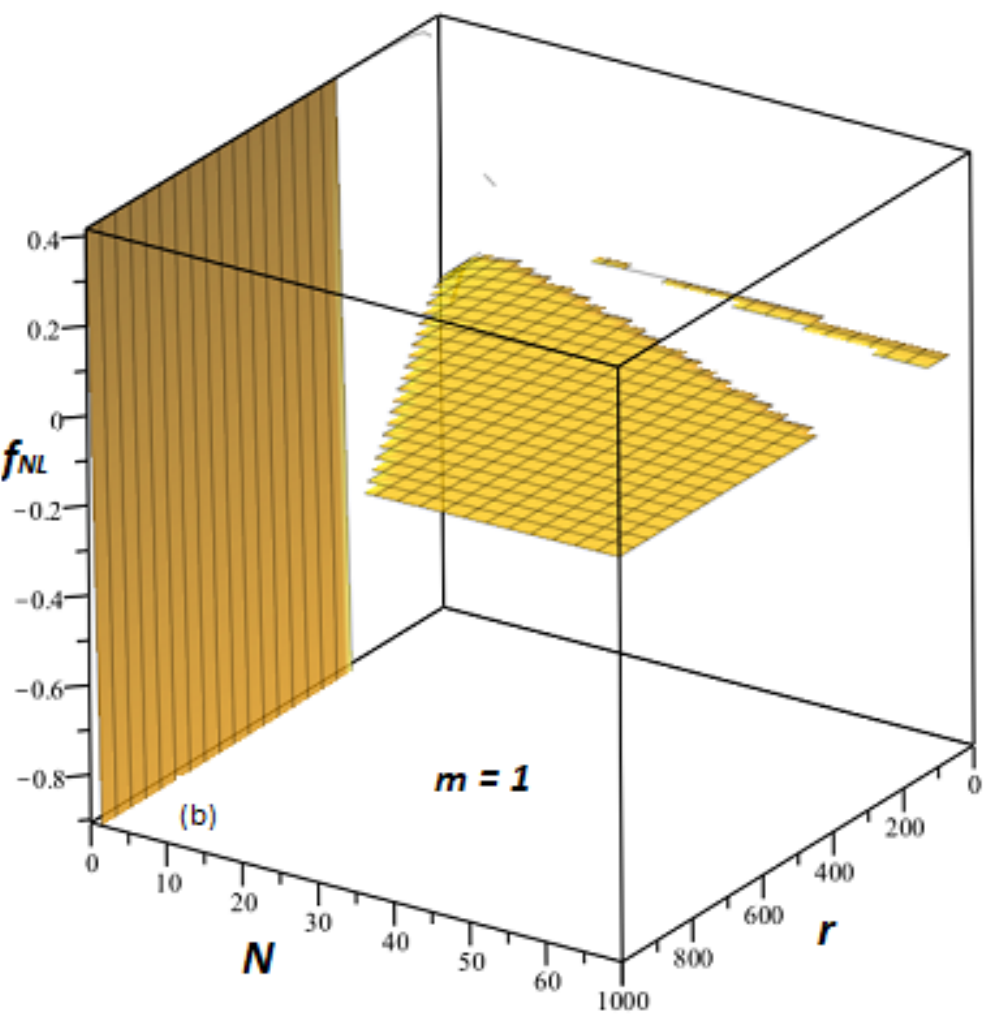}
	\includegraphics[width=.32\textwidth,keepaspectratio]{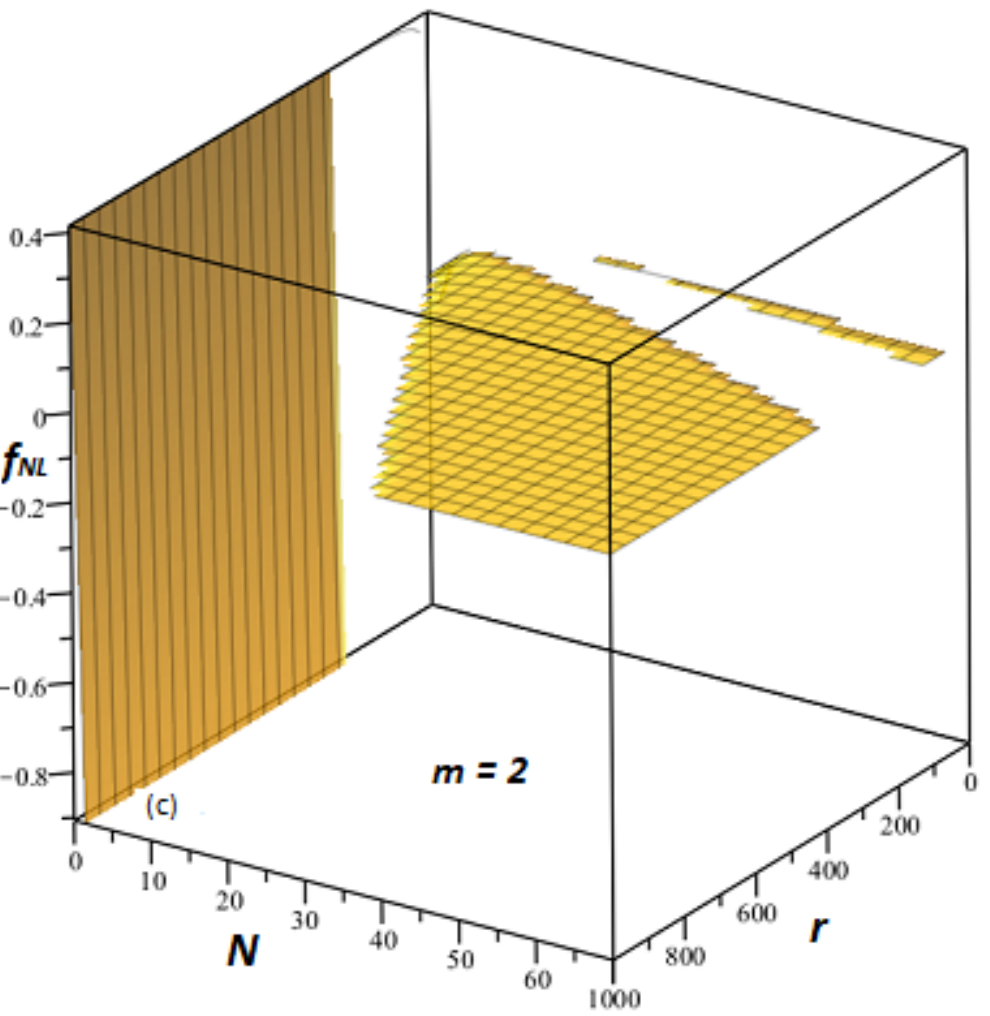}
	\includegraphics[width=.32\textwidth,keepaspectratio]{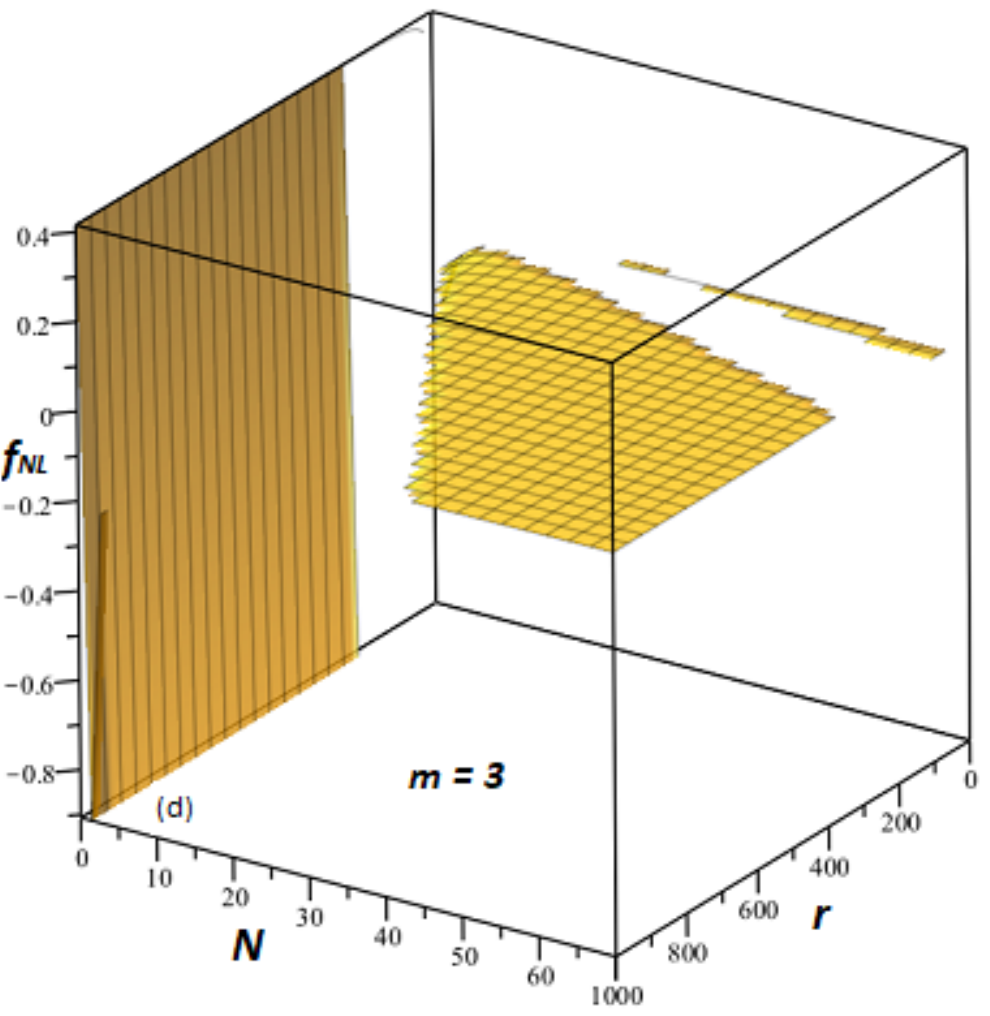}
	\includegraphics[width=.32\textwidth,keepaspectratio]{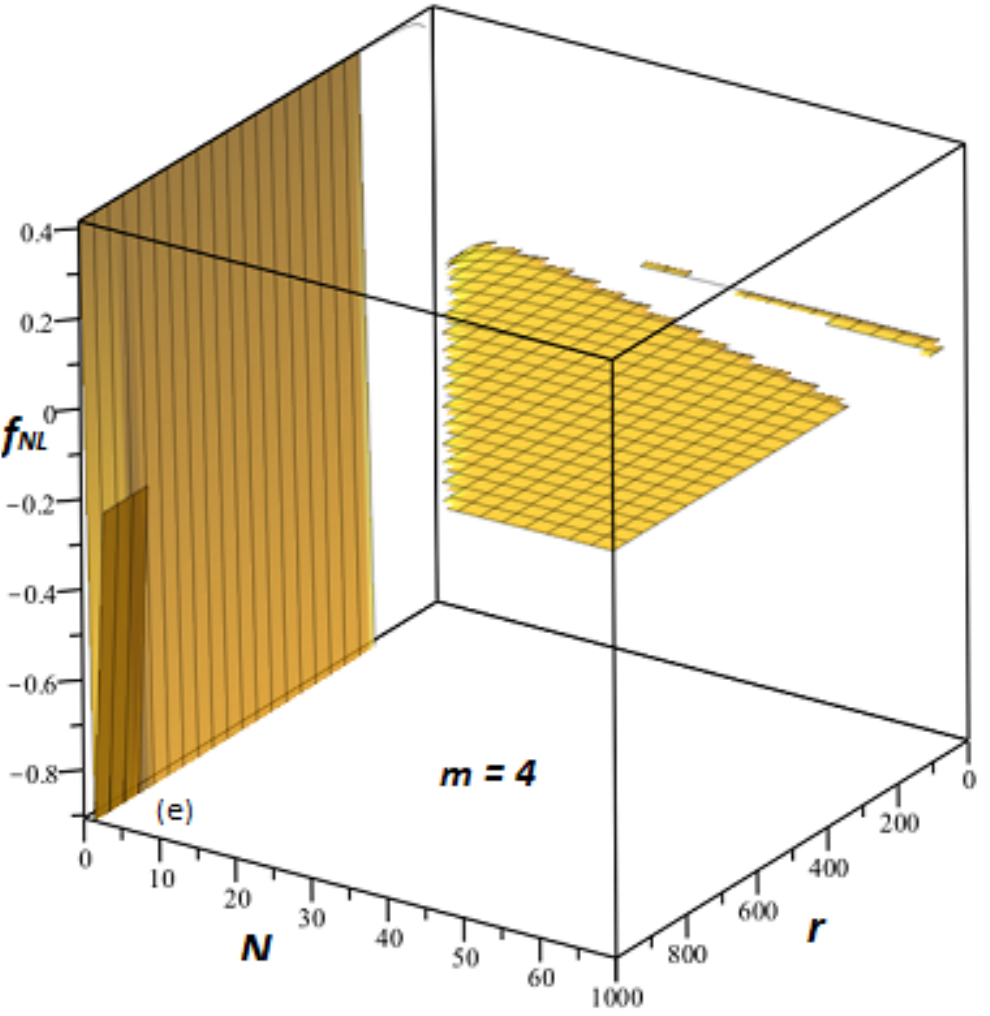}
	\includegraphics[width=.32\textwidth,keepaspectratio]{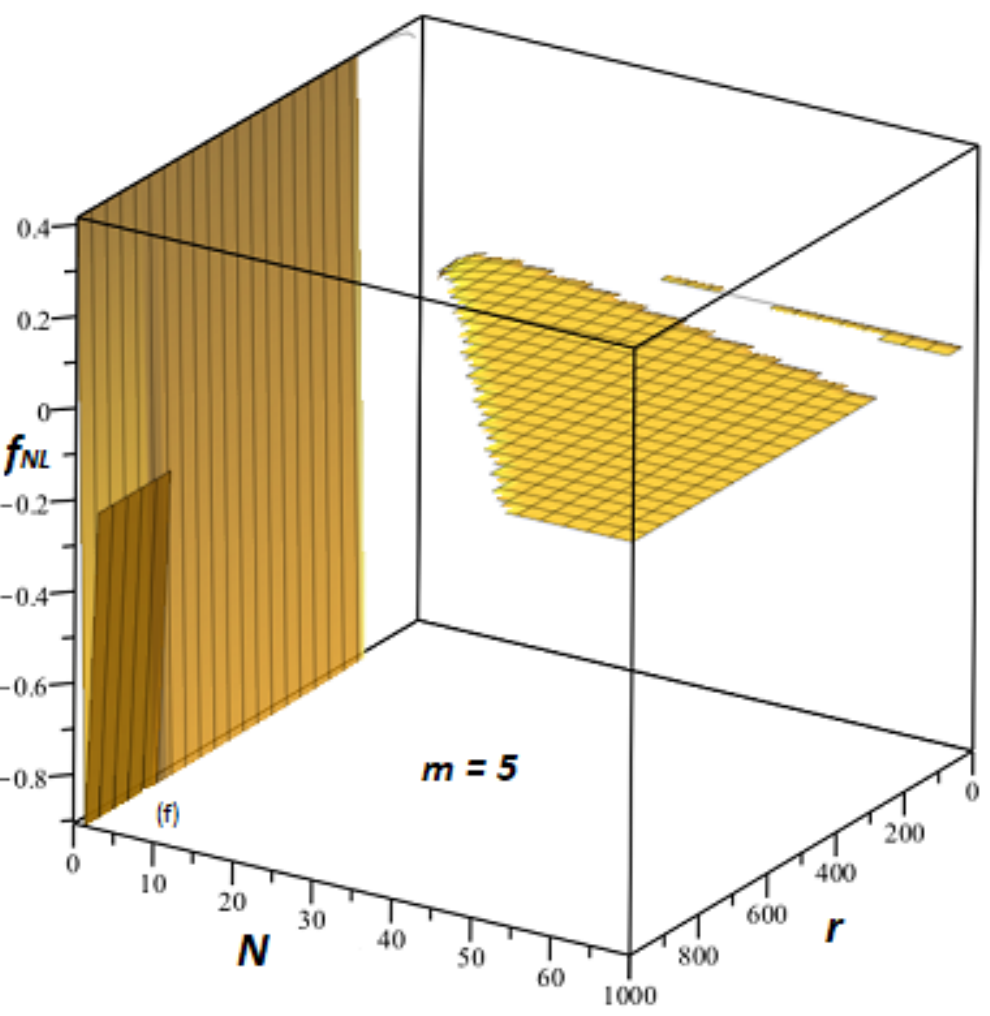}
	\caption{The non-linear parameter $f_{NL}$ versus number of e-folds $N$ and dissipation strength $r$ for different values of $m$ in logamediate model in the presence of the second term of the Eq. (\ref{56}). The figure is plotted for $\alpha=1$, $\beta=6$, $\gamma=1.5$, $\xi_{0}\sim 10^{-8}$ and $T_{r}\sim10^{-5}$ \cite{c4,caption}.}
	\label{fig17}
\end{figure*}
From Eq. (\ref{36}), we find that the scalar field during the inflationary scenario  ($\epsilon<1$) is given by
 \begin{equation}
\varphi>2\sqrt{2(1-\beta)}\Big(\frac{3(1-\beta)}{\alpha\beta(m+2)}\Big)^{\frac{1}{2\beta}}. 
\end{equation}
\begin{figure*}[!hbtp]
	\centering
	\includegraphics[width=.32\textwidth,keepaspectratio]{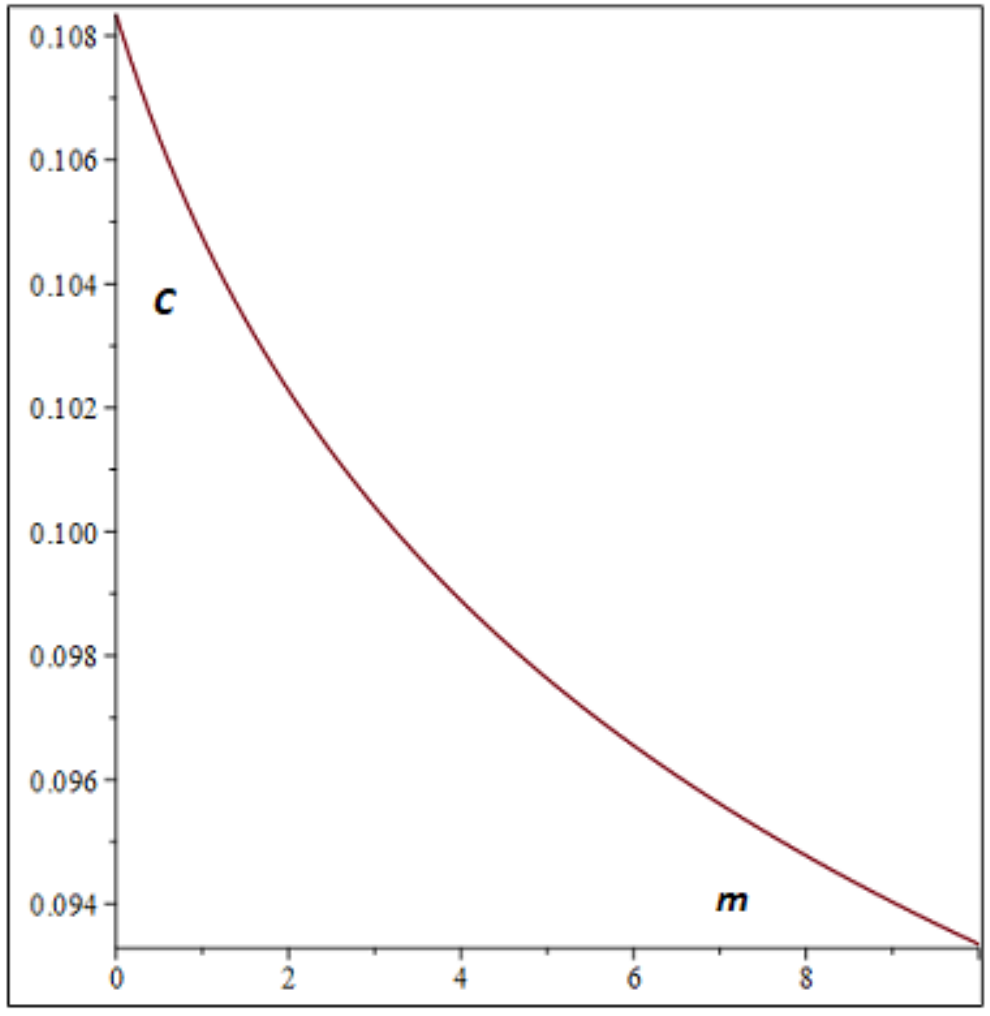}
	\hspace{0.5cm}
	\includegraphics[width=.32\textwidth,keepaspectratio]{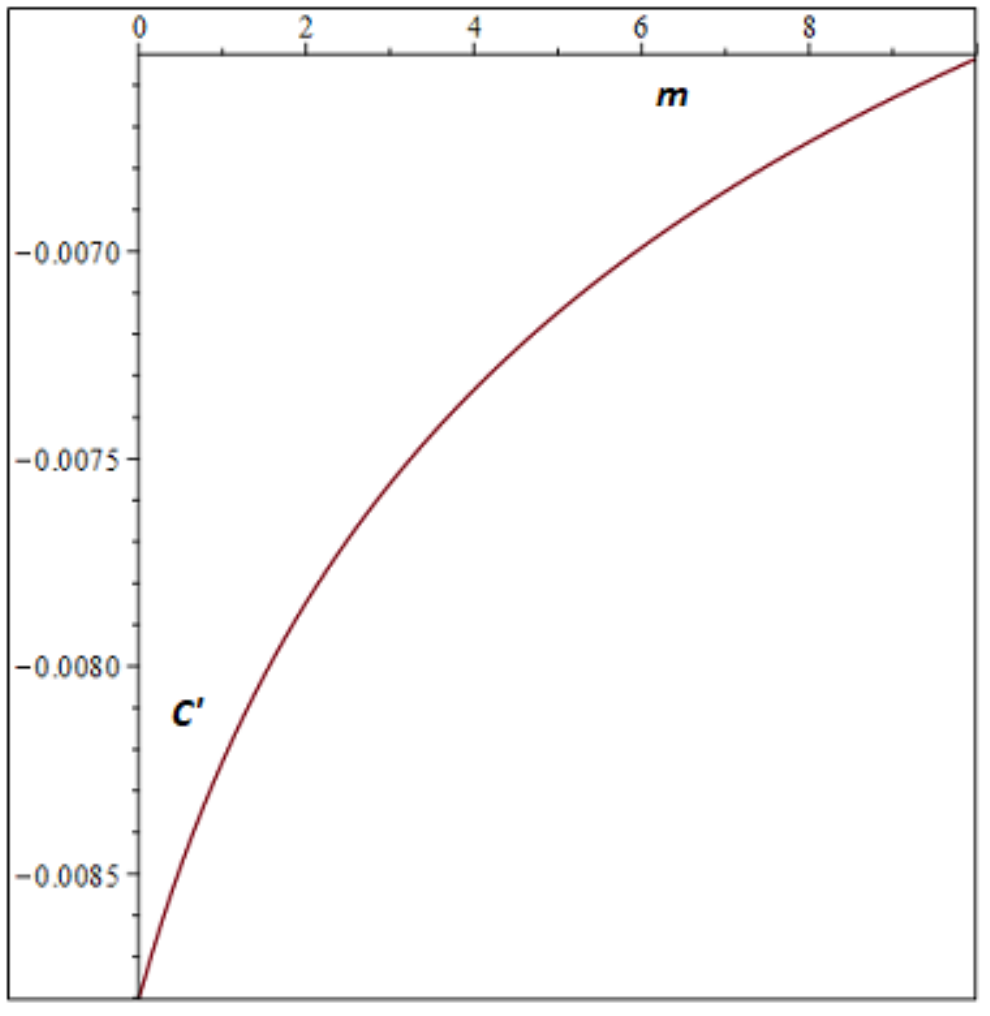}
	\caption{The behaviour of the swampland parameters $c$ and $c'$ versus $m$ in logamediate model (\ref{42}) in the case of $\Gamma$ and $\xi$ as variable parameters when $\alpha=1$, $\beta=6$ and $N=60$.}
	\label{fig18}
\end{figure*}
On the other hand, the spectral index and tensor-to-scalar ratio (\ref{18}) and (\ref{19}) using the Eqs.(\ref{36}) and (\ref{39}) are calculated and shown in (\ref{a4}) and (\ref{a5}). In Figure \ref{fig7}, we present the $n_{s} - r$ constraints coming from the marginalized joint 68\% and 95\% CL regions of the Planck 2018 in combination with BK14+BAO data on the intermediate model (\ref{22}) \cite{cmb}. The dashed and solid lines represent $N=50$ and $N=60$, respectively. The figure is plotted for different values of $m$ when $\alpha=1$, $\beta=0.9$, $\gamma=1.5$, $\xi_{0}\sim10^{-8}$ and $T_{r}\sim10^{-5}$ \cite{c4,caption}. Considering the CMB anisotropies datasets from the Planck alone, we find that the values of spectral index $n_{s}$ and tensor-to-scalar ratio $r$ in the case of $N=50$ and $N=60$ are almost situated in observational regions at the 68\% and 95\% CL, respectively. For $N=50$, the corresponding values of $0.5<m<3$ are compatible with the observations at the 68\% CL while for $N=60$ the case $0.5\leq m<3.5$ show observationally desirable values at the 68\% CL. Also, the results of the case $0.5\leq m<2.5$ for $N=60$ are in good agreement with the observations at 98\% CL. By combination of the BK14 data and the Planck data, the cases $0.5<m<2.5$ and $0.5\leq m\leq2.5$ show observationally acceptable values of $n_{s}$ and $r$ at the 68\% CL for $N=50$ and $N=60$, respectively. Besides the mentioned constraints, the plot tells us that the values of $n_{s}$ and $r$ corresponded to the case $0.5\leq m\leq2$ are compatible with the CMB observations for $N=60$ at the 95\% CL. As a full consideration, we compare the results with the observations coming from the Planck in combination with BK15+BAO data. At the 68\% CL, we find that the the CMB constraints $1<m<2.5$ and $0.5\leq m\leq2.5$ for $N=50$ and $N=60$, respectively. Moreover, we find the constraint $0.5\leq m\leq2$ for $N=60$ at the 95\% CL. On the other hand, we investigate the non-Gaussianity property of the model by calculating the non-linear parameter. For the case of (\ref{32}), the parameter $f_{NL}$ (\ref{21}) takes the form 
\begin{equation}
-\frac{3}{5}f_{NL}=\frac{1}{2}\Big(-\epsilon-\frac{\epsilon r}{1+r}+\eta\Big)+\frac{P_{R}}{2}\Big(-\epsilon-\frac{\epsilon r}{1+r}+\eta\Big)^{3}.
\label{40}
\end{equation}
For the intermediate model by using the Eqs. (\ref{12}) and (\ref{14}), the parameter is obtained as
\begin{equation}
-\frac{3}{5}f_{NL}=\frac{r(1-2\beta)-1}{4(1+r)(N+\beta-1)}+\frac{P_{R}}{2}\bigg(\frac{r(1-2\beta)-1}{2(1+r)(N+\beta-1)}\bigg)^{3},
\label{41}    
\end{equation}
and as before the power spectrum of curvature perturbations $P_{R}$ is given by the Eq.(\ref{15}). In figure (\ref{8}), we present the behaviour of the non-linear parameter $f_{NL}$ versus the number of $e$-folds $N$ and the dissipation strength $r$ for the intermediate model in the absence of the second term of the Eq. (\ref{41}). Here, we have considered the specific cases in which $\alpha=1$, $\beta=0.9$, $\gamma=1.5$, $\xi_{0}\sim10^{-8}$ 
and $T_{r}\sim10^{-5}$, see refs. \cite{c4,caption}. The figure shows that the parameter $f_{NL}$ and also the sign change very fast in both cases of weak ($r\ll1$) and strong ($r\gg1$) dissipation so that finally it reaches zero when inflation ends ($N=70$). Moreover, figure \ref{9} reveals the changes of the non-linear parameter
$f_{NL}$ versus the number of e-folds $N$ and the dissipation strength $r$ for different values of $m$ when the second term of the Eq.(\ref{41}) is taken into account. Obviously, the non-Gaussianity feature of the model can be seen for $m\geq2$ in strong dissipation regime and for higher values of $m$ in weak dissipation regime. Also, we observe that for the strong dissipative regime in which $r\gg 1$ and for
values of the anisotropic parameter $m>2$, the second term of the Eq.(\ref{41}) becomes negligible and then this term does not contribute to the non-lineal parameter $f_{NL}$.

In figure \ref{10}, two swampland parameters $c$ and $c'$ are plotted versus $m$ for the intermediate model in the case of $\Gamma$ and $\xi$ as the variable parameters when $\alpha=1$, $\beta=0.9$ and $N=60$. As the result, we find the swampland conditions $0.0608017900360\leq c\leq0.0608017900375$ and $-0.0110905730086\leq c'\leq-0.0110905730090$ for the obtained constraints of $m$ coming from a full consideration of the observational datasets. Similar the previous case, the range of $c$ and $c'$ is very narrow to satisfy the observations.
\begin{figure*}[!hbtp]
	\centering
	\includegraphics[width=.32\textwidth,keepaspectratio]{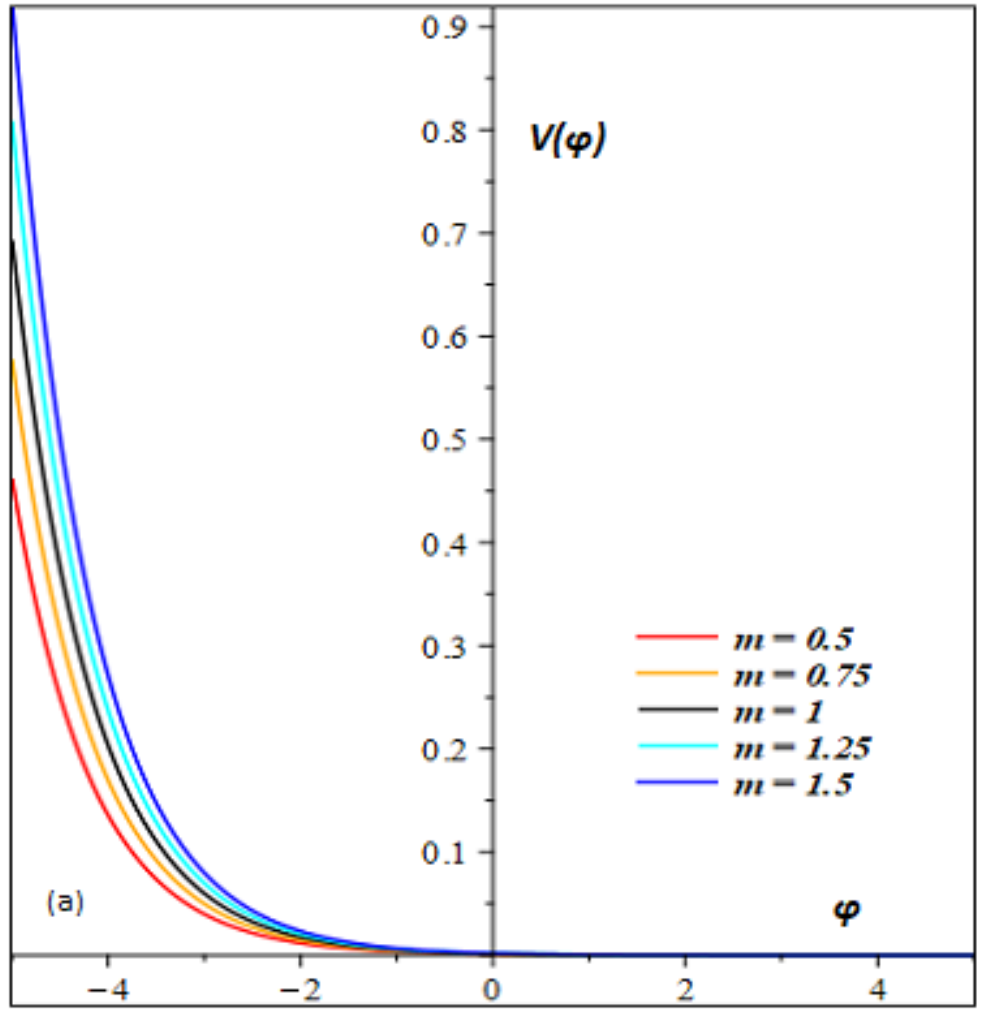}
	\includegraphics[width=.32\textwidth,keepaspectratio]{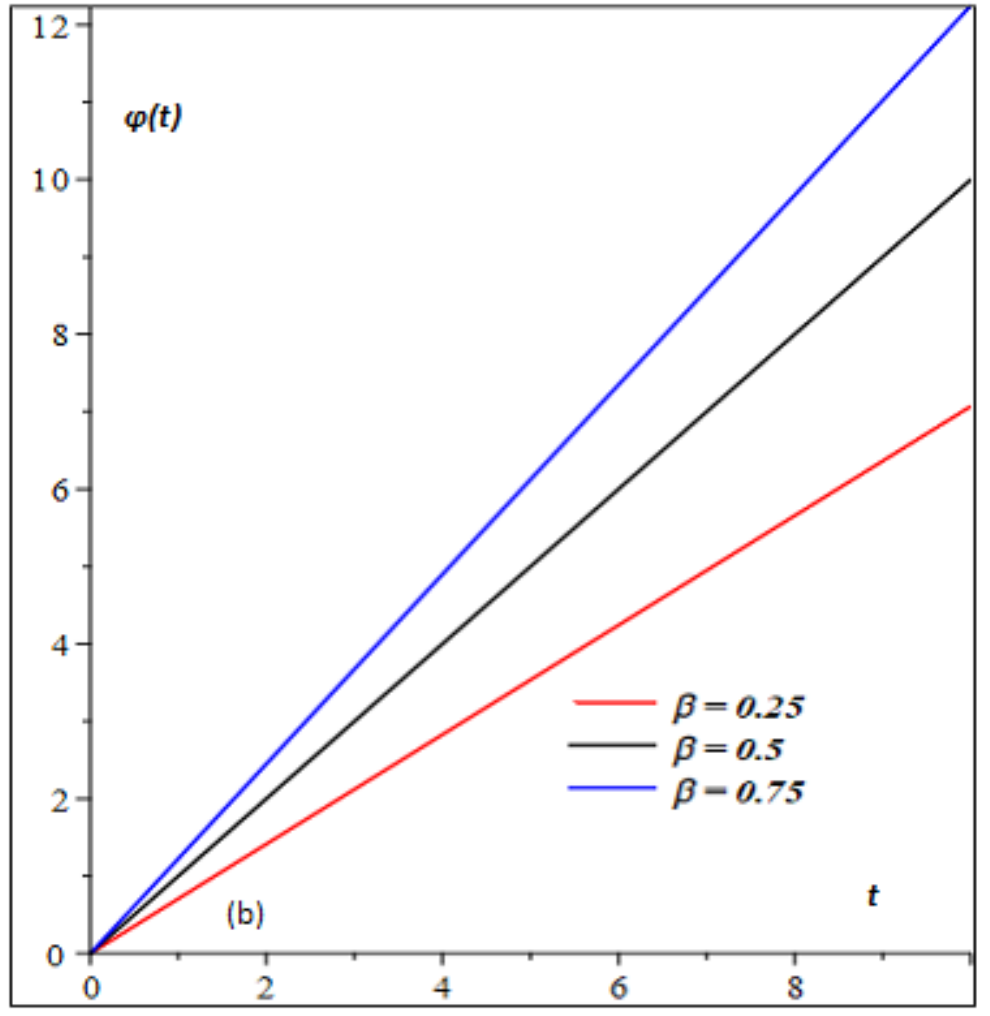}
	\includegraphics[width=.32\textwidth,keepaspectratio]{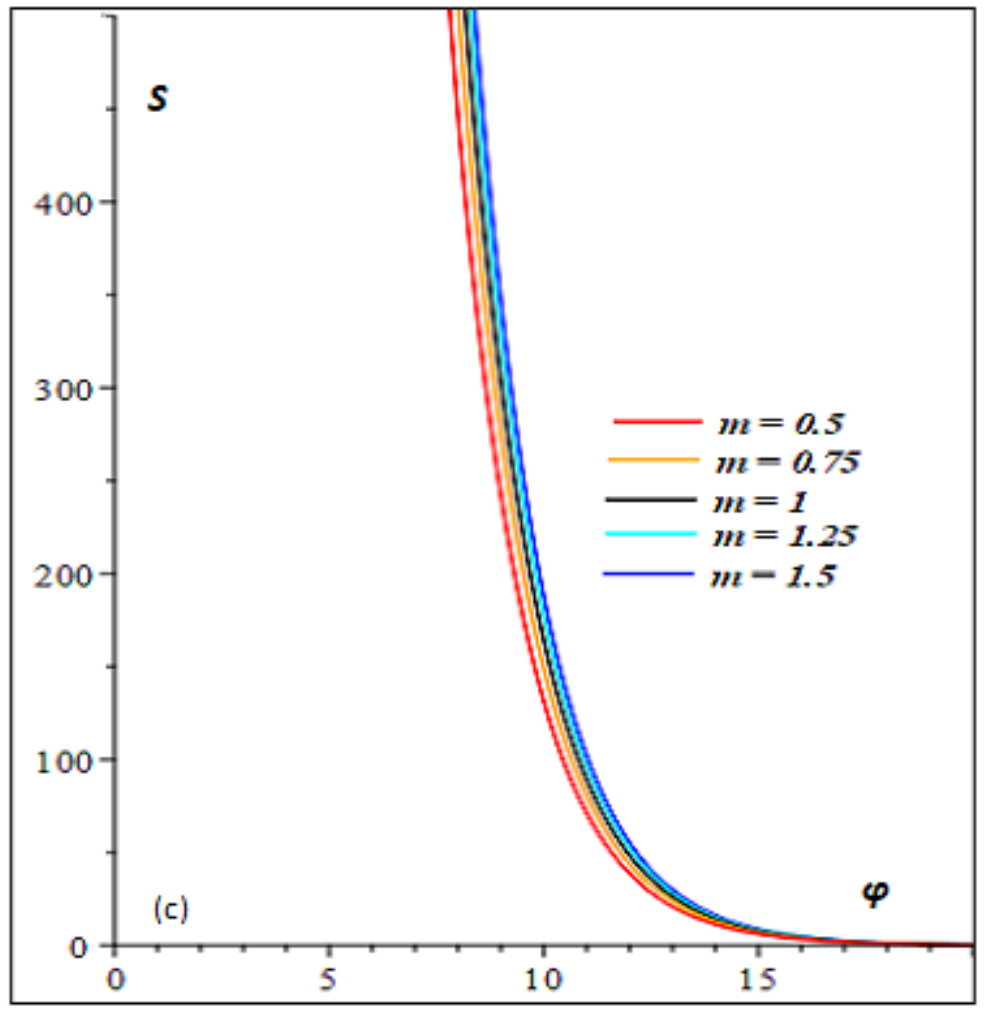}
	\caption{Panel (a): The potential (\ref{58}) plotted versus inflaton $\varphi$ for different values of $m$ when $\alpha=0.03$ and $\beta=0.75$. Panel (b): The evolution of inflaton (\ref{58}) versus cosmic time $t$ for different values of $\beta$. Panel (c): The entropy (\ref{59}) plotted versus inflaton $\varphi$ for different values of $m$ when $\alpha=1$, $\beta=0.75$, $\gamma=1.5$, $\xi_{0}\sim 10^{-8}$ and $T\sim10^{-5}$ \cite{caption}.}
	\label{fig19}
\end{figure*}
\begin{figure*}[!hbtp]
	\centering
	\includegraphics[width=.65\textwidth,keepaspectratio]{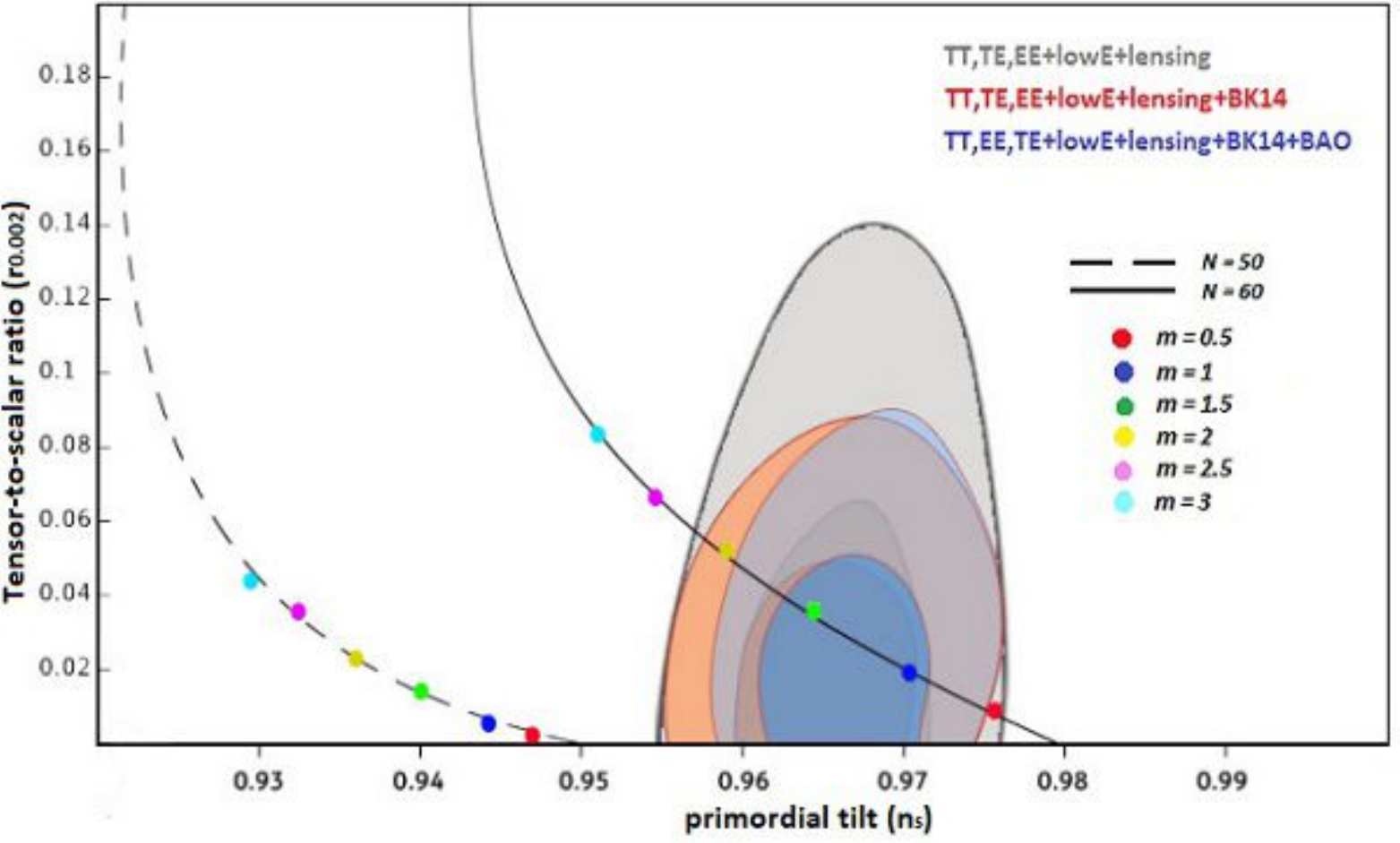}
	\caption{The marginalized joint 68\% and 95\% CL regions for $n_{s}$ and $r$ at $k = 0.002$ Mpc$^{-1}$ from Planck 2018 in combination with BK14+BAO data \cite{cmb} and the $n_{s}-r$ constraints on the exponential model (\ref{57}). The dashed and solid lines represent $N=50$ and $N=60$, respectively. The panel is plotted for different values of $m$ when $\alpha=1$, $\beta=0.35$, $\gamma=1.5$, $\xi_{0}\sim10^{-8}$ and $T_{r}\sim10^{-5}$ \cite{caption}.}
	\label{fig20}
\end{figure*}
\section{Logamediate model}

Now, let us consider a warm inflationary model in the presence of bulk viscous pressure for an anisotropic universe expanded by the logamediate scale factor \cite{c26}
\begin{equation}
b(t)\propto\exp\big(\alpha (\ln t)^{\beta}\big), 
\label{42}
\end{equation}
\begin{figure*}[!hbtp]
	\centering
    \includegraphics[width=.32\textwidth,keepaspectratio]{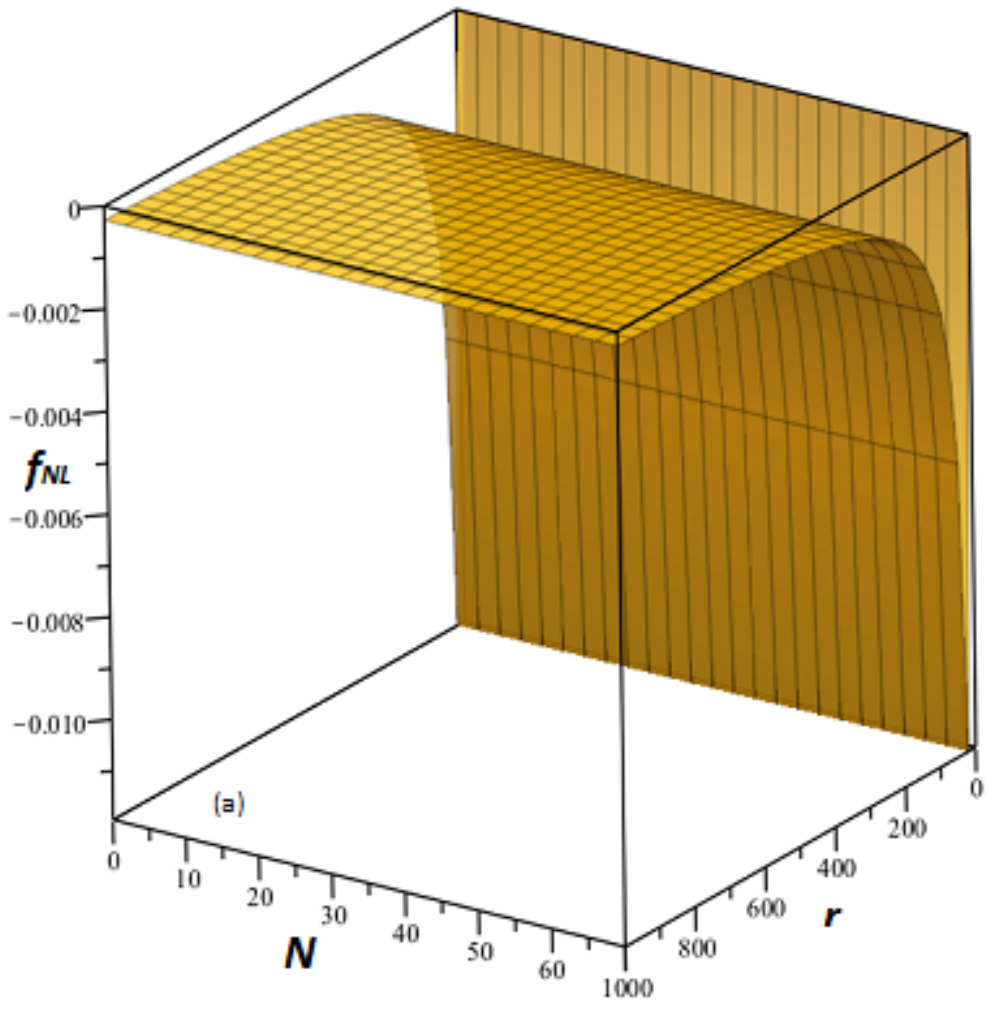}
    \includegraphics[width=.30\textwidth,keepaspectratio]{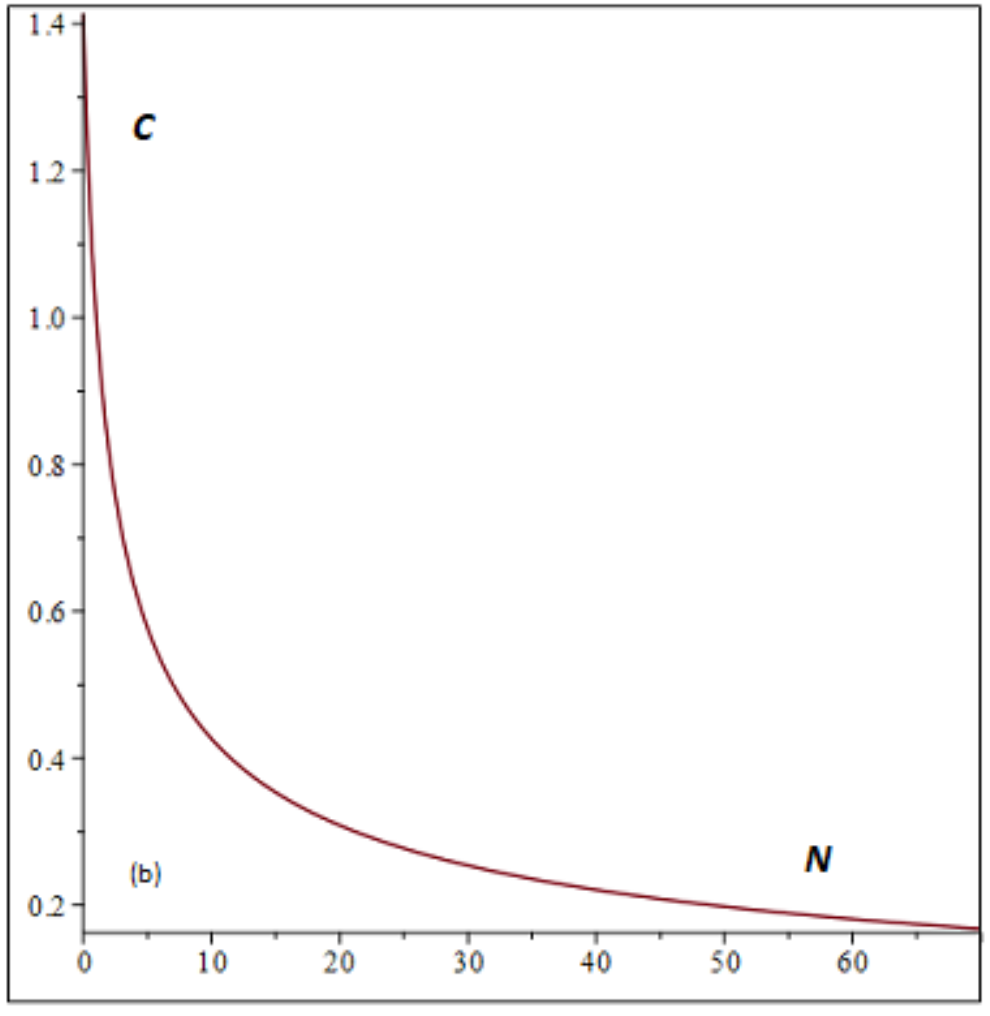}
	\hspace{0.2cm}
	\includegraphics[width=.30\textwidth,keepaspectratio]{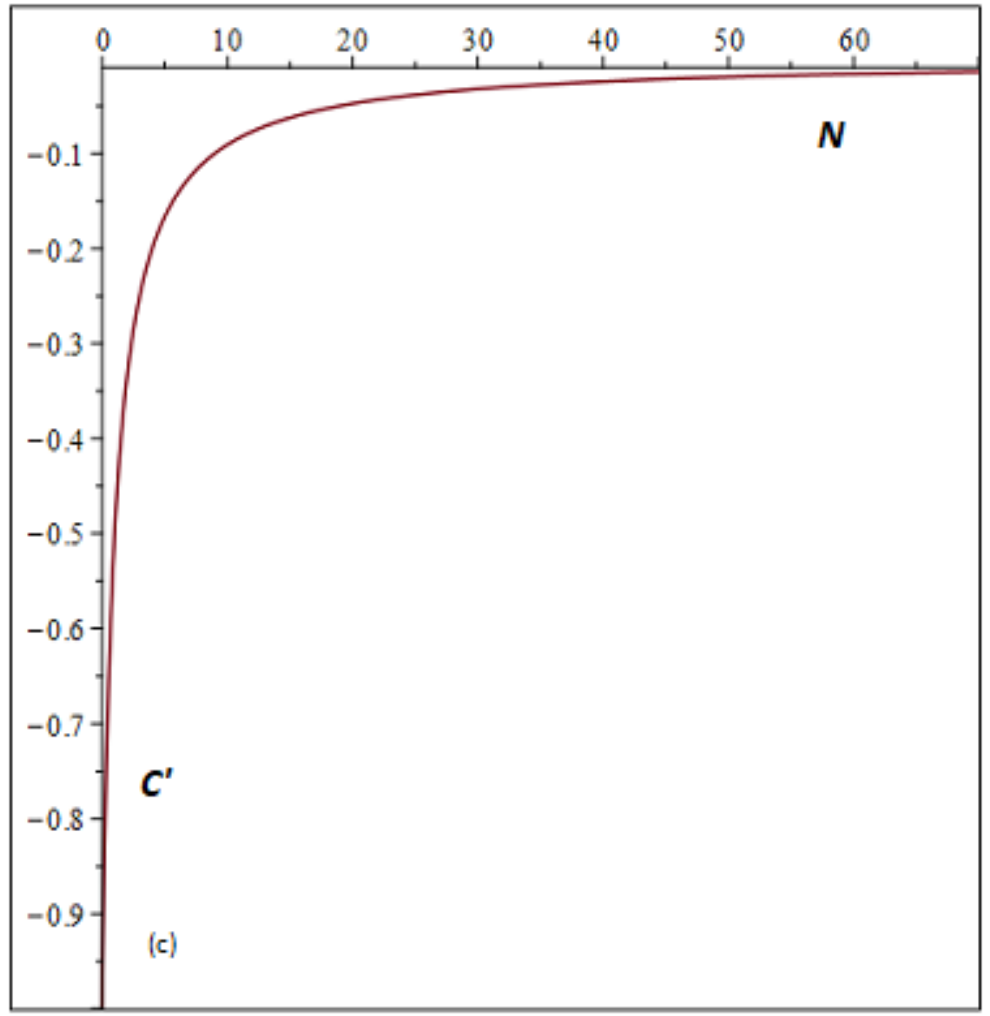}
	\caption{Panel a): The non-linear parameter $f_{NL}$ versus number of e-folds $N$ and dissipation strength $r$ for the  exponential model when $\alpha=1$, $\beta=0.35$, $\gamma=1.5$, $\xi_{0}\sim 10^{-8}$ and $T_{r}\sim10^{-5}$  \cite{caption}. The behaviour of the swampland parameters $c$ and $c'$ versus $N$ for the  exponential model (\ref{57}) (Panels b) and c), respectively).}
	\label{fig21}
\end{figure*}
where $\alpha>0$ and $\beta>1$ are dimensionless constant parameters. Similar to the previous analysis, we study the model in two cases \textit{i.e.} $\Gamma$, $\xi$ as constant and variable parameters, respectively.
\subsection{$\Gamma$, $\xi$ as constant}
Adopting the logamediate model with the case of $\Gamma=\Gamma_0$ and $\xi=\xi_0$ as constant parameters (\ref{23}), we find that the velocity of the scalar field and  the effective potential in terms of cosmic time $t$ 
are given by
\begin{equation}
\dot{\varphi}(t)=\sqrt{\frac{2(1+2m)}{\Gamma_{0}}}\frac{\alpha\beta}{t^{\frac{3}{2}}}(\ln{t})^{\beta-1},\hspace{1cm}V(t)=(1+2m)\frac{(\alpha\beta)^{2}}{t^{2}}(\ln{t})^{2(\beta-1)}.
\label{43}    
\end{equation}
Here, we observe that both the velocity of the scalar field and the potential do not depend the coefficient of bulk viscosity $\xi_0$. Also, we note that from Eq.(\ref{43}) we cannot invert the time $t$ as a function of the scalar field $\varphi$ (transcendental equation), in order to reconstruct the effective potential $V(\varphi)$. Here, the solution of $\varphi=\varphi(t)$ in terms of the time $t$ from Eq.(\ref{43}) is given by an incomplete Gamma function.

The potential (\ref{43}) versus cosmic time $t$ is drawn in panel (a) of figure \ref{11} for different values of $m$ when $\alpha=0.7$ and $\beta=1.25$.  Panel (b) shows the evolution of the velocity associated to scalar field (\ref{43}) versus cosmic time $t$ for different values of $m$ when $\alpha=1$ and $\beta=1.25$. Adding (\ref{43}) to (\ref{11}), we find the energy density of imperfect fluid as
\begin{equation}
\rho(t)=TS=\frac{1}{\gamma}\bigg(\frac{\alpha\beta}{t}\big(\frac{2(1+2m)}{t(m+2)}+\xi_{0}(m+2)\big)(\ln{t})^{\beta-1}\bigg).
\label{44}    
\end{equation}
Panel (c) of figure \ref{11} shows the changes of entropy of imperfect fluid (\ref{44}) versus cosmic time $t$ for different values of $m$ when $\alpha=0.005$, $\beta=1.25$, $\gamma=1.5$, $\xi_{0}\sim10^{-6}$ and $T_{r}\sim10^{-5}$. The slow-roll parameters (\ref{12}) of the model are driven by
\begin{equation}
\epsilon=\frac{3}{\alpha\beta(m+2)}(\ln{t})^{1-\beta},\hspace{1cm}\eta=\frac{9}{2\alpha\beta(m+2)}(\ln{t})^{1-\beta}, 
\label{45}    
\end{equation}
also, the number of e-folds (\ref{14}) is given by
\begin{equation}
N(t)=\frac{\alpha(m+2)}{3}\bigg((\ln{t_{f})^{\beta}}-(\ln{t}_{i})^{\beta}\bigg). 
\label{46}    
\end{equation}
As the case of the intermediate inflation by setting $\epsilon=1$ at the beginning of inflation, we find
\begin{equation}
t_{i}=\exp{\[\Big(\frac{\alpha\beta(m+2)}{3}\Big)^{\frac{1}{1-\beta}}\]}, 
\label{47}    
\end{equation}
and then by combination of (\ref{46}) and (\ref{47}), $t_{f}$ takes the form
\begin{equation}
t_{f}=\exp{\[\bigg(\Big(\frac{3}{\alpha\beta(m+2)}\Big)^{\frac{\beta}{\beta-1}}+\frac{3N}{\alpha(m+2)}\bigg)^{\frac{1}{\beta}}\]}. 
\label{48}    
\end{equation}
The inflationary parameters of the model (\ref{18}) and (\ref{19}) are calculated and shown in (\ref{a7}) and (\ref{a8}). In Figure \ref{fig12}, we find the $n_{s} - r$ constraints coming from the marginalized joint 68\% and 95\% CL regions of the Planck 2018 in combination with BK14+BAO data on the logamediate model (\ref{42}) in the case of $\Gamma$ and $\xi$ as constant parameters. The dashed and solid lines represent $N=50$ and $N=60$, respectively. The figure is drawn for different values of $m$ when $\alpha=0.005$, $\beta=6$, $\gamma=1.5$, $\xi_{0}\sim10^{-6}$, $\Gamma_{0}\sim10^{-3}$ and $T_{r}\sim10^{-5}$. Obviously, the observational constraints on $m$ for different combination of CMB data in both cases of $N=50$ and $N=60$ are almost the same. For the only Planck data, we find that the values of spectral index $n_{s}$ and tensor-to-scalar ratio $r$ related to $1.5\leq m<5$ and $1\leq m\leq4$ compatible with the observations at the 68\% CL for $N=50$ and $N=60$, respectively. At the 95\% CL, the observational constraints are reduced to $1.5<m\leq3$ and $1.5\leq m<3$. By Adding the BK14 data to the Planck data, the cases $1.5\leq m\leq4$ and $1\leq m<4$ show observationally desirable values of $n_{s}$ and $r$ at the 68\% CL for $N=50$ and $N=60$, respectively. Also, the figure reveals that the obtained values of $n_{s}$ and $r$ associated to the cases $1.5\leq m\leq3.5$ and $1.5\leq m<3$ are in good agreement with the observations at the 95\% CL. For a full consideration of CMB data, we combine the Planck data with the data coming from BK14+BAO. At the 68\% CL, one can find the observational constraints $1.5<m\leq4$ and $1<m<4$ for $N=50$ and $N=60$, respectively. While at the 95\% CL, the constraints are turned to $2\leq m\leq3.5$ and $1.5\leq m\leq3$. Using the Eq. (\ref{15}), the non-linear parameter (\ref{30}) of the model is expressed by 
\begin{equation}
-\frac{3}{5}f_{NL}=\frac{(3r+1)}{4(1+r)(1-N)}+\frac{P_{R}}{2}\bigg(\frac{(3r+1)}{2(1+r)(1-N)}\bigg)^{3}.
\label{49}    
\end{equation}
Panel (a) of figure \ref{13} presents the non-linear parameter $f_{NL}$ versus the number of e-folds $N$ and the dissipation strength $r$ for the logamediate model when $\alpha=1$, $\beta=6$, $\gamma=1.5$, $\xi_{0}\sim10^{-6}$, $\Gamma_{0}\sim10^{-3}$ and $T_{r}\sim10^{-5}$. The panel shows that the sign of $f_{NL}$ changes sharply in both dissipations and finally the magnitude of the parameter $f_{NL}$ approaches zero for values of the number of $e-$folds $N>20$. Moreover,  panels (a) and (b) show the behaviour of two swampland parameters $c$ and $c'$ versus $m$ for the logamediate model in the case of $\Gamma$ and $\xi$ as the constant parameters. Here, we have considered the values $\alpha=0.005$, $\beta=6$ and $N=60$. Consequently, one can find the swampland conditions $0.156\leq c\leq0.163$ and $-0.0195\leq c'\leq0.018$ for the constraints of $m$ coming from the full CMB anisotropy datasets. 

\subsection{$\Gamma$, $\xi$ as variable}

For $\Gamma$, $\xi$ as variable parameters (\ref{32}), the inflationary potential and the evolution of inflaton of the logamediate model are calculated as
\begin{equation}
\varphi(t)=2\sqrt{2t},\hspace{1cm}V(\varphi)=\frac{64(1+2m)(\alpha\beta)^{2}}{\varphi^{4}}\Big(\ln \frac{\varphi^{2}}{8}\Big)^{2(\beta-1)},
\label{50}    
\end{equation}
where for simplicity we have used $\Gamma_0=1$. As the previous case, we note that both the scalar field and the reconstructed potential do not depend the coefficient of bulk viscosity. 

The Panel (a) of Figure \ref{fig14} shows the behaviour of the potential (\ref{50}) for different values of $m$ when $\alpha=0.7$ and $\beta=1.25$. Panel (b) displays the evolution of inflaton (\ref{50}) versus cosmic time $t$. Using the Eqs. (\ref{50}) and (\ref{11}), the energy density of imperfect fluid takes the form
\begin{equation}
\rho=TS=\frac{128(1+2m)\alpha\beta\Big(\ln \frac{\varphi^{2}}{8}\Big)^{\beta-1}}{(m+2)\Big(\gamma\varphi^{2}-8(m+2)\xi_{0}\alpha\beta\Big(\ln \frac{\varphi^{2}}{8}\Big)^{\beta-1}\Big)\varphi^{2}}.
\label{51}    
\end{equation}
Similar to the intermediate case, the value of entropy of imperfect fluid (\ref{51}) increases and decreases for $m>1$ and $m<1$ with respect to $m=1$ as the isotropic case, respectively (see panel (c) of Figure \ref{fig14}). The slow-roll parameters (\ref{12}) of the model are given by
\begin{equation}
\epsilon=\frac{3}{\alpha\beta(m+2)}\Big(\ln \frac{\varphi^{2}}{8}\Big)^{1-\beta},\hspace{1cm}\eta=\frac{9}{2\alpha\beta(m+2)}\Big(\ln \frac{\varphi^{2}}{8}\Big)^{1-\beta},  
\label{52}    
\end{equation}
also the number of e-folds (\ref{14}) can be found as
\begin{equation}
N=\frac{\alpha(m+2)}{3}\bigg(\Big(\ln \frac{\varphi_{f}^{2}}{8}\Big)^{\beta}-\Big(\ln \frac{\varphi_{i}^{2}}{8}\Big)^{\beta}\bigg).   
\label{53}    
\end{equation}
The value of inflaton when inflation begins ($\epsilon=1$) $\varphi_{i}$ is obtained as
\begin{equation}
\varphi_{i}=2\sqrt{2}\exp{\[\frac{1}{2}\Big(\frac{\alpha\beta(m+2)}{3}\Big)^{\frac{1}{1-\beta}}\]}, 
\label{54}    
\end{equation}
also its value at the end of inflation $\varphi_{f}$ in terms of the number of e-folds $N$ is
\begin{equation}
\varphi_{f}=2\sqrt{2}\exp{\[\frac{1}{2}\bigg(\Big(\frac{3}{\alpha\beta(m+2)}\Big)^{\frac{\beta}{\beta-1}}+\frac{3N}{\alpha(m+2)}\bigg)^{\frac{1}{\beta}}\]}.  
\label{55}    
\end{equation}
During the inflationary scenario in which $\epsilon<1$, the scalar field satisfies
\begin{equation}
\varphi<2\sqrt{2}\exp{\[\frac{1}{2}\Big(\frac{\alpha\beta(m+2)}{3}\Big)^{\frac{1}{1-\beta}}\]}. 
\end{equation}
We calculate the spectral parameters (\ref{18}) and (\ref{19}) shown in (\ref{a10}) and (\ref{a11}). Figure \ref{fig15} presents the $n_{s} - r$ constraints coming from the marginalized joint 68\% and 95\% CL regions of the Planck 2018 in combination with BK14+BAO data \cite{cmb} on the logamediate model (\ref{42}) in the case of $\Gamma$ and $\xi$ as variable parameters. The dashed and solid lines represent $N=50$ and $N=60$, respectively. The figure is drawn for different values of $m$ when $\alpha=1$, $\beta=6$, $\gamma=1.5$, $\xi_{0}\sim10^{-8}$ and $T_{r}\sim10^{-5}$. Regrading the Planck data only, the plot shows the observational constraints $1\leq m\leq3.5$ and $0.5\leq m\leq2.5$ at the 68\% CL for $N=50$ and $N=60$, respectively while at the 95\% CL, the constraints are reduced to $1<m\leq2.5$ and $0.5<m\leq1.5$. A combination of the Planck with the BK14 data shows the values of $n_{s}$ and $r$ associated to $1\leq m<3.5$ and $0.5\leq m<2.5$ at the 68\% CL are situated in the observational regions for $N=50$ and $N=60$, respectively. At the 95\% CL, we find the spectral values corresponded to $1<m\leq2.5$ and $0.5<m<1.5$ compatible with the observations. For the Planck data combined with BK14+BAO, the plot reveals that the cases $1\leq m<3.5$ and $0.5<m\leq2.5$ show observationaly acceptable values of $n_{s}$ and $r$ at the 68\% CL for $N=50$ and $N=60$, respectively. The constraints are turned to $1<m\leq2.5$ and $0.5<m<1.5$ at the 95\% CL. Moreover, using the form of power spectrum $P_{R}$ (\ref{15}) and Eqs.(\ref{12}) and (\ref{14}), the non-linear parameter of the model (\ref{40}) is written
\begin{equation}
-\frac{3}{5}f_{NL}=\frac{(1-r)}{4(1+r)(1-N)}+\frac{P_{R}}{2}\bigg(\frac{(1-r)}{2(1+r)(1-N)}\bigg)^{3}.
\label{56}    
\end{equation}
Figure \ref{16} presents the variation of the non-linear parameter $f_{NL}$ versus the number of $e$-folds $N$ and the dissipation strength $r$ for the logamediate model in the absence of the second term of the Eq.(\ref{56}) when $\alpha=1$, $\beta=6$, $\gamma=1.5$, $\xi_{0}\sim10^{-8}$ and $T_{r}\sim10^{-5}$. From the panel, we find that sign and magnitude of the parameter $f_{NL}$ changes very fast in both cases of weak and strong dissipations. Also, figure \ref{17} shows the changes of the non-linear parameter
$f_{NL}$ versus the number of e-folds $N$ and the dissipation strength $r$ for different values of $m$ when the second term of the Eq. (\ref{56}) is considered. We realize that the non-Gaussianity property of the model starts to appear from $m=0.5$ in only a strong dissipation regime and it is emerged in both dissipation regimes for bigger values of $m$.

In figure \ref{18}, two swampland parameters $c$ and $c'$ are drawn versus $m$ for the logamediate model in the case of $\Gamma$ and $\xi$ as the variable parameters when $\alpha=1$, $\beta=6$ and $N=60$. Considering the observation constraints on $m$, the swampland conditions are found as $0.104<c<0.107$ and $-0.0085<c'<0.0080$.

\section{Exponential model}

Besides the intermediate and logamediate regimes, we are concerned to investigate warm inflation with bulk viscous pressure for an anisotropic universe described by an exponential scale factor. In this context, following Refs. \cite{Exp,y} we can write for the scale factor 
\begin{equation}
b(t)=b_{0}\exp\big(\alpha e^{-\beta t}\big), 
\label{57}
\end{equation}
where $\alpha>0$ and $\beta>0$ are dimensionless constants. From the Eq. (\ref{11}), scalar field and corresponding reconstructed potential take the forms
\begin{equation}
\varphi(t)=\sqrt{2\beta}t,\hspace{1cm}V(\varphi)=(1+2m)(\alpha\beta)^{2}e^{-\sqrt{2\beta}\varphi}. 
\label{58}    
\end{equation}
Here we have assumed 
the special case in which  $\Gamma$, $\xi$ correspond to variable parameters defined by Eq.(\ref{32}) and as before we have considered the parameter $\Gamma_0=1$. In Figure \ref{fig19}, the obtained potential and evolution of inflaton are plotted in panels (a) and (b), respectively. Combining the Eqs. (\ref{11}) and (\ref{58}), the energy density of imperfect fluid is driven as
\begin{equation}
\rho=TS=\frac{2(1+2m)\alpha\beta^{2}e^{-\sqrt{2\beta}\frac{\varphi}{2}}}{(m+2)\Big(\gamma-(m+2)\alpha\beta\xi_{0} e^{-\sqrt{2\beta}\frac{\varphi}{2}}\Big)}.
\label{59}
\end{equation}
Panel (c) of Figure \ref{fig19} presents the entropy of imperfect fluid versus inflaton for different values of $m$. The slow-roll parameters (\ref{12}) of the model can be found as
\begin{equation}
\epsilon=\eta=\frac{3}{\alpha(m+2)}e^{\sqrt{2\beta}\frac{\varphi}{2}},    
\label{60}    
\end{equation}
and also, the number of e-folds (\ref{14}) is given by
\begin{equation}
N=\frac{\alpha(m+2)}{3}\Big(e^{-\sqrt{2\beta}\frac{\varphi_{i}}{2}}-e^{-\sqrt{2\beta}\frac{\varphi_{f}}{2}}\Big).    
\label{61}    
\end{equation}
From $\epsilon=1$, we find the value of the inflation at the end of inflation as
\begin{equation}
\varphi_{f}=\frac{2}{\sqrt{2\beta}}\ln{\Big(\frac{\alpha(m+2)}{3}\Big)},
\label{62}    
\end{equation}
and then by combining (\ref{61}) and (\ref{62}), we have
\begin{equation}
\varphi_{i}=\frac{2}{\sqrt{2\beta}}\ln{\Big(\frac{\alpha(m+2)}{3(N+1)}\Big)}.   
\label{63}    
\end{equation}
During inflation the scalar field becomes $\varphi<\frac{2}{\sqrt{2\beta}}\ln{\Big(\frac{\alpha(m+2)}{3}\Big)}$, since $\ddot{a}>0$ or $\epsilon<1$.
Using the Eqs. (\ref{60}) and (\ref{63}), we find the inflationary parameters (\ref{18}) and (\ref{19}) shown in (\ref{a13}) and (\ref{a14}). In Figure \ref{20}, we present the $n_{s} - r$ constraints coming from the marginalized joint 68\% and 95\% CL regions of the Planck 2018 in combination with BK14+BAO data on the exponential model (\ref{57}) in the case of $\Gamma$ and $\xi$ as variable parameters \cite{cmb}. The dashed and solid lines represent $N=50$ and $N=60$, respectively. The figure is plotted for different values of $m$ when $\alpha=1$, $\beta=0.35$, $\gamma=1.5$, $\xi_{0}\sim10^{-8}$ and $T_{r}\sim10^{-5}$. For $N=50$, we are not able to find any observational constraints on $m$ since the values of $n_{s}$ and $r$ are situated in a observationally disfavoured regions. For $N=60$, the observational constraints on $m$ for all three datasets are almost similar each other. Concerning the only Planck data, we find the constraints $0.5\leq m<2.5$ and $1\leq m<2$ at the 68\% and 95\% CL, respectively. By combination of the Planck and the BK14 datasets, the cases $0.5<m\leq2$ and $1\leq m\leq1.5$ show observationally favoured values of $n_{s}$ and $r$ at the 68\% and 95\% CL, respectively. In a full consideration of datasets \text{i.e.} Planck+BK14+BAO, the constraints are obtained as $0.5<m<2$ and $1\leq m\leq1.5$ at the 68\% and 95\% CL, respectively. The non-linear parameter (\ref{40}) of the model is given by
\begin{equation}
-\frac{3}{5}f_{NL}=\frac{3(1+2m)\beta r}{2(1+r)^{2}(m+2)^2}+\frac{P_{R}}{2}\bigg({\frac{3(1+2m)\beta r}{(1+r)^{2}(m+2)^2}}\bigg)^{3},
\label{64}    
\end{equation}
that is plotted in panel (a) of figure \ref{21} versus the number of e-folds $N$ and the dissipation strength $r$ when $\alpha=1$, $\beta=0.35$, $\gamma=1.5$, $\xi_{0}\sim10^{-8}$ and $T_{r}\sim10^{-5}$. Here we note that the parameter $f_{NL}\sim 0$, for the strong dissipative regime and for the weak dissipative regimen $f_{NL}<0$. Also, panels (b) and (c) present the behaviour of two swampland parameters $c$ and $c'$ versus $N$ for the exponential model in the case of $\Gamma$ and $\xi$ as variable parameters.

\section{Conclusion}

Besides cold inflation, we deal with an interesting approach to inflation which inflaton decays to the particles and radiation during the inflationary era. Then, the existence of the interactions between inflaton and other matters generates a thermal bath of particles continuously during inflation so that the universe smoothly links to the radiation-dominated phase without the reheating process. As the inflationary solutions of warm inflation, we can work with the intermediate and logamediate models while in the cold inflation regime, they are ruled out by the observations in the framework of the General Relativity. In this paper, we have studied warm inflation for an anisotropic universe filled with inflaton, radiation and bulk viscous pressure and also described by Bianchi I metric. The analysis has been carried out for three different solutions (\textit{i.e.} the intermediate, logamediate and exponential models) of such universe in two different cases $\Gamma$ and $\xi$ as constant and variable parameters. In this analysis, we have  unified the theoretical foundations from the swampland criteria  and the observational foundations considering the Planck and Bicept2/Keck array \cite{cmb,bicepnew} constraints on the parameters space of the models and also the predictions of the models from the spectrum parameters. Results can be summarized as follows:
\begin{itemize}
\item \textbf{Intermediate model}
\begin{itemize}
      \item\textbf{$\Gamma$, $\xi$ as constant parameters.} From the Planck alone and also in combination with BK14, we have realized that the obtained values of $n_{s}$ and $r$ are out of the observational regions in the case of $N=50$ while for $N=60$ we have found the observational constraint $0.5\leq m<4$ at the 68\% CL. A full consideration of the observational datasets (Planck+BK14+BAO) does not show any constraints on $m$.
      \item\textbf{$\Gamma$, $\xi$ as variable parameters.} From the Planck alone and for $N=50$, we have obtained the constraint $0.5<m<3$ at the 68\% CL. For $N=60$, we have found the observational constraints $0.5\leq m<3.5$ and $0.5\leq m<2.5$ at the 68\% and 95\% CL, respectively. From the Planck in combination with BK14, we have found the values of $n_{s}$ and $r$ in the cases $0.5<m<2.5$ and $0.5\leq m\leq2.5$ are in good agreement with the observations at the 68\% CL for $N=50$ and $N=60$, respectively. At the 95\% CL, we only have the constraint  $0.5\leq m\leq2$ for $N=60$. From a full observational datasets, we have obtained the CMB constraints $1< m<2.5$ and $0.5\leq m\leq2.5$ at the 68\% CL for $N=50$ and $N=60$, respectively. At the 95\% CL, we have $0.5\leq m\leq2$ for $N=60$ at the 95\% CL. 
\\
\\
In relation to the non-lineal parameter $f_{NL}$, we have obtained that in the cases of $\Gamma$ and $\xi$ constant or variable parameters, the contribution of the  second term of the Eq. (\ref{31}) becomes negligible in the specific case of the strong dissipative regime, when we consider ever larger values of the anisotropic parameter $m$. In the context of the theoretical foundations, we have found that in both cases ($\Gamma$ and $\xi$ constant or variable parameters) the range for the parameters associated to the swampland criteria $c$ and $c'$
is very narrow.
\end{itemize}
\item \textbf{Logamediate model}
\begin{itemize}
      \item\textbf{$\Gamma$, $\xi$ as constant parameters.} 
      In this scenario,  we have not been able to reconstruct the effective potential $V(\varphi)$ and we have found that  the evolution of the scalar field in terms of the time $\varphi(t)$ corresponds to an  incomplete Gamma function. From the Planck alone, we have the constraints $1.5\leq m<5$ and $1\leq m\leq4$ at the 68\% CL for $N=50$ and $N=60$, respectively. At the 95\% CL, these constraints are turned to $1.5<m\leq3$ and $1.5\leq m<3$. From the combination of the Planck and BK14, we have found that the cases of $1.5\leq m\leq4$ and $1\leq m<4$ show observationally favoured values of $n_{s}$ and $r$ at the 68\%CL for $N=50$ and $N=60$, respectively. At the 95\% CL, the constraints are reduced to $1.5\leq m\leq3.5$ and $1.5\leq m<3$. From the full observational datasets, we have found the observational $1.5<m\leq4$ and $1<m<4$ at the 68\% CL for $N=50$ and $N=60$, respectively. At the 95\% CL, these constraints are changed to $2\leq m\leq3.5$ and $1.5\leq m\leq3$.
      
      In relation to the non-lineal parameter, we have obtained that 
      the sign of $f_{NL}$ changes sharply in both dissipative regimes and finally the magnitude of the parameter $f_{NL}$ approaches zero for values of the number of $e-$folds $N>20$.

      \item\textbf{$\Gamma$, $\xi$ as variable parameters.} From the Planck alone, we have found the cases $1\leq m\leq3.5$ and $0.5\leq m\leq2.5$ show observationally desirable values of $n_{s}$ and $r$ at the 68\% CL fro $N=50$ and $N=60$, respectively. At the 95\% CL, the constraints are reduced to $1<m\leq2.5$ and $0.5<m\leq1.5$. From a combination the Planck and BK14, the observational constraints $1\leq m<3.5$ and $0.5\leq m<2.5$ at the 68\% CL for $N=50$ and $N=60$, respectively. At the 95\% CL, the constraints are turned to $1<m\leq2.5$ and $0.5<m<1.5$. From a full observational datasets, the cases of $1\leq m<3.5$ and $0.5<m\leq2.5$ are in good agreement with the observations at the 68\% CL for $N=50$ and $N=60$, respectively. At the 95\% CL, the constraints are reduced to $1<m\leq2.5$ and $0.5<m<1.5$.
      
In this stage, we have obtained that the non-Gaussianity property of the model from the parameter $f_{NL}$ starts to appear from $m=0.5$ in the case of  strong dissipation regime and it is emerged in both dissipation regimes for bigger values of $m$. In the context of the swampland conditions, we have found that in both cases ($\Gamma$ and $\xi$ constant or variable parameters) the parameter $c$ presents a range very narrow values, while the parameter $c'$ much larger.
\end{itemize}
\newpage
\item \textbf{Exponential model} 
\begin{itemize}
      \item\textbf{$\Gamma$, $\xi$ as variable parameters.} 
      In this scenario, we have obtained that the reconstruction of the effective potential corresponds to an exponential exponential and the scalar field evolves as $\varphi(t)\propto\, t$. For $N=50$, all obtained values of $n_{s}$ and $r$ are inconsistent with the observations. From the Planck only and for $N=60$, we have found the constraints $0.5\leq m<2.5$ and $1\leq m<2$ at the 68\% and 95\% CL, respectively. From the Planck combined with BK14, we have the constraints $0.5<m\leq2$ and $1\leq m\leq1.5$. For a full consideration of the CMB data, the constraints are turned to $0.5<m<2$ and $1\leq m\leq1.5$.
      
In relation to  the parameter $f_{NL}$, we have found that the non-lineal parameter $f_{NL}\sim 0$,  for the strong dissipative regime and for the weak dissipative regimen $f_{NL}<0$. Also, we have obtained that the range for the parameters $c$ and $c'$ associated to swampland criteria results to be bigger in relation to the intermediate and logamediate expansions.  
\end{itemize}
\end{itemize}

\bibliographystyle{ieeetr}
\bibliography{biblo}
\appendix
\section{The spectral parameters}
\subsection{Intermediate model}
\subsubsection{$\Gamma$, $\xi$ as constant}
\begin{eqnarray}
&\!&\!n_{s}=\frac{1}{\sqrt{-\frac{(1+2m)(\beta-1)}{\Gamma_{0}}}\alpha\beta\mathcal{T}(m+2)\Big(8\beta\gamma\alpha(m+2)\mathcal{T}^{\frac{2\beta}{2\beta-1}}+\xi\gamma(m+2)^{2}\mathcal{T}^{\frac{2}{2\beta-1}}+8(\beta-1)(\gamma-\frac{1}{2})(m+\frac{1}{2})\Big)}\times\nonumber\\&\!&\!
\times\Bigg\{\sqrt{\frac{27(1-\beta)^{3}}{4\Gamma_{0}}}\bigg(16\alpha\beta\gamma(m+2)-\xi\gamma(m+2)^{2}\mathcal{T}^{\frac{1}{2\beta-1}}-8(\beta-1)(\gamma-\frac{1}{2})(m+\frac{1}{2})\mathcal{T}^{\frac{1}{1-2\beta}}\bigg)+\sqrt{\frac{(1-\beta)(1+2m)}{\Gamma_{0}}}\times\nonumber\\&\!&\!
\times\bigg(8\alpha\beta\gamma(m+2)\mathcal{T}^{\frac{4\beta-1}{2\beta-1}}+\xi\gamma(m+2)^{2}\mathcal{T}^{\frac{2\beta+1}{2\beta-1}}+8(\beta-1)(\gamma-\frac{1}{2})(m+\frac{1}{2})\mathcal{T}\bigg)\Bigg\},
\label{a1}
\end{eqnarray}
\begin{equation}
R_{0}=\frac{2(1-\beta)\sqrt{\Gamma_{0}\alpha^{7}\beta^{7}(m+2)(1+2m)^{3}}\mathcal{T}^{\frac{7\beta-9}{2\beta-1}}}{3T_{r}}\exp{\bigg(\frac{3(1-\beta)}{4\beta-2}\ln{\Big(\frac{8(1+2m)(1-\beta)\alpha^{2}\beta^{2}\mathcal{T}^{2}}{\Gamma_{0}(2\beta-1)^{2}}\Big)}\bigg)}\coth[{\frac{k_{0}}{2T}}],
\label{a2}
\end{equation}
where
\begin{equation}
\mathcal{T}=\bigg(\frac{3}{m+2}\Big(\frac{1-\beta}{\alpha\beta}+\frac{N}{\alpha}\Big)\bigg)^{\frac{2\beta-1}{2\beta}}.
\label{a3}
\end{equation}
\subsubsection{$\Gamma$, $\xi$ as variable}
\begin{eqnarray}
&\!&\!n_{s}=\frac{1}{\sqrt{2-2\beta}\beta\alpha(m+2)\mathcal{T}^{2}}\Bigg\{\bigg(-4(\beta-1)\beta\xi\alpha(m+2)(m+\frac{1}{2})(\gamma-\frac{1}{4})\mathcal{T}^{2\beta+2}+2\beta\gamma^{2}\alpha(m+2)\mathcal{T}^{4+2\beta}-\nonumber\\&\!&\!
-2\alpha^{2}\beta^{2}\xi(m+2)^{2}\mathcal{T}^{4\beta+2}+(\beta-1)(m+\frac{1}{2})\Big(\alpha^{2}\beta^{2}\xi^{2}\mathcal{T}^{4\beta}+2\gamma(\gamma-\frac{1}{2})\mathcal{T}^{4}\Big)\bigg)^{-1}\times\Bigg(27\sqrt{-\frac{6(\beta-1)^{3}\mathcal{T}^{-4\beta+2}}{4(1+2m)}}\times\nonumber\\&\!&\!
\times\bigg((\beta-1)(m+\frac{1}{2})\frac{1}{3}\alpha^{2}\beta^{2}\big(3(m+2)^{2}\xi^{2}+8\gamma(\gamma-1)\big)\mathcal{T}^{4\beta+1}-4(\beta-1)\xi\beta(m+2)(m+\frac{1}{2})(\gamma-\frac{1}{4})\mathcal{T}^{2\beta+3}-\nonumber\\&\!&\!
-\frac{8}{3}(\beta-1)((\gamma-1)m+3\gamma-2)\beta^{3}\xi\alpha^{3}(m+\frac{1}{2})\mathcal{T}^{6\beta-1}+\frac{8}{3}(\beta-1)\beta^{4}\alpha^{4}\xi^{2}(m+2)(m+\frac{1}{2})\mathcal{T}^{8\beta-3}-\frac{8}{12}\Big(4\alpha^{2}\beta^{2}\xi\times\nonumber\\&\!&\!
\times(m+2)^{2}\mathcal{T}^{4\beta+3}-4(m+2)\beta\gamma\alpha\mathcal{T}^{2\beta+5}-3(\beta-1)(m+\frac{1}{2})(\gamma-\frac{1}{2})\mathcal{T}^{5}\Big)\bigg)+\sqrt{2-2\beta}\bigg(9(\beta-1)\beta^{2}\xi^{2}(\beta-\frac{4}{3})\alpha^{2}\times\nonumber\\&\!&\!
\times(m+2)^{3}(m+\frac{1}{2})\mathcal{T}^{2\beta +2}-4\beta^{2}\xi\alpha^{2}(m+2)^{2}\Big(\big((\gamma-\frac{1}{4}\big)m+5\gamma-\frac{1}{8}\big)\beta+(-\gamma+\frac{1}{4})-\frac{13}{2}\gamma+\frac{1}{8}\Big)\mathcal{T}^{4+2\beta}
\bigg)+(\beta-1)\times\nonumber\\&\!&\!
\times\beta^{3}\xi^{2}\alpha^{3}(m+2)^{3}(m+\frac{1}{2})\mathcal{T}^{4\beta+2}+18(\beta-1)(\beta-\frac{4}{3})(m+\frac{1}{2})\gamma(\gamma-\frac{1}{2})\mathcal{T}^{6-2\beta}-2\beta\alpha(m+2)\Big(\alpha^{2}\beta^{2}\gamma\xi(m+2)^{2}\mathcal{T}^{4\beta+4}-\nonumber\\&\!&\!
-\beta\gamma^{2}\alpha(m+2)\mathcal{T}^{2\beta+6}+18\big(\xi(m+\frac{1}{2})(\gamma-\frac{1}{4})\beta^{2}+\beta(-\frac{7}{3}\xi(m+\frac{1}{2})(\gamma-\frac{1}{4})-\frac{\gamma}{18}((\gamma-\frac{1}{2})m+\frac{19}{2}\gamma-\frac{1}{4})\mathcal{T}^{2})+\frac{4}{3}\xi\times\nonumber\\&\!&\!
\times(m+\frac{1}{2})(\gamma-\frac{1}{4})+\frac{\gamma}{18}((\gamma-\frac{1}{2})m+\frac{25}{2}\gamma-\frac{1}{4})\mathcal{T}^{2}\big)\mathcal{T}^{4}\Big)\Bigg)\Bigg\},
\label{a4}
\end{eqnarray}
\begin{eqnarray}
&\!&\!R(k_{0})=-\frac{\coth[{\frac{k_{0}}{2T}}]}{(6m+3)\beta\alpha T_{r}}\Bigg(2(m+2)(-1+\beta)\mathcal{T}^{-2\beta}\sqrt{\frac{(1+2m)^{6}\alpha^{11}\beta^{11}\mathcal{T}^{-22+22\beta}}{(m+2)}}\exp{\bigg(-\frac{22(\beta-1)}{\gamma(m+2)(-2+\beta)(-3+2\beta)}}\times\nonumber\\&\!&\!\times\Big((-\frac{3}{2}+\beta)(m+2)(-2+\beta)\gamma\ln(\sqrt{(2-2\beta)}\mathcal{T})+\frac{6}{11}(\gamma-1)(-\frac{3}{2}+\beta)(m+\frac{1}{2})\beta\alpha(-1+\beta)\mathcal{T}^{2\beta-4}-\frac{3}{11}(-2+\beta)\Big(\xi\times\nonumber\\&\!&\!\times(m+\frac{1}{2})\beta^{2}\alpha^{2}(-1+\beta)\mathcal{T}^{4\beta-6}-\frac{11}{3}\ln{(2)}(-\frac{3}{2}+\beta)(m+2)\gamma\Big)\Big)\bigg)\Bigg),
\label{a5}
\end{eqnarray}
where
\begin{eqnarray}
\mathcal{T}=\bigg(\frac{3}{m+2}\Big(\frac{1-\beta}{\alpha\beta}+\frac{N}{\alpha}\Big)\bigg)^{\frac{1}{2\beta}}.
\label{a6}
\end{eqnarray}
\subsection{Logamediate model}
\subsubsection{$\Gamma$, $\xi$ as constant}
\begin{eqnarray}
&\!&\!n_{s}=\frac{1}{2\alpha\beta(m+2)\Big(8\beta\gamma\alpha(m+2)\mathcal{T}^{\beta}+\mathcal{T}\big(\xi_{0}\gamma(m+2)^{2}e^{\mathcal{T}}+(-8\gamma+4)m-4\gamma+2\big)\Big)}\Bigg(-72\alpha\beta\sqrt{\frac{3e^{-\mathcal{T}}}{\Gamma_{0}}}\Big(-\frac{1}{3}\sqrt{2}e^{-\mathcal{T}}\times\nonumber\\&\!&\!\times\big(\alpha\beta\gamma(m+2)\mathcal{T}^{\beta}-(m+\frac{1}{2})(\gamma-\frac{1}{2})\mathcal{T}\big)+\gamma(m+2)\mathcal{T}(-\frac{\xi_{0}(m+2)\sqrt{2}}{24}+\sqrt{\frac{\Gamma_{0}e^{\mathcal{T}}}{1+2m}})\Big)\Bigg),
\label{a7}
\end{eqnarray}
\begin{equation}
R_{0}=\frac{2\sqrt{\Gamma_{0}e^{-9\mathcal{T}}\alpha^{7}\beta^{7}(m+2)(1+2m)^{3}}\mathcal{T}^{\frac{7}{2}(\beta-1)}}{3T_{r}}\exp{\bigg(-\frac{(-1)^{\beta}3^{\beta-1}2^{\frac{1-2\beta}{2}}\sqrt{\frac{\Gamma_{0}}{1+2m}}\Gamma(2-\beta,-\frac{3}{2}\mathcal{T})}{\alpha\beta}\bigg)}\coth[{\frac{k_{0}}{2T}}],
\label{a8}
\end{equation}
where
\begin{equation}
\mathcal{T}=\bigg(\Big(\frac{3}{(m+2)\alpha\beta}\Big)^{\frac{\beta}{\beta-1}}+\frac{3N}{(m+2)\alpha}\bigg)^{\frac{1}{\beta}}.
\label{a9}
\end{equation}
\subsubsection{$\Gamma$, $\xi$ as variable}
\begin{eqnarray}
&\!&\!n_{s}=\frac{e^{-\frac{5\mathcal{T}}{2}}}{\alpha\beta(m+2)}\Bigg\{\bigg(\Big(-4(m+2)\alpha(\gamma^{2}e^{\frac{2\mathcal{T}}{2}}+e^{\mathcal{T}}(\gamma-\frac{1}{4})\xi(m+\frac{1}{2}))\beta\mathcal{T}^{1+\beta}+(m+2)^{2}(2\gamma e^{\mathcal{T}}+\xi(m+\frac{1}{2}))\alpha^{2}\beta^{2}\xi\mathcal{T}^{2\beta}+\nonumber\\&\!&\!
+2\gamma(\gamma-\frac{1}{2}\mathcal{T}^{2}e^{2\mathcal{T}}(m+\frac{1}{2}))\Big)\bigg)^{-1}\times\bigg(-\frac{27}{2}\sqrt{\frac{3e^{\mathcal{T}}\mathcal{T}^{2-2\beta}}{1+2m}}\Big(-\frac{1537}{288}e^{\mathcal{T}}(m+2)\alpha^{3}\beta^{3}\xi(\gamma-\frac{768}{1537}(m+ \frac{1}{2})\mathcal{T}^{-1+3\beta}+\frac{769}{288}\times\nonumber\\&\!&\!
\times(m+2)^{2}\alpha^{4}\beta^{4}\xi^{2}(m+\frac{1}{2})\mathcal{T}^{-2+4\beta}-4(m+2)\alpha((\gamma-\frac{1}{4})\xi(m+\frac{1}{2})e^{3\mathcal{T}}+\frac{2}{3}\gamma^{2}e^{4\mathcal{T}})\beta\mathcal{T}^{1+\beta}+\alpha^{2}\beta^{2}((m+2)^{2}\xi^{2}+\frac{8}{3}\gamma^{2}-\nonumber\\&\!&\!-\frac{8}{3}\gamma)(m+\frac{1}{2})e^{2\mathcal{T}}+\frac{8}{3}\gamma e^{3\mathcal{T}}\xi(m+2)^{2}\mathcal{T}^{2\beta}+2\gamma(\gamma-\frac{1}{2})e^{4\mathcal{T}}\mathcal{T}^{2}(m+\frac{1}{2})\Big)-24\gamma e^{-\frac{9}{2}\mathcal{T}}(\gamma-\frac{1}{2})(m+\frac{1}{2})\mathcal{T}^{3-\beta}+(m+2)\times\nonumber\\&\!&\!
\times\alpha\beta\Big(-4\xi e^{\frac{7}{2}\mathcal{T}}(m+2)\alpha((\gamma-\frac{1}{4})m+\frac{13}{2}\gamma-\frac{1}{8})\beta\mathcal{T}^{1+\beta}-\frac{1}{2}\beta^{2}\gamma\alpha^{2}(m+2)^{2}\mathcal{T}^{2\beta}-12\mathcal{T}^{2}(\gamma-\frac{1}{4}(m+\frac{1}{2})\Big)-2\gamma e^{\frac{9}{2}\mathcal{T}}\times\nonumber\\&\!&\!
\times\big((m + 2)\gamma\alpha\beta\mathcal{T}^{1+\beta}-\mathcal{T}^{2}((\gamma-\frac{1}{2})m+\frac{25}{2}\gamma-\frac{1}{4})\big)+(m+2)\alpha e^{\frac{5}{2}\mathcal{T}}\beta(-12\mathcal{T}^{1+\beta}+(m+2)\beta\alpha\mathcal{T}^{2\beta})\xi^{2}(m+\frac{1}{2})\bigg)\Bigg\},
\label{a10}
\end{eqnarray}
\begin{eqnarray}
R(k_{0})=\frac{\coth[{\frac{k_{0}}{2T}}]}{(6m+3)\beta\alpha T_{r}}\Bigg(2(m+2)\mathcal{T}^{1-\beta}\sqrt{\frac{(1+2m)^{6}\alpha^{11}\beta^{11}e^{-11\mathcal{T}}\mathcal{T}^{-11+11\beta}}{m+2}}\exp{\bigg((2\gamma(m+2))^{-1}\Big(-36(m+\frac{1}{2})\beta^{2}\alpha^{2}9^{-\beta}(m+2)\xi\Gamma(2\beta-1,3\mathcal{T})+22\gamma(m+2)\ln{(e^{\frac{\mathcal{T}}{2}})}+12(m+\frac{1}{2})\beta\alpha(-1+\gamma)2^{-\beta}\Gamma(\beta,2\mathcal{T})+33\ln{(2)}\gamma(m+2)\Big)\bigg)}\Bigg),
\label{a11}
\end{eqnarray}
where
\begin{equation}
\mathcal{T}=\bigg(\Big(\frac{3}{(m+2)\alpha\beta}\Big)^{\frac{\beta}{\beta-1}}+\frac{3N}{(m+2)\alpha}\bigg)^{\frac{1}{\beta}}.
\label{a12}
\end{equation}
\subsection{Exponential model}
\begin{eqnarray}
&\!&\!n_{s}=\frac{(4\alpha\mathcal{T}(m+2))^{-1}}{\Big(\xi(\xi(m+\frac{1}{2})\beta +2\gamma)\beta\mathcal{T}^{2}(m+2)^{2}\alpha^{2}-4\mathcal{T}(m+2)((\gamma-\frac{1}{4})\xi(m+\frac{1}{2})\beta+\frac{\gamma^{2}}{2})\alpha+2(m+\frac{1}{2})(\gamma-\frac{1}{2})\gamma\Big)}\times\nonumber\\&\!&\!
\times\Bigg(-144\sqrt{\frac{3}{(1+2m)\beta}}\bigg(-\gamma\sqrt{\beta}\big(\gamma\mathcal{T}(m+2)\alpha-\frac{3}{4}(m+\frac{1}{2})(\gamma-\frac{1}{2})\big)+\beta^{\frac{3}{2}}\alpha\mathcal{T}\Big(\xi^{2}(m+\frac{1}{2})\beta^{3}\mathcal{T}^{3}(m+2)^{2}\alpha^{3}-\nonumber\\&\!&\!-2\xi(m+\frac{1}{2})\beta^{2}\mathcal{T}^{2}(\gamma-\frac{1}{2})(m+2)\alpha^{2}+\frac{3}{8}\mathcal{T}\alpha(((m+2)^{2}\xi^{2}+\frac{8}{3}\gamma^{2}-\frac{8}{3}\gamma(m+\frac{1}{2})\beta+\frac{8}{3}\xi\gamma(m+2)^{2})- \frac{3}{2}(\gamma-\frac{1}{4})\times\nonumber\\&\!&\!
\times\xi(m+\frac{1}{2})(m+2))\Big)\bigg)+4(-9+\alpha(m+2)\mathcal{T})\Big(\xi(\xi(m+\frac{1}{2})\beta+2\gamma)\beta\mathcal{T}^{2}(m+2)^{2}\alpha^{2}-4\mathcal{T}(m+2)((\gamma-\frac{1}{4})\xi\times\nonumber\\&\!&\!
\times(m+\frac{1}{2})\beta+\frac{\gamma^{2}}{2})\alpha+2(m+\frac{1}{2})(\gamma-\frac{1}{2})\gamma\Big)\Bigg),
\label{a13}
\end{eqnarray}
\begin{equation}
R(k_{0})=\frac{2(m+2)\sqrt{\frac{(1+2m)^{6}\alpha^{11}\beta^{11}\mathcal{T}^{11}}{m+2}}}{(3+6m)\alpha\mathcal{T}^{\frac{13}{2}}T_{r}}\exp{\Big(-\frac{3\mathcal{T}(\alpha\beta\xi\mathcal{T}(m+2)-2\gamma+2)(m+\frac{1}{2})\beta^{2}\alpha}{\gamma(m+2)}\Big)}\coth[{\frac{k_{0}}{2T}}],
\label{a14}
\end{equation}
where
\begin{equation}
\mathcal{T}=\frac{3(N+1)}{\alpha(m+2)}.
\label{a15}
\end{equation}
\end{document}